\documentclass[12pt,draftclsnofoot,onecolumn]{IEEEtran}
\newcommand\Deltabar{\underline{\Delta}}
\newcommand\sbar{\underline{s}}
\newcommand\bbar{\underline{b}}

\newcommand\R{\mathrm{R}}
\newcommand{\ones}{\mathbf 1}
\newcommand{\reals}{{\mbox{\bf R}}}
\newcommand{\integers}{{\mbox{\bf Z}}}
\newcommand{\symm}{{\mbox{\bf S}}}  % symmetric matrices

\newcommand{\nullspace}{{\mathcal N}}
\newcommand{\range}{{\mathcal R}}
\newcommand{\Rank}{\mathop{\bf Rank}}
\newcommand{\Tr}{\mathop{\bf Tr}}
\newcommand{\diag}{\mathop{\bf diag}}
\newcommand{\card}{\mathop{\bf card}}
\newcommand{\rank}{\mathop{\bf rank}}
\newcommand{\conv}{\mathop{\bf conv}}
\newcommand{\prox}{\mathbf{prox}}

\newcommand{\Expect}{\mathop{\bf E{}}}
\newcommand{\Prob}{\mathop{\bf Prob}}
\newcommand{\Co}{{\mathop {\bf Co}}} % convex hull
\newcommand{\dist}{\mathop{\bf dist{}}}
\newcommand{\epi}{\mathop{\bf epi}} % epigraph
\newcommand{\Vol}{\mathop{\bf vol}}
\newcommand{\dom}{\mathop{\bf dom}} % domain
\newcommand{\intr}{\mathop{\bf int}}
\newcommand{\sign}{\mathop{\bf sign}}

\newcommand{\cf}{{\it cf.}}
\newcommand{\eg}{{\it e.g.}}
\newcommand{\ie}{{\it i.e.}}
\newcommand{\etc}{{\it etc.}}
\usepackage{dsfont}
\usepackage{moreverb}
\usepackage{epsfig}
\usepackage{amsmath,amssymb,amsthm,mathrsfs,amsfonts,dsfont}
\usepackage{adjustbox,lipsum}
\usepackage{algorithm}
\usepackage{algpseudocode}
\usepackage{amsfonts}
\usepackage{amssymb}
\usepackage{amsmath}
\usepackage{amsthm}
\usepackage{subfigure}
\usepackage{multirow}
\usepackage{rotating}
\usepackage{graphicx}
\usepackage{tabularx}
\usepackage{array}
\usepackage{anyfontsize}
\usepackage{color,soul}
\usepackage{graphicx,dblfloatfix}
\usepackage{blindtext}
\usepackage{amsmath}
\usepackage{amsthm,amssymb,amsmath,bm}
\usepackage{amsfonts}
\usepackage{amssymb}
\usepackage{amsmath}
\usepackage{cite}
\usepackage{graphicx}
\usepackage{fancyhdr}
\usepackage[font={small}]{caption}
\usepackage{tabularx}
\usepackage{tcolorbox}
\usepackage{cite}
\usepackage{soul}
\usepackage{setspace}
\usepackage{enumitem}

\DeclareMathOperator*{\argmax}{arg\,max}
\DeclareMathOperator*{\argmin}{arg\,min}
\newtheorem{theorem}{Theorem}
\newtheorem{lemm}{Lemma}
\newtheorem{Definition}{Definition}
\newtheorem{remark}{Remark}
\newtheorem{Pro}{Proposition}

\def\blue{\textcolor{blue}}
\def\red{\textcolor{red}}
\def\green{\textcolor{green}}
\def\black{\textcolor{black}}
\def\brown{\textcolor{brown}}
\def\yellow{\textcolor{yellow}}
\def\orange{\textcolor{orange}}

\usepackage[hidelinks]{hyperref}

\allowdisplaybreaks

\graphicspath{{Figure/}}
\usepackage{pifont}
\newcommand{\cmark}{\ding{51}}%
\newcommand{\xmark}{\ding{55}}%

\begin{document}

% \title{{(1-SubTCOM) AoI Minimization in Status Update Control with Energy Harvesting Sensors}\\
% {(2) Age of Information Minimization in a Status Update IoT Network under Transmission and Energy Limitations}\\
% {\red{**(3) On-Demand AoI Minimization in an IoT Energy Harvesting Network under Transmission Limitation **--**Multi-sensor Multi-user**}}}
 
% 1) On-Demand AoI Minimization in an IoT Energy Harvesting Network under Transmission Limitation \brown{--**Multi-sensor Multi-user**}
% \\
% 2)
 
%\title{On-Demand AoI Minimization in a Multi-sensor Multi-user IoT Energy Harvesting Network with Transmission Constraint }

%\title{On-Demand AoI Minimization in an IoT Network under Energy Harvesting and Transmission Limitations}

\title{On-Demand AoI Minimization in Resource-Constrained Cache-Enabled IoT Networks with Energy Harvesting Sensors
}

%% Original, Conference style 
\iffalse
\author{\IEEEauthorblockN{Mohammad Hatami$^1$, Markus Leinonen$^1$, and Marian Codreanu$^2$}\\
\IEEEauthorblockA{$^{1}$Centre for Wireless Communications, University of Oulu, Finland \\
$^{2}$Department of Science and Technology, Link\"{o}ping University, Sweden \\
Email:  mohammad.hatami@oulu\hspace{0.125em}.fi, markus.leinonen@oulu\hspace{0.125em}.fi, marian.codreanu@liu\hspace{0.125em}.se\\
}}
\fi

%% Modified, Journal style 
\author{Mohammad Hatami\IEEEauthorrefmark{1}, %~\IEEEmembership{Student Member,~IEEE,} 
Markus Leinonen\IEEEauthorrefmark{1}, %~\IEEEmembership{Member,~IEEE,}
Zheng Chen\IEEEauthorrefmark{2}, %,~\IEEEmembership{Member,~IEEE}
Nikolaos Pappas\IEEEauthorrefmark{2}, %,~\IEEEmembership{Member,~IEEE}
and Marian Codreanu\IEEEauthorrefmark{2} %,~\IEEEmembership{Member,~IEEE}
\thanks{
\IEEEauthorrefmark{1}%Mohammad Hatami and Markus Leinonen are with 
Centre for Wireless Communications -- Radio Technologies, University of Oulu, Finland. e-mail: firstname.lastname@oulu.fi%mohammad.hatami@oulu.fi, markus.leinonen@oulu.fi.

\IEEEauthorrefmark{2}%Marian Codreanu is with 
Department of Science and Technology, Link\"{o}ping University, Sweden. e-mail: firstname.lastname@liu.se%marian.codreanu@liu.se.

%Preliminary results of this work were presented in \cite{Hatami-etal-20}. (We use this kind of sentence if your conference paper were presented) <-- This is not true now; not presented yet. You have the ArXiv preprint though, which is "published/publicly available". 

%\yellow{The work has been financially supported in part by Infotech Oulu, the Academy of Finland (grant 323698), and Academy of Finland 6Genesis Flagship (grant 318927). The work of M. Leinonen has also been financially supported in part by the Academy of Finland (grant 340171 and 319485). M. Codreanu would like to acknowledge the support of the European Union's Horizon 2020 research and innovation programme under the Marie Sk\l{}odowska-Curie Grant Agreement No. 793402 (COMPRESS NETS).}
}
}

% make the title area
\maketitle

% As a general rule, do not put math, special symbols or citations in the abstract
\vspace{-5mm} %% COMPRESSION TWEAK
\begin{spacing}{1.33} %%COMPRESSION TWEAK
\begin{abstract}
We consider a resource-constrained IoT network, where multiple users make on-demand requests to a cache-enabled edge node to send status updates about various random processes, each monitored by an energy harvesting sensor. The edge node serves users' requests by deciding whether to command the corresponding sensor to send a fresh status update or retrieve the most recently received measurement from the cache. Our objective is to find the best actions of the edge node to minimize the average age of information (AoI) of the received measurements upon request, i.e., \textit{average on-demand AoI}, subject to \textit{per-slot transmission and energy constraints}. First, we  derive a Markov decision process model and propose an iterative algorithm that obtains an optimal policy. Then, we develop an \textit{asymptotically optimal low-complexity} algorithm -- termed \textit{relax-then-truncate} -- and prove that it is optimal as the number of sensors goes to infinity. Simulation results illustrate that the proposed relax-then-truncate approach significantly reduces the average on-demand AoI compared to a request-aware greedy (myopic) policy and also depict that it performs close to the optimal solution even for moderate numbers of sensors.
\end{abstract}
\end{spacing}%%COMPRESSION TWEAK
\vspace{-2mm} 

%\red{To serve users' requests, the edge node \st{uses the system state information to} decides whether to command the corresponding sensor to send a fresh status update or to retrieve the most recently received measurement from the cache.}

%\begin{IEEEkeywords}
%Internet of Things (IoT), age of information (AoI), energy harvesting, reinforcement learning (RL), value iteration, dynamic programming, Q-learning.
%\end{IEEEkeywords}

\sloppy

% \IEEEpeerreviewmaketitle
% \newpage
% \brown{**Meeting's Discussions can be written here:**}

% \newpage
% Cont.

% \newpage

%%%%%%%%%%%%%%%%%%%%%%%%%%%%%%%%%%%%%%%%%%%%
\section{Introduction}
Internet of Things (IoT) is a key technology in providing ubiquitous, intelligent networking solutions to create a smart society. In IoT sensing networks, the sensors measure physical quantities (e.g., speed or pressure) and send the measurements to a destination for further processing. IoT networks are subject to stringent energy limitations, due to battery-powered sensors. This energy scarcity is often counteracted by \textit{energy harvesting} (EH) technology, relying on, e.g., solar or RF ambient sources. Moreover, reliable control actions in emerging time-critical IoT applications (e.g., drone control and industrial monitoring) require high \textit{freshness} of information received by the destination. Such destination-centric information freshness can be quantified by the \textit{age of information} (AoI) \cite{AoI_Orginal_12,sun2019age}. These call for designing effective \textit{AoI-aware status updating} procedures for IoT networks to provide the end users with timely status of remotely observed processes while accounting for the limited energy resources of EH sensors.

We consider a resource-constrained IoT sensing network that consists of multiple EH sensors, a cache-enabled edge node, 
% \yellow{(a gateway)},
and multiple users. Users are interested in timely status information about the random processes associated with physical quantities (e.g., speed or temperature), each measured by a sensor. We consider \textit{request-based} status updating where the users demand for the status of physical quantities from the edge node which acts as a gateway between the users and the sensors. The edge node is equipped with a \textit{cache} that stores the most recently received \textit{status update packet} from each sensor. Upon receiving request(s) for the status of a physical quantity, the edge node has two options to serve the requesting user(s): {either} command the corresponding sensor to send a fresh measurement, i.e., a status update packet, {or} use the aged measurement from the cache. The former enables the edge node to serve the user(s) with fresh measurements, yet consuming energy from the sensor's battery. The latter prevents the activation of the sensors for every request so that the sensors can utilize a sleep mode to save considerable amount of energy \cite{niyato2016novel}, but the data received by the users becomes stale. Thus, there is an inherent \textit{trade-off} between the AoI at the users and conservation of the sensors' energy in the {finite} batteries.

In particular, the considered status updating network is subject to the following \textit{energy and transmission constraints}.
First, since the sensors rely only on the energy harvested from the environment, the sensors' batteries may be empty and thus they cannot send an update for each request. This \textit{energy causality} induces an inherent \textit{per-slot energy constraint}.  
% so that the sensors cannot send an update at every slot, i.e., \textit{energy causality constraint}.
% \st{Second, finite battery capacity for each EH sensor induces another constraint on the available energy at the sensors.}
Second, motivated by the limited amount of radio resources (e.g., bandwidth, time-frequency resource blocks), only a limited number of sensors can send fresh status updates to the edge node at each time slot, imposing a \textit{per-slot transmission constraint}.

The objective of our network design is to keep the {freshness of information} at the users as small as possible, subject to the constraints in the system.
{To this end, we use the concept of \textit{on-demand AoI} \cite{hatami2020aoi} that quantifies the freshness of information at the users restricted to the users' request instants.}
% To quantify the \st{AoI seen} \blue{freshness of information} at the users, we use the concept of \textit{on-demand AoI} \cite{hatami2020aoi} that measures the freshness of information at the users restricted to the users' request instants.
% The \textit{on-demand AoI} metric is defined to quantify the AoI seen at the users, i.e., the freshness of information at the users restricted to the users' request instants \cite{hatami2020aoi}. 
% \brown{Here, we extended the on-demand AoI to account for the number of requests at each time slot.} 
We aim to find an \textit{optimal policy}, i.e., the best action of the edge node at each time slot that minimizes the average on-demand AoI over all the sensors and the users subject to {the per-slot transmission and energy constraints}.
%\st{the energy causality constraint at the sensors and the per-slot transmission constraint}.
%--- Difference between the AoI optimization and on-demand AoI minimization
It is worth emphasizing that the on-demand AoI minimization is different from the conventional AoI optimization in that the freshness of information \textit{is only important when user(s) request the information}, i.e., an optimal policy for the on-demand AoI minimization problem adapts to the request pattern.
% , while in traditional AoI minimization, it does not depend on the request process.
% The conventional AoI optimization is an special case of the on-demand AoI minimization formulation whenever the value of each physical quantity is requested by (only) one user at each time slot.

We first cast the problem as a Markov decision process (MDP) and propose an 
% iterative
algorithm that obtains an optimal policy. Moreover, since the complexity of finding an optimal policy increases exponentially in the number of sensors, we propose an asymptotically optimal low-complexity algorithm -- termed \textit{relax-then-truncate} --
and show that it performs close to the optimal solution.

\subsection{Contributions}
The main contributions of our paper are as follows:
\begin{itemize}
    \item We consider on-demand AoI minimization problem in a multi-user multi-sensor IoT EH network subject to per-slot transmission and energy constraints. We formulate the problem as an MDP and propose an iterative algorithm that finds an optimal policy. 
    % However, the complexity of finding an optimal policy increases exponentially in the the number of sensors. 
    \item To deal with massive IoT scenarios, we propose a sub-optimal low-complexity algorithm whose complexity increases linearly in the number of sensors. In particular, we relax the per-slot transmission constraint into a time average constraint, model the relaxed problem as a constrained MDP (CMDP), obtain an optimal relaxed policy, and propose an online truncation procedure to ensure that the transmission constraint is satisfied at each time slot.
    \item  We analytically find an upper bound for the difference between the average cost obtained by the proposed {relax-then-truncate} approach and the average cost obtained by an optimal policy. Then, we show that the relax-then-truncate approach is asymptotically optimal as the number of sensors goes to infinity.
    \item  Numerical experiments are conducted to analyze the performance of the proposed relax-then-truncate approach and show that it significantly reduces the average on-demand AoI as compared to a request-aware greedy  policy. Interestingly, the proposed algorithm performs close to the optimal solution for moderate numbers of sensors.
\end{itemize}

Our considered model is highly relevant to resource-constrained IoT scenarios with a massive number of devices. \textit{To the best of our knowledge, this work is the first one that proposes an asymptotically optimal low-complexity algorithm for minimizing on-demand AoI in an IoT network with multiple EH sensors}.

\subsection{Related Work}\label{sec_related_works}
% Scheduling problem for age-optimal status update systems 
AoI-optimal scheduling has attracted a significant amount of interest from the research community over the last few years \cite{Hsu_modiano2020aoimultiuser_tcm,Kadota_Modiano2018Broadcast,Maatouk2021OptimalityWhittle,kriouile2021global,Ceran2019HARQ,Ceran2021MultiUserCMDP,tang2020CMDPTRUNCATE,zakeri2021minimizing,bacinoglu2015age_oneunitenergy,ceran2021learningEH,tunc2019optimal,leng2019AoIcognitive,Zheng2021OptHet,Stamatakis2019control,Elvina2021SourceDiversityEH,AbdElmagid2020AoIWPC,Zhao2020StatusCorEH,hatami2020aoi,Hatami-etal-20,hatami2021spawc,yin2019only,Li2021waiting,holm2020freshness,chiariotti2021query}.
Particularly, a popular approach
% A commonality in these works
is to model the problem as an MDP and find an optimal policy by using model-based reinforcement  learning (RL) methods based on dynamic programming \cite{Hsu_modiano2020aoimultiuser_tcm,Ceran2019HARQ,Ceran2021MultiUserCMDP,tang2020CMDPTRUNCATE,zakeri2021minimizing,tunc2019optimal,leng2019AoIcognitive,Stamatakis2019control,Elvina2021SourceDiversityEH,AbdElmagid2020AoIWPC,hatami2020aoi,holm2020freshness,chiariotti2021query}, e.g., relative value iteration algorithm (RVIA), and/or model-free RL methods \cite{Ceran2019HARQ,Ceran2021MultiUserCMDP,ceran2021learningEH,Zhao2020StatusCorEH,hatami2020aoi,Hatami-etal-20}, e.g., (deep) Q-learning. 

% \red{**Whittle index for RMAB **}

% \brown{--- ** MDP/CMDP + AoI** ---}
% {The authors of \cite{zhou2019joint} derived optimal sampling and updating policies that minimize the average AoI at the destination under an average energy constraint in an IoT monitoring system.} 
In \cite{Hsu_modiano2020aoimultiuser_tcm}, the authors proposed AoI-optimal scheduling algorithms for a broadcast network where a base station is updating the users on random information arrivals under a transmission capacity constraint.
In \cite{Kadota_Modiano2018Broadcast}, the authors developed low-complexity scheduling algorithms, including a Whittle's index policy, and derived performance guarantees for a broadcast network.
% with a base station sending information to the clients through unreliable channels. 
In \cite{Maatouk2021OptimalityWhittle,kriouile2021global}, the optimality of the Whittle's index policy has been investigated for the AoI minimization problem where a central entity schedules a number of users among the total available users for transmission over unreliable channels.
% \brown{It was shown  that the Whittle's index policy has analytically provable optimality  for the AoI minimization problem where a central entity schedules a number users among the total available users for transmission over unreliable channels. In \cite{Maatouk2021OptimalityWhittle}, the authors established the optimality in the neighborhood of a specific system’s state and assume that the number of states is finite. In \cite{kriouile2021global}, the authors provided a new mathematical approach to establish the optimality in the many-users regime for specific network settings. Their approach is based on intricate techniques, and unlike previous works in the literature, it is free of any mathematical assumptions.}
In \cite{Ceran2019HARQ}, the authors studied AoI-optimal scheduling under a constraint on the average number of transmissions 
% at the source node
where the source sends status updates to a destination (user) over an error-prone channel. 
% \blue{with retransmissions}.
The authors in \cite{Ceran2021MultiUserCMDP} extended \cite{Ceran2019HARQ} to a multi-user setting,
% The work \cite{Ceran2019HARQ} has been extended to a multi-user setting in \cite{Ceran2021MultiUserCMDP},
where the source has to decide not only when to transmit but also to which user.
% \brown{Scheduling the transmission of time-sensitive information from a source node to multiple users over error-prone communication channels is studied with the goal of minimizing the long-term average age of information (AoI) at the users. A long-term average resource constraint is imposed on the source, which limits the average number of transmissions.} 
In \cite{tang2020CMDPTRUNCATE}, the authors proposed an asymptotically optimal algorithm for the AoI-optimal scheduling problem under both bandwidth and average power constraints in a wireless network with time-varying channel states.  In \cite{zakeri2021minimizing}, the authors studied AoI minimization problem in a multi-source relaying system under per-slot transmission and average resource constraints.

% zhou2019joint,
Different from \cite{Ceran2019HARQ,Ceran2021MultiUserCMDP,tang2020CMDPTRUNCATE,Kadota_Modiano2018Broadcast,Maatouk2021OptimalityWhittle,kriouile2021global,Hsu_modiano2020aoimultiuser_tcm,zakeri2021minimizing}, another line of research 
\cite{bacinoglu2015age_oneunitenergy,ceran2021learningEH,tunc2019optimal,leng2019AoIcognitive,Zheng2021OptHet,Stamatakis2019control,Elvina2021SourceDiversityEH,AbdElmagid2020AoIWPC,Zhao2020StatusCorEH}
focused on the class of problems where the sources are powered by \textit{energy harvested from the environment}, i.e., investigating AoI-optimal scheduling policies subject to the energy causality constraint at the source(s).
% , under various system settings. 
% \brown{The objective of this line of research was to investigate age-optimal offline/online policies for update packet transmissions subject to the energy causality constraint at the source under various assumptions regarding the battery size, transmission time of update packets and channel modeling.}
The works \cite{bacinoglu2015age_oneunitenergy,ceran2021learningEH,tunc2019optimal,leng2019AoIcognitive,Zheng2021OptHet,Stamatakis2019control,Elvina2021SourceDiversityEH,AbdElmagid2020AoIWPC} studied AoI-optimal scheduling in single-sensor EH networks
where the sensor is sending time-sensitive information to the user(s). 
The authors of \cite{bacinoglu2015age_oneunitenergy} derived age-optimal sampling instants for the sensor by assuming known EH statistics.
% Particularly, the work \cite{bacinoglu2015age_oneunitenergy} assumed known EH statistics.
In \cite{ceran2021learningEH}, the authors studied AoI-optimal policies under an erasure channel with retransmissions where the channel and EH statistics are either known or unknown.
% The  work \cite{ceran2021learningEH} investigated age-optimal scheduling policies for an EH sensor transmitting time-sensitive data over a noisy channel to a receiver.
In \cite{tunc2019optimal}, AoI-optimal scheduling was studied where the sensor takes advantage of multiple available transmission modes.
The work \cite{leng2019AoIcognitive} investigated AoI-optimal scheduling in a cognitive radio EH system. 
% In \cite{Zheng2021OptHet}, the authors studied age-optimal scheduling under stability constraints in a two-user multiple access channel \blue{where a grid-connected node and an EH node transmit updates to a common destination.}
%
{In \cite{Zheng2021OptHet}, the authors studied age-optimal scheduling under stability constraints in a multiple access channel with two heterogeneous nodes (including an EH node) transmitting to a common destination.}
% The first node is connected to a power grid and it has randomly arriving data packets. Another energy harvesting (EH) sensor monitors a stochastic process and sends status updates to the destination.} 
%
In \cite{Stamatakis2019control}, {the sensor monitors a stochastic process and tracks its evolution and thereby,  a modified definition of AoI is proposed to account for the discrepancy in the remote destination.} 
% \st{the sensor monitors a stochastic process that is in either a normal or an alarm state of operation}.
In \cite{Elvina2021SourceDiversityEH}, the monitoring node (sensor) collects status updates from multiple heterogeneous information sources.
{In \cite{AbdElmagid2020AoIWPC}, the authors studied AoI-optimal scheduling for a wireless powered communication system under the costs of generating status updates at the sensor nodes.}
In \cite{Zhao2020StatusCorEH}, the authors developed a deep RL algorithm for minimizing the average age of correlated information in an IoT network with multiple correlated EH sensors whose status updates are processed by a data fusion center.

% \st{Majority of the existing works, including all the above ones, investigate the AoI minimization in cases where the updates are relevant to the monitoring entity at all time moments.}
{The majority of the literature that considers} AoI minimization, including all the above ones, assume that the time-sensitive information of the source(s) is needed at the destination \textit{at all time moments}.
% is \textit{always} useful at the destination.
However, in many applications, {a user demands for fresh status updates only when it needs such timely information}. To account for such information freshness driven by users' requests, we introduced the concept of on-demand AoI in \cite{hatami2020aoi,Hatami-etal-20}. {In these works and a follow-up work \cite{hatami2021spawc},} the main focus was on-demand AoI minimization in an IoT network with multiple \textit{decoupled} EH sensors, {as opposed to tackling the transmission-constrained status updating problem herein.}
% \yellow{In \cite{Hatami-etal-20,hatami2020aoi}, we introduced a concept of on-demand AoI and developed MDP-based learning methods to find optimal policies that minimize the average on-demand AoI in an IoT EH network.}
%Here, we extend the on-demand AoI definition to account for the number of requests for each physical quantity at each time slot, incorporate the request information as a part of the state to improve the average on-demand AoI, and investigate optimal scheduling under per-slot transmission constraints.
Only a few works have studied a concept similar to the on-demand AoI. In \cite{yin2019only}, the authors introduced the idea of effective AoI (EAoI) under a generic request-response model where a server serves the users with time-sensitive information. 
%\st{They elaborated on the fact that minimizing the time average EAoI is in general different from minimizing the time average AoI.}
In \cite{Li2021waiting}, the authors studied an information update system where a user pulls information from servers.
However, in contrast to our paper, the works \cite{yin2019only,Li2021waiting} do not consider energy limitation at the source nodes.
% \blue{In \cite{pappas2020average}, the authors analyzed the average AoI in a cache enabled status updating system where external requests arrive for status of a remote source, which is monitored by an EH sensor.}
In \cite{holm2020freshness,chiariotti2021query}, the authors introduced the AoI at query (QAoI) and developed an MDP-based policy iteration method to find an optimal policy that minimizes the average QAoI considering an energy-constrained sensor that is queried to send updates to an edge node under limited transmission opportunities. 
% \blue{The model in \cite{chiariotti2021query} extends the framework that was presented in \cite{holm2020freshness}, which only considered a simple scenario
% with a constant channel and periodic queries.}
The QAoI metric \cite{holm2020freshness,chiariotti2021query} is equivalent to our on-demand AoI when particularized to the single-user single-sensor case.
\section{System Model and Problem Formulation}\label{sec_systemmodel}

% \brown{*The whole system model parts (Sections A---C) might need some changes in the writing style -- Too similar to our TCOM paper*}

% System model: 
%\textbf{\textcolor{red}{we denote $<$something$>$ by $<$notation$>$ or Let $<$ notation$>$ denote $<$something$>$}}

%%%%%%%%%%%%%%%%%%%
\subsection{Network Model}\label{sec_network}
\begin{figure}[t!]
\centering
\includegraphics[width=.63\columnwidth]{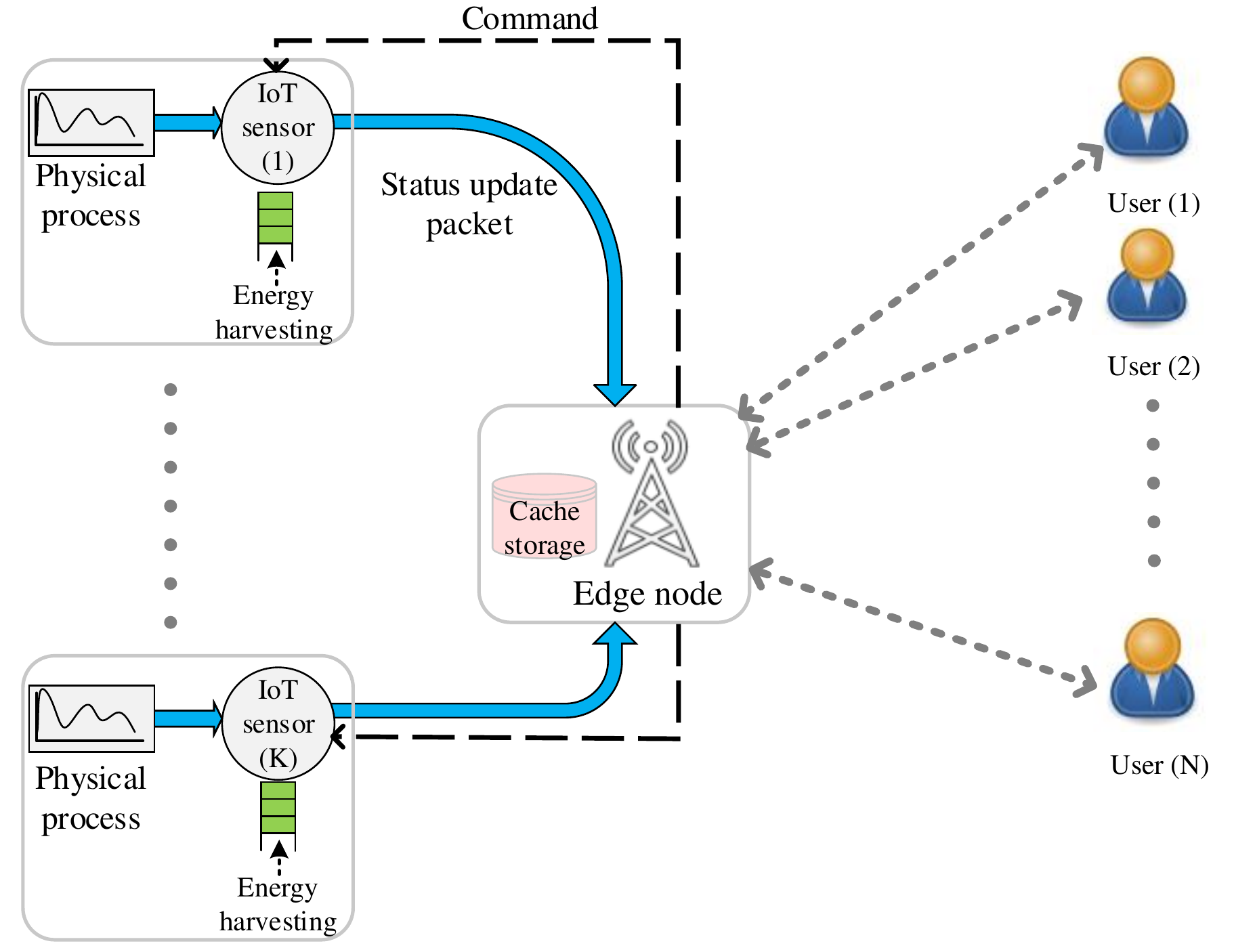}
\vspace{-3mm}
\caption{A multi-user multi-sensor IoT sensing network consisting of $K$ EH sensors, an edge node,
%\yellow{(i.e., the gateway)},
and $N$ users. The end users are interested in timely status update information of the physical processes measured by the sensors.}
% \footnote{I do not know yet how to add the resource limitation in the figure. Is that even  necessary?!}
\label{fig_systemmodel}
\vspace{-7mm}
\end{figure}

% \begin{figure}[h]
% \centering
% \includegraphics[width=.55\columnwidth]{fig_systemmodel5.pdf}\vspace{-3mm}
% \caption{\red{*Will be merged into Fig1*}--\red{**If it is used the channel must be removed in the figure**} The link between {each sensor} and the edge node consists of an error-free binary command link from the edge node to each sensor and an error-prone wireless communication link from each sensor to the edge node.}\vspace{-7mm}
% \label{fig_systemmodel2}
% \end{figure}

%% Network operation
We consider a multi-user multi-sensor IoT sensing network that consists of a set ${\mathcal{K}=\{1,\ldots,K\}}$ of $K$ energy harvesting (EH) sensors, an edge node (a gateway), and a set ${\mathcal{N}=\{1,\ldots,N\}}$ of $N$ users, as depicted in Fig.~\ref{fig_systemmodel}. Users are interested in timely status information about random processes associated with physical quantities $f_k$, e.g., speed or temperature, each of which is independently measured by sensor $k\in\mathcal{K}$. 
% \yellow{To keep the users updated about the status of each $f_k$, the sensors send \textit{status update packets}, each containing the measured value of the monitored process and a time stamp representing the time when the sample was generated.} 
We consider \textit{request-based} status updating, where the users send requests \textit{on demand} for obtaining status of quantities $f_k$, ${k\in\mathcal{K}}$.
{When a request for the physical quantify $f_k$ is generated at the user side, the associated sensor $k$ may send a \textit{status update packet} that contains the measured value of the monitored process and a time stamp of the generated sample.}
% \blue{\st{, in an \textit{on-demand} fashion}}.
We assume that there is no direct link between the users and the sensors, i.e., the users receive the status updates only via the edge node.

%% Time slot, Request model 
We consider a time-slotted system with slots indexed by ${t \in \mathbb{N}}$. At the beginning of slot $t$, users send requests for the status of physical quantities $f_k$ to the edge node. Let $r_{k,n}(t) \in \{0,1\}$, $t=1,2,\dots$, denote the random process of requesting the 
% \st{value} 
status of $f_k$ by user $n$; $r_{k,n}(t) = 1$ if the status  of $f_k$ is requested by user $n \in \mathcal{N}$ at slot $t$ and  $r_{k,n}(t)=0$ otherwise. The requests are independent across the users, sensors, and time slots. Let $p_{k,n}$ be the probability that the  status of $f_k$ is requested by user $n$ at each slot, i.e., $\mathrm{Pr}\{r_{k,n}(t) = 1\} = p_{k,n}$. Note that there can be multiple users requesting for $f_k$ at each slot; $r_k(t) = \sum_{n=1}^{N}r_{k,n}(t) \in \{0,1,\ldots,N\}$ indicates the number of requests for $f_k$ at slot $t$. We assume that all requests that arrive at the beginning of slot $t$ are handled by the edge node during the same slot $t$. To this end, we assume that all the communication links, i.e., the sensor-edge and edge-user links, are error-free\footnote{The main focus of this paper is to devise effective status updating procedures in a multi-user multi-sensor IoT network under both \textit{energy and transmission limitations}. Thus, the possibly error-prone communication links are left for future study.}.

%% Edge node, cache, command action 
% \st{To facilitate the status updating procedure,}
The edge node is equipped with a \textit{cache} of size $K$ that stores the most recently received status update packet from each sensor. Upon receiving a request for the status of $f_k$ %\st{by at least one user} 
at slot $t$, the edge node has two options to serve the request: 1) command sensor $k$ to send a fresh status update, or 2) use the previous measurement from the cache.
% \begin{enumerate}
% \item command sensor $k$ to send a {fresh} status update,
% \item use the previous measurement from the cache. 
% \end{enumerate}
Let $a_k(t) \in \{0,1\}$ be the \textit{command action of the edge node} at slot $t$; $a_k(t)=1$ if the edge node commands sensor $k$ to send an update and $a_k(t)=0$ otherwise.

%%% Transmission constraint
We consider that, due to limited amount of radio resources (e.g.,
% \st{bandwidth,}
time-frequency resource blocks), no more than $M\leq K$ sensors can transmit status updates to the edge node {within} each slot. This \textit{transmission constraint} imposes a limitation to the number of commands as
%the edge node can command no more than  $M\leq K$ sensors at each time slot\blue{, i.e.,}
%\red{\st{. This imposes a constraint, and also \textit{inter-sensor coupling}, on the command actions as}} 
\begin{equation}\label{eq_bw_st}
    \textstyle\sum_{k=1}^{K}a_k(t) \leq M,\,\forall{t}.
\end{equation}
%This effectively acts as a \textit{transmission constraint} for the communications between the sensors and the edge node in the system, where $M$ \blue{represents the maximum number of sensors that can send status updates at each time slot}, called \textit{transmission budget} hereinafter.
{We refer to $M$ as the \textit{transmission budget} hereinafter.}

% We assume that all the requests that arrive at the beginning of slot $t$ are handled during the same slot $t$. 
% We also assume that the sensors have independent communication links to the edge node. Accordingly, at each slot $t$, the command actions $a_k(t)$, $k\in\mathcal{K}$, are independent across $k$.
% \yellow{Note that while the communications between the edge node and the users are assumed to be  error-free, the transmissions from the sensors to the edge node are prone to errors\footnote{Typically, the edge node accesses to sufficient power 
%(e.g., a base station connected to a fixed power grid), whereas the sensors rely only on the energy harvested from the environment.}, as detailed in Section~\ref{Comm_link}.}}}

%% Energy harvesting, battery evolution, sensor's action 
% The sensors rely on the energy harvested from the environment. Sensor $k$ stores the harvested energy into a battery of finite size $B_k$ (units of energy). Let $b_k(t)$ denote the battery level of sensor $k$ at the beginning of slot $t$. Thus, ${b_k(t) \in \{0,\ldots,B_{k}\}}$.

%%%%%%%%%%%%%%%%%%%
\subsection{Energy Harvesting Sensors}\label{EH_model}
%% EH in general, define battery "bkt" 
We assume that the sensors harvest energy from the environment for sustainable operation. 
We model the energy arrivals at the sensors as independent Bernoulli processes with intensities $\lambda_k$, $k\in\mathcal{K}$. 
This characterizes the discrete nature of the energy arrivals in a slotted-time system, i.e., at each slot, a sensor either harvests one unit of energy or not (see, e.g., \cite{bacinoglu2015age_oneunitenergy,Stamatakis2019control,pappas2020average}).
Let $e_k(t) \in \left\lbrace 0 ,1\right\rbrace $, $t = 1,2,\dots$, denote the \textit{energy arrival process} of sensor $k$. Thus, the probability that sensor $k$ harvests one unit of energy during one slot is $\lambda_k$\footnote{The scenario where the sensors access to sufficient power (e.g., connected to a fixed power grid) so that they can send a status update at every slot is a special case of our system model, i.e., a unit energy arrival at every slot, $\lambda_k = 1$.}, i.e., ${\Pr\{ e_k(t) = 1 \} = \lambda_k}$, $\forall t$.
Sensor $k$ stores the harvested energy into a battery of finite size $B_k$ (units of energy). Formally, let $b_k(t)$ denote the battery level of sensor $k$ at the beginning of slot $t$, where ${b_k(t) \in \{0,\ldots,B_{k}\}}$.

%% Limited energy, define "dkt". Define cost of transmission. 
We assume that transmitting a status update from each sensor to the edge node consumes one unit of energy (see, e.g., \cite{hatami2020aoi,bacinoglu2015age_oneunitenergy,Stamatakis2019control,Zhao2020StatusCorEH,pappas2020average}).
% (see e.g., \cite{arafa2019timely,bacinoglu2015age_oneunitenergy,wu2017optimal_oneunitenergy,arafa2019age,michelusi2013transmission}).
Once sensor $k$ is commanded by the edge node (i.e., $a_k(t)=1$), sensor $k$ sends a status update if it has at least one unit of energy (i.e., $b_k(t) \geq 1$). Formally, let random variable $d_k(t) \in \left\lbrace 0 ,1\right\rbrace$ denote the \textit{action of sensor $k$} at slot $t$; $d_k(t)=1$ if sensor $k$ sends a status update to the edge node and $d_k(t)=0$ otherwise. {Accordingly, the sensor's action, the edge node's action, and the sensor's battery level are interrelated as}
%the action of sensor $k$ (i.e., $d_k(t)$) and the command action of the edge node (i.e., $a_k(t)$) can be expressed as
\begin{equation}\label{eq_d}
d_k(t) =  a_k(t) \mathds{1}_{\{b_k(t) \geq 1\}},
\end{equation}
where $\mathds{1}_{\{\cdot\}}$ is the indicator function. Note that $d_k(t)$ in \eqref{eq_d} characterizes the energy consumption of sensor $k$ at slot $t$. It is also worth noting that by \eqref{eq_d}, we have $d_k(t)\le{a_k(t)}$, and consequently, \eqref{eq_bw_st} implies that
$\sum_{k=1}^{K}d_k(t) \leq M$ for all slots; hence, the name transmission constraint for \eqref{eq_bw_st}. 
% \red{We remark that sending status updates by sensor $k$ is subject to the energy causality constraint that reads as
% \begin{equation}\label{eq_energy_causality}
%     d_k(t) \leq b_k(t).
% \end{equation}
% }

% \brown{The energy causality constraint that should be satisfied can be expressed as}

%%%% Now I should have bkt, dkt, TX-cost, everything for the EH equation.
%% EH model: 1) define ekt. 2) define battery evolution f(bkt,dkt,TXcost,ekt)  
% \red{\st{We model the energy arrivals at the sensors as independent Bernoulli processes with intensities $\lambda_k$, $k\in\mathcal{K}$. 
% This characterizes the discrete nature of the energy arrivals in a slotted-time system, i.e., at each time slot, a sensor either harvests one unit of energy or not (see, e.g.,}} \cite{Stamatakis2019control}).
% \red{\st{Let $e_k(t) \in \left\lbrace 0 ,1\right\rbrace $, $t = 1,2,\dots$, denote the \textit{energy arrival process} of sensor $k$. Thus, the probability that sensor $k$ harvests one unit of energy during one time slot is $\lambda_k$, i.e., ${\Pr\{ e_k(t) = 1 \} = \lambda_k}$, $\forall t$.}} 
% \footnote{\blue{It is worth noting that the case without energy harvesting (always charged sensor) can be a special case of our model, where there is always one energy arrival in every slot.}}
%\st{The energy harvested during slot $t$ can be used only in a later slot}

Finally, using the defined quantities $b_k(t)$, $d_k(t)$, and $e_k(t)$, the evolution of the battery level of sensor $k$ is expressed as
\begin{equation}\label{eq_battery_evo}
b_k(t+1) = \min\left\lbrace  b_k(t)+e_k(t)-d_k(t) , B_k \right\rbrace.
\end{equation}
\subsection{On-demand Age of Information}\label{sec_AoI}
To measure the freshness of information seen by the users in our request-based status updating system, we use the notion of age of information (AoI) \cite{AoI_Orginal_12} and define \textit{on-demand AoI} 
%\st{that measures the AoI seen at the users} 
\cite{hatami2020aoi}.
% The AoI is a destination-centric metric that quantifies the freshness of information of a remotely observed random process.
% Since we consider a request-based status updating system, we need an appropriate metric to quantify the freshness of information at the users. 
% By using the notion of age of information (AoI), which is a destination-centric metric that quantifies the freshness of information of a remotely observed random process \cite{AoI_Orginal_12}, we define \textit{on-demand AoI} that measures the AoI seen at the users \cite{hatami2020aoi}.
In contrast to AoI that measures the freshness of information at every slot, on-demand AoI quantifies \textit{the freshness of information at the users' request instants (only).}
%\st{accounts for timely information at the users restricted to the users' request instants}.

Let  $\Delta_k(t)$ be the AoI about the physical quantity $f_k$ at the edge node at the beginning of slot  $t$, i.e., the number of slots elapsed since the generation of the most recently  received status update packet from sensor $k$.
Let $u_k(t)$ denote the most recent slot in which the edge node received a status update packet from sensor $k$, i.e., $u_k(t) = \max \{t'| t'<t, d_k(t') = 1 \}$. Thus, the AoI about $f_k$ is given by the random process
% \begin{equation}\label{eq_AoI_def}
$\Delta_k(t) \triangleq t - u_k(t)$.
% \end{equation}
%\brown{{More precisely, $\Delta_k(t) = t - u_k(t)$, where $u_k(t)$ represents the most recent time slot in which the edge node received a status update packet from sensor $k$, i.e., $u_k(t) = \max \{t'| t'<t, h_k(t') = 1 \}$.}}

We make a common assumption (see e.g., \cite{hatami2020aoi,Maatouk2021OptimalityWhittle,Ceran2019HARQ,Ceran2021MultiUserCMDP,zakeri2021minimizing,ceran2021learningEH,tunc2019optimal,leng2019AoIcognitive, Zhao2020StatusCorEH, Elvina2021SourceDiversityEH, AbdElmagid2020AoIWPC, Stamatakis2019control,Hatami-etal-20}) that $\Delta_k(t)$ is upper-bounded by a finite value $\Delta^{\mathrm{max}}$, i.e., $\Delta_k(t) \in \{1, 2,\ldots ,\Delta^{\mathrm{max}}\}$. 
%This is reasonable, because once $\Delta_k(t)$ reaches a high value $$\Delta^{\mathrm{max}}$$, the available measurement about {quantity} $f_k$ becomes excessively stale/expired, so further counting would be irrelevant.
{Besides tractability, this accounts for the fact that once the available measurement about $f_k$ becomes excessively stale, further counting would be irrelevant.}
%When $\Delta_k(t)$ reaches $\Delta_{k,\mathrm{max}}$, it means that the available measurement  about process $k$ at the edge node is too stale/irrelevant to be of any use.}
At each slot, the AoI about $f_k$ drops to one if the edge node receives a status update from the corresponding sensor, or otherwise, it increases by one. Accordingly, the evolution of  $\Delta_k(t)$ can be written as
\begin{equation}\label{eq_AoI}
\Delta_k(t+1)=
\begin{cases}
1,&\text{if} ~d_k(t)=1, \\
\min \{\Delta_k(t)+1,\Delta^{\mathrm{max}}\},&\text{if}~d_k(t)=0,
\end{cases}
\end{equation}
{which can be expressed compactly as  $\Delta_k(t+1)=\min \{ (1-d_k(t)) \Delta_k(t)+1,\Delta^{\mathrm{max}}\}$.}

% \brown{I think it's better to define the on-demand AoI here than in the next subsection.}

% \blue{The AoI definition in \eqref{eq_AoI_def} measures the freshness of information at every slot, whereas, since we consider a request-based status updating, we need an appropriate metric that accounts for the freshness of information at the users restricted to the users' request instants.}
We define on-demand AoI for a sensor-user pair $(k,n)$ at slot $t$ as the sampled version of \eqref{eq_AoI} where the sampling is controlled by the request process $r_{k,n}(t)$, i.e., 
\begin{equation}\label{eq-persensor-peruser_cost}
{\Delta^\mathrm{OD}_{k,n}(t) \triangleq r_{k,n}(t) \Delta_k(t+1) = r_{k,n}(t) \min \{ (1-d_k(t)) \Delta_k(t)+1,\Delta^{\mathrm{max}}\} }.
\end{equation}
% Note that owing to the multiplicative factor $r_{k,n}(t)$, the on-demand AoI definition in \eqref{eq-persensor-peruser_cost} considers the freshness of information whenever it is requested by a user, i.e., \eqref{eq-persensor-peruser_cost} is a sampled version of \eqref{eq_AoI_def} where the sampling is controlled by the request process $r_{k,n}(t)$.
{In \eqref{eq-persensor-peruser_cost}, since the requests come at the beginning of slot $t$ and the edge node sends {measurements} to the users at the end of the same slot, $\Delta_k(t+1)$ is the AoI about $f_k$ seen by the users.}

% In \eqref{eq-persensor-peruser_cost}, $\Delta_k(t+1)$ is the AoI about $f_k$ seen by the requesting user $n$, because the requests 
% % \st{come at the beginning of slot $t$}
% occur at the beginning of each slot and the edge node sends status updates to the requesting users by the end of the same slot.

% \brown{Moreover, since the requests come at the beginning of slot $t$ and the edge node sends values to the users at the end of the same slot, $\Delta_k(t+1)$ is the effective AoI about $f_k$ seen by the users.}

% Since the requests come at the beginning of slot $t$ and the edge node sends values to the users by the end of the same slot, thus, $\Delta_k(t+1)$ is the effective AoI about $f_k$ seen by the requesting user $n$.

% given $r_{k,n}(t) = 1$, i.e., the value of $f_k$ is requested by user $n$ at slot $t$.}
%%%%%%%%%%%%%%%%%%%%%%%%%%%%%%%
\subsection{State Space, Action Space, Policy, and Cost Function} \label{sec_state_action_cost_def}

\subsubsection{State}{Let $s_k(t) \in \mathcal{S}_k$ denote the state associated with sensor $k$
% \st{(the per-sensor state)}
at slot $t$, which is defined as $s_k(t) = \left(r_k(t),b_k(t), \Delta_k(t)\right)$, where $r_k(t) \in \{0,1,\ldots,N\}$ indicates {the number of requests} for $f_k$, $b_k(t) \in \{0, 1,\ldots, B_k\}$ is the battery level, and $\Delta_k(t) \in \{1,2,\ldots,\Delta^{\mathrm{max}}\}$ is the AoI about $f_k$ at the edge node; $\mathcal{S}_k$ is the per-sensor state space %which contains all the combinations of the state variables
with dimension $ |\mathcal{S}_k| = (N+1)(B_k+1)\Delta^{\mathrm{max}}$. The state of the system at slot $t$ is expressed as $\mathbf{s}(t) = \left(s_1(t), \dots, s_K(t)\right) \in \mathcal{S}$, $\mathcal{S} = \mathcal{S}_1 \times \dots \times \mathcal{S}_K$; the state space $\mathcal{S}$ has a finite dimension $|\mathcal{S}| = \prod_{k =1}^{K}(N+1)(B_k+1)\Delta^{\mathrm{max}}$.}

\subsubsection{Action}\label{sec-action-def}
As discussed in Section~\ref{sec_network}, the edge node decides at each slot whether to command sensor $k$ to send a fresh status update {(and update the cache)} or not, i.e., $a_k(t) \in \mathcal{A}_k =  \{0,1\}$, where $\mathcal{A}_k$ is the per-sensor action space. The action of the edge node at slot $t$ is given by a $K$-tuple
%expressed as 
$\mathbf{a}(t) = \big(a_1(t), \dots, a_K(t)\big) \in \mathcal{A}$ with action space
${\mathcal{A}=\big\{(a_1,\ldots,a_K) \mid a_k\in \mathcal{A}_k
,\;\sum_{k=1}^{K}a_k\le{M}\big\}}$. The action space dimension is $|\mathcal{A}| = \sum_{m = 0}^M \binom{K}{m}$.
It is worth stressing that the action space $\mathcal{A}$ considers the transmission constraint \eqref{eq_bw_st} in its definition. Additionally, we define the \textit{relaxed} action space that does not consider the transmission constraint \eqref{eq_bw_st} as $\mathcal{A}_{\mathrm{R}} = \mathcal{A}_1 \times \cdots \times \mathcal{A}_K = \{0,1\}^K$, which has the dimension $|\mathcal{A}_{\mathrm{R}}| = 2^K$.
% { 
% $\tilde{\mathcal{A}}$
% $\underline{\mathcal{A}}$
% $\hat{\mathcal{A}}$
% $\mathcal{A}_{\mathrm{Relaxed}} = \mathcal{A}_1 \times \dots \times \mathcal{A}_K$ }
%%%%%%%%%%%%%%%%%%%

%%%%%%%%%%%%%%%%%%%%%%%%%%%
\subsubsection{Policy}\label{sec_policy_def}
A policy $\pi$ is a rule that determines the action by observing the state.
Particularly, {a randomized policy is a mapping from state $\mathbf{s} \in \mathcal{S}$ to a \textit{probability distribution} ${\pi(\mathbf{a}|\mathbf{s}) : \mathcal{S} \times \mathcal{A} \rightarrow \left[ 0 , 1\right]}$, $\sum_{\mathbf{a} \in \mathcal{A}} \pi(\mathbf{a}|\mathbf{s}) = 1$,  of choosing each possible action $\mathbf{a} \in \mathcal{A}$.} A deterministic policy is a special case of the randomized policy where, in each state $\mathbf{s}$, $\pi(\mathbf{a}|\mathbf{s}) = 1$ for some $\mathbf{a}$; with a slight abuse of notation, we use $\pi(\mathbf{s})$ to denote the action taken in state $\mathbf{s}$  by a deterministic policy $\pi$.
% A deterministic policy $\pi \in \Pi$ is a mapping from a state space $\mathcal{S}$ to the action space $\mathcal{A}$, i.e., $\pi: \mathcal{S} \rightarrow \mathcal{A}$, where the set $\Pi$ includes all possible $\pi$; the action taken in state $s \in \mathcal{S}$ under policy $\pi$ is denoted by $\pi(s)$.
In addition, we define a (relaxed) policy as $\pi_{\mathrm{R}}: \mathcal{S} \times \mathcal{A}_{\R} \rightarrow \left[ 0 , 1\right]$ and a per-sensor policy as $\pi_k: \mathcal{S}_k \times  \mathcal{A}_k \rightarrow \left[ 0 , 1\right]$. 
% Note that $\Pi \subseteq \Pi_{\mathrm{R}}$**}

% \brown{the class of station-
% ary randomized and stationary deterministic policies, respec-
% tively.}

%%%%%%%%%%%%%%%%%%%%%%%%%%%
\subsubsection{Cost Function}\label{sec_cost}
% \st{Since the users demand for fresh status data as per their requests,} 
We consider a cost function that incurs a penalty with respect to the staleness of a status update requested and received by a user.
Accordingly, we define the cost associated with user $n$ and sensor $k$ at slot $t$ as the {on-demand AoI} for the sensor-user pair $(k,n)$, i.e., $\Delta^{\mathrm{OD}}_{k,n}(t)$ defined in \eqref{eq-persensor-peruser_cost}.
Then, we define the per-sensor cost at slot $t$ as 
\begin{equation}\label{persensor_cost}
c_k(t) =  \textstyle\sum_{n=1}^{N} \Delta^{\mathrm{OD}}_{k,n}(t) = \sum_{n=1}^{N} r_{k,n}(t) \Delta_k(t+1) = r_{k}(t) \Delta_k(t+1).
\end{equation}
\begin{remark}
% \vspace{-2mm}
\normalfont
Note that, due to the multiplicative factor $r_k(t)$, \eqref{persensor_cost} accounts for the number of requests for each physical quantity at each slot, i.e., the more the requests for $f_k$, the more important the corresponding freshness becomes.
% also reveals the fundamental difference between the traditional AoI metric and the on-demand AoI metric.
Particularly, when the status of $f_k$ is not requested by any user at slot $t$, i.e., $r_k(t) = 0$, the immediate cost becomes $c_k(t) = 0$.
% \vspace{-4mm}
% {Moreover, since the requests come at the beginning of slot $t$ and the edge node sends values to the users by the end of the same slot, $\Delta_k(t+1)$ is the effective AoI about $f_k$ seen by the users.} 
% \brown{The conventional/traditional AoI optimization is an special case of the above formulation whenever $r_k(t) = 1$,~$\forall t$.}
\end{remark}
\subsection{Problem Formulation}\label{subsec_problem_formulation}
For the considered system, the energy and transmission constraints pose limitations on when and how often a new status update can be generated at each sensor, which in turn affect the on-demand AoI.
% \st{ AoI seen at the users upon requesting for updates about status information}.
Our objective is to keep the on-demand AoI as small as possible, subject to the constraints in the system. 
% \st{Note that the on-demand AoI minimization is different from the conventional AoI optimization in that the freshness of information \textit{is only important when user(s) request the information}, i.e., an optimal policy is able to adapt to the request pattern, while in traditional AoI minimization, it does not depend on the request process.}

%%%%%%%%%%%%%%%%%%%%%%%%

% {A policy, denoted by $\pi$, is a mapping from the state space to the action space, i.e., $\pi: \mathcal{S} \rightarrow \mathcal{A}$; the action taken in state $s$ under policy $\pi$ is denoted by $\pi(s)$.}
Formally, for a given policy $\pi$, we define {the average cost as \textit{the average on-demand AoI over all sensors and users,}
% \red{-- the average cost --} as
i.e., }
\begin{equation}\label{eq_average_cost}
\bar{C}_{\pi} \triangleq \lim_{T\rightarrow\infty} \frac{1}{NKT}\sum_{t=1}^{T}\sum_{k=1}^{K} \mathbb{E}_{\pi}[c_{k}(t) \mid \mathbf{s}(0)],
\end{equation}
where $\mathbb{E}_{\pi}[\cdot]$ is the (conditional) expectation when the policy $\pi$ is applied to the system
% where $\mathbb{E}_{\pi}[\cdot]$ denotes the expectation conditioned that the edge node follows policy $\pi$
%the expected value of $c_k(t)$ given that the edge node follows policy $\pi$, 
and $\mathbf{s}(0) = \big(s_1(0),\ldots,s_K(0)\big)$ is the initial state\footnote{It is shown in Proposition~\ref{lemma_Bellman} that the minimum average cost is independent of the initial state,
% notational simplicity
thus, we omit the initial state henceforth.}. We aim to find an optimal policy $\pi^\star$ that achieves the minimum average cost, i.e., 
\begin{equation}\label{average_cost_sp1}
(\textbf{P1})~~\pi^\star \in~ \textstyle{\argmin_{\pi}}~\bar{C}_{\pi}.
\end{equation}

% \brown{**Here we can also briefly explain what are the technical challenges for solving this problem**.}
% We aim to find  the best action of the edge node at each time slot, i.e., $a(t)$, $t = 1,2,\ldots$, called an \textit{optimal policy}, that minimizes the long-term average cost, defined as
% \begin{equation}\label{eq_average_cost_persensor}
% \bar{C}=\lim_{T\rightarrow\infty} \frac{1}{T}\textstyle\sum_{t=1}^{T} \mathbb{E} [c(t)],
% \end{equation}
% where the expectation is taken over all system dynamics. 

% \red{** Check the format of the optimization problem**}
% Accordingly, the optimization problem with the per-slot constraint can be written as
% \begin{equation}\label{average_cost_sp1}
% (\textbf{P1})~~\pi^* =  \argmin_{\pi}~\bar{C}_{\pi}.
% \end{equation}

% \red{**The above constraint in \eqref{st_opt1} can be written as $\sum_{k = 1}^{K} \pi_k(s_k(t)) \leq M,~\forall{t}$. **}

% where $\{a_k(t)\}_{k\in\mathcal{K}}$, $t=1,2,\dots$ are the optimization variables.

%%%%%%%%%%%%%%%%%%%%%%%%%%%%%%%
\section{MDP Modeling and Optimal Policy}\label{sec-MDP-optimalpolicy}

In this section, we model the problem (\textbf{P1}) as an MDP and propose a value iteration algorithm that finds an optimal policy $\pi^\star$.
% and  
%propose a \red{**structural-based**} value iteration algorithm that finds such an optimal policy 
% find an optimal policy using \textit{relative value iteration algorithm (RVIA)}. Particularly, t

%%%%%%%%%%%%%%%%%
\subsection{MDP Modeling}\label{sec-MDP}
% \sloppy We model the problem (\textbf{P1}) as an MDP that is defined by
\sloppy {The MDP is defined by the tuple} ${\big(\mathcal{S}, \mathcal{A},\Pr(\mathbf{s}(t + 1)|\mathbf{s}(t), \mathbf{a}(t)), c(\mathbf{s}(t), \mathbf{a}(t))\big)}$,
% whose elements are explained below.
where the state space $\mathcal{S}$ and the action space $\mathcal{A}$ were defined in Section~\ref{sec_state_action_cost_def}.
% \brown{The state transition probability $\mathcal{P}_k(s_k(t+1) \mid s_k(t), a_k(t))$ maps a state-action pair at slot $t$ onto a distribution of states at  slot $t + 1$; the probability of transition from current state {($s_k(t)$) $s =\{r, b, \Delta\}$} to next state ($s_k(t+1)$) $s^\prime = \{r^\prime, b^\prime , \Delta^\prime\}$  under action $a_k(t) = a$ is given by}
% \subsubsection{State Transition Probability}
\sloppy The cost function $c(\mathbf{s}(t),\mathbf{a}(t))$ represents the cost of taking action
$\mathbf{a}(t)$ in state $\mathbf{s}(t)$, which  is 
% calculated as
{given by} ${c(\mathbf{s}(t),\mathbf{a}(t)) = {\frac{1}{NK}} \sum_{k=1}^{K} c_k(s_k(t),a_k(t))}$, where the per-sensor cost $c_k(s_k(t),a_k(t))$ is calculated  using \eqref{persensor_cost}, {i.e., $c_k(s_k(t), a_k(t)) =  r_k(t) \min \left\{ \left(1-a_k(t)\mathds{1}_{\{b_k(t) \geq 1\}}\right) \Delta_k(t)+1,\Delta^{\mathrm{max}}\right\}$.}
% \yellow{Particularly, the per-sensor cost of taking action $a_k(t)$ in state $s_k(t) = \left(r_k(t),b_k,\Delta_k(t)\right)$ is given by  $c_k(s_k(t), a_k(t)) =  r_k(t) \min \left\lbrace \left(1-a_k(t)\mathds{1}_{\{b_k(t) \geq 1\}}\right) \Delta_k(t)+1,\Delta^{\mathrm{max}}\right\rbrace$.}
% \begin{equation}\label{eq_persensor_cost}
%  c_k(s_k(t), a_k(t)) =  r_k(t) \min \left\lbrace \left(1-a_k(t)\mathds{1}_{\{b_k(t) \geq 1\}}\right) \Delta_k(t)+1,\Delta^{\mathrm{max}}\right\rbrace.
% \end{equation}}
% \brown{ZHENG: Here you can add that we impose  an artificial constraint on the maximum age. When maximum value is large enough, it doesn't affect the solution of the problem.}
The state transition probability $\Pr(\mathbf{s}(t+1)|\mathbf{s}(t),\mathbf{a}(t))$ maps a state-action pair at slot $t$ onto a distribution of states at  slot $t + 1$. The probability of transition from
 current 
state $\mathbf{s}(t) = \left(s_1(t),\dots,s_K(t)\right)$ to 
 next
state $\mathbf{s}(t+1)= \left(s_1(t+1),\dots,s_K(t+1)\right)$ under action $\mathbf{a}(t) = \left(a_1(t),\dots,a_K(t)\right)$ 
is factorized as
\begin{equation}
    \Pr\left( \mathbf{s}(t+1) \mid \mathbf{s}(t), \mathbf{a}(t) \right)
    \overset{(a)}{=} \textstyle\prod_{k = 1}^{K} \Pr \left(s_k(t+1) \mid s_k(t), a_k(t) \right),
\end{equation}
where $(a)$ follows from the fact that given action $\mathbf{a}$, the state associated with each sensor (i.e., the per-sensor state) evolves independently  from the other sensors. Above, the per-sensor state transition probability $\Pr \left(s_k(t+1) \mid s_k(t), a_k(t) \right)$ gives the probability of transition from per-sensor state $s_k(t) = \left(r_k,b_k,\Delta_k\right)$ to next per-sensor state $s_k(t+1) = \left(r^\prime_k,b^\prime_k,\Delta^\prime_k\right)$ under action $a_k(t) = a_k$, and it is expressed as
\begin{equation}\label{eq_persensor_STP}
\begin{array}{lll}
    & \Pr \left(s_k(t+1) \mid s_k(t), a_k(t) \right )  \triangleq \Pr \left(r^\prime_k,b^\prime_k,\Delta^\prime_k \mid r_k,b_k,\Delta_k, a_k \right ) \\ & \overset{(a)}{=}  \underbrace{\Pr\left(r^\prime_k\mid r_k,b_k,\Delta_k, a_k\right)}_{\overset{(b)}{=} \Pr(r^\prime_k)} 
    \underbrace{\Pr \left(b^\prime_k\mid r_k, b_k,\Delta_k, a_k,r^\prime_k \right)}_{\overset{(c)}{=} \Pr \left(b^\prime_k\mid b_k, a_k \right)} \underbrace{\Pr \left(\Delta^\prime_k\mid r_k, b_k,\Delta_k, a_k,r^\prime_k,b^\prime_k \right)}_{\overset{(d)}{=} \Pr \left(\Delta^\prime_k \mid b_k, \Delta_k,a_k \right)} \\ & = \Pr(r^\prime_k) \Pr \left(b^\prime_k\mid b_k, a_k \right) \Pr \left(\Delta^\prime_k \mid b_k, \Delta_k,a_k \right),
\end{array}
\end{equation}
where $(a)$ follows from the chain rule, $(b)$ follows from the independence between the request process and the other random variables, $(c)$ follows because, given current battery level $b_k$ and action $a_k$, next battery level $b_k^\prime$ is independent of the requests and the current AoI, and $(d)$  follows since given $b_k$, $\Delta_k$, and $a_k$, the next value of AoI $\Delta^\prime_k$ can be obtained deterministically (see \eqref{eq_AoI}). The probabilities in {\eqref{eq_persensor_STP}} are calculated in the following.

The random variable $r^\prime_k = \sum_{n}r^\prime_{k,n}$ is a sum of independent Bernoulli trials that are not necessarily identically distributed. Therefore, it has a Poisson binomial distribution {\cite{Wang1993Poisson-Binomial}}
% ,Fernandez2010Poisson-Binomial}}
as
% \begin{subequations}
% \begin{small}
% \hspace{-5mm}
% \begin{array}{ll}
\begin{equation}\label{eq_persensor_STP_r}
    \Pr (r^\prime_k) = \left\lbrace 
    \begin{array}{ll}
    \prod_{n=1}^{N}{(1-p_{k,n}),} & r^\prime_k = 0,\\
    \sum_{n=1}^{N}p_{k,n}\prod_{m\neq n}{(1-p_{k,m}),} & r^\prime_k = 1,\\
    \ldots & \ldots\\
    \prod_{n=1}^{N}{p_{k,n},} & r^\prime_k = N,\\
    0, & \mbox{otherwise.}
    \end{array}
\right.
\end{equation}
At each slot, sensor $k$ consumes one unit of energy for sending a status update (i.e., when $a_k(t) = 1$ and $b_k(t) \geq 1$) and harvests one unit of energy with probability $\lambda_k$, thus, we have
% \begin{equation}\label{eq_persensor_STP_battery}
%     \Pr (b^\prime_k\mid  b_k < B_k , a_k = 0) = \left\lbrace 
%     \begin{array}{ll}
%     {\lambda_k}, & b^\prime_k = b_k + 1,\\{1-\lambda_k,} & b^\prime_k = b_k ,\\0, & \mbox{otherwise.}
%     \end{array}
%     \right.
% \end{equation}
\begin{equation}\label{eq_persensor_STP_battery}
\begin{array}{ll}
    &\Pr (b^\prime_k\mid  b_k < B_k , a_k = 0) = \left\lbrace 
    \begin{array}{ll}
    {\lambda_k}, & b^\prime_k = b_k + 1,\\{1-\lambda_k,} & b^\prime_k = b_k ,\\0, & \mbox{otherwise.}
    \end{array}
    \right.\\
    &\Pr (b^\prime_k\mid b_k = B_k , a_k = 0) = \mathds{1}_{\{b_k^\prime=B_k\}},\\
    &\Pr (b^\prime_k \mid b_k = 0 , a_k = 1) = \left\lbrace 
    \begin{array}{ll}
    {\lambda_k,} & b^\prime_k = 1,\\
    {1-\lambda_k,} & b^\prime_k = 0, \\
    0, & \mbox{otherwise.}
    \end{array}
\right.
\end{array}
\end{equation}
\begin{equation}
\begin{array}{ll}
&\Pr (b^\prime_k\mid  b_k\geq1 , a_k = 1) = \left\lbrace
    \begin{array}{ll}
    {\lambda_k,} & b^\prime_k = b_k,\\
    {1-\lambda_k,} & b^\prime_k = b_k - 1,\\
    0, & \mbox{otherwise.}
    \end{array}
\right.
\end{array}\notag
\end{equation}
% \begin{equation}
% \begin{array}{llll}
% &\Pr (b^\prime_k\mid  b_k < B_k , a_k = 0) = \left\lbrace 
%     \begin{array}{ll}
%     {\lambda_k,} & b^\prime_k = b_k + 1,\\
%     {1-\lambda_k,} & b^\prime_k = b_k ,\\
%     0, & \mbox{otherwise.}
%     \end{array}
% \right.\\
% &\Pr (b^\prime_k\mid b_k = B_k , a_k = 0) = \mathds{1}_{\{b_k^\prime=B_k\}},\\
% &\Pr (b^\prime_k \mid b_k = 0 , a_k = 1) = \left\lbrace 
%     \begin{array}{ll}
%     {\lambda_k,} & b^\prime_k = 1,\\
%     {1-\lambda_k,} & b^\prime_k = 0, \\
%     0, & \mbox{otherwise.}
%     \end{array}
% \right.\\
% &\Pr (b^\prime_k\mid  b_k\geq1 , a_k = 1) = \left\lbrace
%     \begin{array}{ll}
%     {\lambda_k,} & b^\prime_k = b_k,\\
%     {1-\lambda_k,} & b^\prime_k = b_k - 1,\\
%     0, & \mbox{otherwise.}
%     \end{array}
% \right.
% \end{array}
% \end{equation}
According to \eqref{eq_AoI} and \eqref{eq_d}, given current battery level $b_k$, AoI $\Delta_k$, and action $a_k$, the next value of AoI $\Delta^\prime_k$ can be obtained deterministically. Thus, we have
\begin{equation}\label{eq_persensor_STP_AoI}
\begin{array}{lll}
&\Pr (\Delta^\prime_k\mid  b_k, \Delta_k, a_k = 0) = \mathds{1}_{\{\Delta^\prime_k= \min\{\Delta_k+1,\Delta^{\mathrm{max}}\}\}},\\
&\Pr (\Delta^\prime_k \mid  \!b_k \geq 1 , \!\Delta_k, a_k \!=\! 1) \!=\! \mathds{1}_{\{\Delta^\prime_k= 1\}},\\
&\Pr (\Delta^\prime_k\mid  b_k = 0 , \Delta_k, a_k = 1 ) = \mathds{1}_{\{\Delta^\prime_k= \min\{\Delta_k+1,\Delta^{\mathrm{max}}\}\}}.
\end{array}
\end{equation}

\subsection{Optimal Policy}\label{sec-optimalpolicy}
% \brown{$*$ vs $\star$: which one is better? $\pi^\star$ OR $\pi^*$}
In this section, we propose an iterative algorithm that obtains an optimal policy $\pi^\star$ for (\textbf{P1}). To this end, we first define the accessibility condition for an MDP and prove that our MDP modeling in Section~\ref{sec-MDP} satisfies this condition. 
% Having this condition satisfied,
Then, we present a {proposition} that characterizes an optimal policy $\pi^\star$ for (\textbf{P1}). 
% Finally, we propose an iterative algorithm that obtains $\pi^\star$.

% An MDP is communicating if for any two states s, s there
% exists a deterministic policy π such that s is reachable from s
% with positive probability following π. A stronger concept is the
% unichain property, which we define for the more general class
% of CMDPs: a finite CMDP is unichain if any feasible policy
% (i.e., a policy that satisfies the resource constraint) induces a
% finite-state Markov chain that contains a single recurrent class
% and possibly, some transient states. We will show below that
% our MDP is communicating (cf. Theorem 1) and that it is
% unichain under the ARQ protocol (cf. Theorem 2).

% \brown{We say that the Weak Accessibility (WA for short) condition holds if the set of states can be partitioned into two subsets $\mathcal{S}_t$ and $\mathcal{S}_c$ such that: (a) all states in $\mathcal{S}_t$ are transient under every stationary policy and (b) For every two states $i$ and $j$ in $\mathcal{S}_t$, $j$ is accessible from $i$.}
% \brown{An MDP is weakly communicating if and only if there exists a randomized stationary policy which induces a Markov chain with a single closed irreducible class and a set of states which is transient under all stationary policies \cite[Ch.~8]{puterman2014markov},\cite{bertsekas2007dynamic}.}

\vspace{-2mm}
\begin{Definition}
\normalfont
An MDP is \textit{weakly communicating} (or \textit{weakly accessible}) if the set of states can be partitioned into two subsets $\mathcal{S}_\textrm{t}$ and $\mathcal{S}_\textrm{c}$ such that: (i) all states in $\mathcal{S}_\textrm{t}$ are transient under every stationary policy and (ii) every two states in $\mathcal{S}_\textrm{c}$ can be reached from each other under some stationary policy \cite[Definition 4.2.2]{bertsekas2007dynamic}. In particular, an MDP is communicating (or accessible) if every two states can be reached from each other under some stationary policy.
\end{Definition}\vspace{-2mm}
\vspace{-3mm}\begin{Pro}\label{theorem_communicating_mdp}
The MDP defined {in Section~\ref{sec-MDP}} 
%\st{by the tuple $(\mathcal{S}, \mathcal{A},\Pr(\mathbf{s}(t + 1)|\mathbf{s}(t), \mathbf{a}(t)), c(\mathbf{s}(t), \mathbf{a}(t)))$} 
is weakly communicating.
\end{Pro}\vspace{-5mm}
\begin{proof}
The proof is presented in Appendix~\ref{sec_appendix_theorem_communicating_mdp}.
\end{proof}\vspace{-2mm}
% The optimal average cost achieved by an optimal policy $\pi^\star$, denoted by $\bar{C}^\star$ {(i.e., $\bar{C}^\star = \bar{C}_{\pi^\star}$), is independent of the initial state $\mathbf{s}(0)$ and satisfies the following Bellman's equation
% \begin{equation}\label{eq_ballman}
%   \bar{C}^\star + h(\mathbf{s}) = \min_{\mathbf{a} \in \mathcal{A}} \left[c(\mathbf{s},\mathbf{a}) + \textstyle\sum_{\mathbf{s}^\prime \in \mathcal{S}} \Pr(\mathbf{s}^\prime|\mathbf{s},\mathbf{a}) h(\mathbf{s}^\prime)\right],~ \mathbf{s}\in \mathcal{S},
% \end{equation}
% where $h(\mathbf{s})$ is the relative value function.
% Then, an optimal action taken in state $\mathbf{s}$ is expressed as $\pi^\star(\mathbf{s}) = \argmin_{\mathbf{a}\in\mathcal{A}} \left[c(\mathbf{s},\mathbf{a}) + \textstyle\sum_{\mathbf{s}^\prime \in \mathcal{S}} \Pr(\mathbf{s}^\prime|\mathbf{s},\mathbf{a}) h(\mathbf{s}^\prime)\right]$.}
\vspace{-2mm}
\begin{Pro}\label{lemma_Bellman}
The optimal average cost achieved by an optimal policy $\pi^\star$, denoted by $\bar{C}^\star$ (i.e., $\bar{C}^\star = \bar{C}_{\pi^\star}$), is independent of the initial state $\mathbf{s}(0)$ and
satisfies the Bellman's equation, i.e., there exists $h(\mathbf{s})$, $\mathbf{s} \in \mathcal{S}$, such that 
\begin{equation}\label{eq_ballman}
  \bar{C}^\star + h(\mathbf{s}) = \min_{\mathbf{a} \in \mathcal{A}} \left[c(\mathbf{s},\mathbf{a}) + \textstyle\sum_{\mathbf{s}^\prime \in \mathcal{S}} \Pr(\mathbf{s}^\prime|\mathbf{s},\mathbf{a}) h(\mathbf{s}^\prime)\right],~ \mathbf{s}\in \mathcal{S}.
\end{equation}
Further, an optimal action taken in state $\mathbf{s}$ is given by 
\begin{equation}
{\pi^\star(\mathbf{s}) \in \argmin_{\mathbf{a}\in\mathcal{A}} \left[c(\mathbf{s},\mathbf{a}) + \textstyle\sum_{\mathbf{s}^\prime \in \mathcal{S}} \Pr(\mathbf{s}^\prime|\mathbf{s},\mathbf{a}) h(\mathbf{s}^\prime)\right]},~ \mathbf{s}\in \mathcal{S}.
\end{equation}
% where $\bar{C}^\star$ is the optimal average cost achieved by an optimal policy $\pi^\star$ (i.e., $\bar{C}^\star = \bar{C}_{\pi^\star}$) and $\bar{C}^\star$ is independent of the initial state $\mathbf{s}(0)$. In addition, an optimal action taken in state $\mathbf{s}$ is expressed as $\pi^\star(\mathbf{s}) = \argmin_{\mathbf{a}\in\mathcal{A}} \left[c(\mathbf{s},\mathbf{a}) + \textstyle\sum_{\mathbf{s}^\prime \in \mathcal{S}} \Pr(\mathbf{s}^\prime|\mathbf{s},\mathbf{a}) h(\mathbf{s}^\prime)\right]$
\end{Pro}
% The optimal policy π∗ exists and is obtained by
% \vspace{-2mm}\begin{Pro}\label{lemma_Bellman}
% The optimal average cost achieved by an optimal policy $\pi^\star$, denoted by $\bar{C}^\star$ {(i.e., $\bar{C}^\star = \bar{C}_{\pi^\star}$)}, is independent of the initial state $\mathbf{s}(0)$ and satisfies the Bellman's equation\blue{, i.e., there exist $h(s)$, $s \in \mathcal{S}_k$ such that }
% \begin{equation}\label{eq_ballman}
%   \bar{C}^\star + h(\mathbf{s}) = \min_{\mathbf{a} \in \mathcal{A}} \left[c(\mathbf{s},\mathbf{a}) + \textstyle\sum_{\mathbf{s}^\prime \in \mathcal{S}} \Pr(\mathbf{s}^\prime|\mathbf{s},\mathbf{a}) h(\mathbf{s}^\prime)\right],~ \mathbf{s}\in \mathcal{S},
% \end{equation}
% where $h(\mathbf{s})$ is \blue{a relative value function}.
% % and $q^\star(\mathbf{s},\mathbf{a})$ is the optimal action-value function, which is given by
% % \begin{equation}
% %     {q^\star(\mathbf{s},\mathbf{a}) = c(\mathbf{s},\mathbf{a}) + \textstyle\sum_{\mathbf{s}^\prime \in \mathcal{S}} \Pr(\mathbf{s}^\prime|\mathbf{s},\mathbf{a}) \blue{h}(\mathbf{s}^\prime).}
% % \end{equation}
% Then, an optimal action taken in state $\mathbf{s}$ is expressed as $\pi^\star(\mathbf{s}) = \argmin_{\mathbf{a}\in\mathcal{A}} \left[c(\mathbf{s},\mathbf{a}) + \textstyle\sum_{\mathbf{s}^\prime \in \mathcal{S}} \Pr(\mathbf{s}^\prime|\mathbf{s},\mathbf{a}) h(\mathbf{s}^\prime)\right]$.
% \end{Pro}
\vspace{-5mm}
\begin{proof}
By Proposition~\ref{theorem_communicating_mdp}, the weak accessibility condition holds and thus, the proof follows directly from \cite[Prop. 4.2.1]{bertsekas2007dynamic} and \cite[Prop. 4.2.3]{bertsekas2007dynamic}.
\end{proof}\vspace{-2mm}

% \newpage

% Blank
% \newpage

% The optimal state-value function $v^\star(\mathbf{s})$, $\mathbf{s}\in \mathcal{S}$, and consequently, an optimal policy $\pi^\star$, can be computed by turning the Bellman's optimality equation \eqref{eq_ballman} into an iterative procedure, called relative value iteration algorithm (RVIA). Particularly, the state-value function in state $\mathbf{s}$ at iteration $i = 1,2,\ldots$, denoted by $v^{(i)}(\mathbf{s})$, is updated as
% \begin{equation}\label{eq_v_itr}
% v^{(i)}(\mathbf{s}) =\min_{\mathbf{a}\in\mathcal{A}} q^{(i)}(\mathbf{s},\mathbf{a}) - v^{(i-1)}(\mathbf{s}_{\mathrm{ref}}),~\mathbf{s}\in \mathcal{S},
% \end{equation}
% where $q^{(i)}(\mathbf{s},\mathbf{a}) =  \sum_{\mathbf{s}' \in \mathcal{S}} \mathcal{P} (\mathbf{s}^\prime|\mathbf{s},\mathbf{a}) [c(\mathbf{s},\mathbf{a}) + v^{(i-1)}(\mathbf{s}^\prime)]$ and  $\mathbf{s}_{\mathrm{ref}} \in \mathcal{S}$ is an arbitrary reference state.
% For any initialization $v^{(0)}(\mathbf{s})$, the sequence $\{v^{(i)}(\mathbf{s})\}_{i=1,2,\ldots}$ converges to the optimal state-value function $v^\star(\mathbf{s})$, i.e., $\lim_{i\rightarrow \infty} v^{(i)}(\mathbf{s}) = v^\star (\mathbf{s})$, $\forall \mathbf{s} \in \mathcal{S}$, \cite[Section~4.3]{bertsekas2007dynamic}.

An optimal policy $\pi^\star$ can be found by turning the Bellman's optimality equation \eqref{eq_ballman} into an iterative procedure, called relative value iteration algorithm (RVIA) \cite[Section~8.5.5]{puterman2014markov}. Particularly, at each iteration $i = 0,1,\ldots$, we have
\begin{equation}\label{eq_v_itr}
\begin{array}{ll}
&V^{(i+1)}(\mathbf{s}) =\min_{\mathbf{a}\in\mathcal{A}} \left[c(\mathbf{s},\mathbf{a}) + \textstyle\sum_{\mathbf{s}' \in \mathcal{S}} \Pr(\mathbf{s}^\prime \mid \mathbf{s},\mathbf{a}) h^{(i)}(\mathbf{s}^\prime)\right], \\& h^{(i+1)}(\mathbf{s}) = V^{(i+1)}(\mathbf{s}) - V^{(i+1)}(\mathbf{s}_{\mathrm{ref}}),
\end{array}
\end{equation}
where ${\mathbf{s}_{\mathrm{ref}} \in \mathcal{S}}$ is an arbitrary reference state. 
For any initialization $V^{(0)}(\mathbf{s})$, the sequences $\{h^{(i)}(\mathbf{s})\}_{i=1,2,\ldots}$ and $\{V^{(i)}(\mathbf{s})\}_{i=1,2,\ldots}$ converge, i.e., $\lim_{i\to\infty}h^{(i)}(\mathbf{s}) = h(\mathbf{s})$ and ${\lim_{i\to\infty}V^{(i)}(\mathbf{s}) = V(\mathbf{s})}$, $\forall \mathbf{s}\in\mathcal{S}$.
% Let us define $V(\mathbf{s}) = \lim_{i\to\infty}V^{(i)}(\mathbf{s})$.
% to $h(\mathbf{s})$ (i.e., $\lim_{i \rightarrow \infty}h^{(i)}(\mathbf{s}) = h(\mathbf{s})$) that satisfies \eqref{eq_ballman} for all $\mathbf{s} \in \mathcal{S}$.
{Thus, $h(\mathbf{s}) = V(\mathbf{s}) - V(\mathbf{s}_{\mathrm{ref}})$ satisfies \eqref{eq_ballman} 
% for all $\mathbf{s} \in \mathcal{S}$,
and $\bar{C}^\star = V(\mathbf{s}_{\mathrm{ref}})$. Functions $V$ and $h$ are (sometimes) called value function and relative value function, respectively. 
% \blue{\st{It is worth noting that the values of $V(\mathbf{s})$ and $h(\mathbf{s})$ at each state are not unique and depend on the choice of $\mathbf{s}_{\mathrm{ref}}$.} 
It is worth noting that any function $h$ satisfying \eqref{eq_ballman} is unique up to an additive factor, i.e., 
% if $h$ and $\bar{h}$ satisfy \eqref{eq_ballman} then $h(s) - \bar{h}(s) = \mathrm{constant}$, for all $s\in \mathcal{S}_k$.
if $h$ satisfies \eqref{eq_ballman}, then $h + \alpha$, where $\alpha$ is any constant, also satisfies \eqref{eq_ballman}.}
The proposed RVIA is presented in Algorithm~\ref{alg_value_itr}, where $\theta$ is a small constant for the RVIA termination criterion.

\begin{algorithm}[t!]
\caption{RVIA that obtains optimal policy $\pi^\star$.}\label{alg_value_itr}
\begin{algorithmic}[1]
\State \textbf{Initialize} $V(\mathbf{s})\!\leftarrow\!0$, $h(\mathbf{s})\!\leftarrow\!0$, \!\!\! $\forall \mathbf{s} \in \mathcal{S}$, choose a reference state $\mathbf{s}_{\mathrm{ref}} \in \mathcal{S}$ and a small $\theta > 0$
\Repeat
\For{$\mathbf{s} \in \mathcal{S}$}
\State 
$V_{\mathrm{tmp}}(\mathbf{s}) \leftarrow \min_{\mathbf{a}\in\mathcal{A}} [c(\mathbf{s},\mathbf{a}) + \sum_{\mathbf{s}^\prime \in \mathcal{S}} \Pr(\mathbf{s}^\prime \mid \mathbf{s},\mathbf{a}) h(\mathbf{s}^\prime)]$
\EndFor
\State $\delta \leftarrow \max_{\mathbf{s}\in\mathcal{S}} (V_{\mathrm{tmp}}(\mathbf{s}) - V(\mathbf{s})) - \min_{\mathbf{s}\in\mathcal{S}} (V_{\mathrm{tmp}}(\mathbf{s}) - V(\mathbf{s}))$
\State $V(\mathbf{s}) \leftarrow V_{\mathrm{tmp}}(\mathbf{s})$, for all $\mathbf{s}\in\mathcal{S}$
\State $h(\mathbf{s}) \leftarrow V(\mathbf{s}) - V(\mathbf{s}_{\mathrm{ref}})$, for all $\mathbf{s}\in\mathcal{S}$.
\Until{$\delta < \theta$}
\State $\pi^\star(\mathbf{s}) = \argmin_{\mathbf{a} \in \mathcal{A}}[c(\mathbf{s},\mathbf{a}) + \sum_{\mathbf{s}^\prime \in \mathcal{S}} \Pr(\mathbf{s}^\prime \mid \mathbf{s},\mathbf{a})  h(\mathbf{s}^\prime)]$, for all $\mathbf{s}\in\mathcal{S}$
\end{algorithmic}
\end{algorithm}

It is important to point out that the state space $\mathcal{S}$ and action space $\mathcal{A}$ grow exponentially with respect to the number of sensors $K$, and thus, the complexity of the RVIA presented in Algorithm~\ref{alg_value_itr} grows exponentially in $K$. Accordingly, finding an optimal policy $\pi^\star$ is practical (tractable) only for small numbers  of sensors.
To this end, we next propose a low-complexity sub-optimal algorithm whose complexity increases only linearly in $K$. 
% and show that the proposed algorithm is \textit{asymptotically optimal} as the number of sensors goes to infinity.

% \newpage

%%%%%%%%%%%%%%%%%%%%%%%%%%%%%%%%%%%%%%%%%%%%%
\section{Low-Complexity Algorithm Design: Relax-then-Truncate Approach}\label{sec-lowcomplxity_alg}
In this section, to  handle  massive IoT scenarios, we propose a low-complexity algorithm {that provides} a sub-optimal solution to problem (\textbf{P1}). The key observation is that the \textit{per-slot} constraint \eqref{eq_bw_st} couples the actions $a_k(t)$, ${k \in \mathcal{K}}$, which results in the exponential complexity of finding an optimal policy for (\textbf{P1}), as explained in Section~\ref{sec-MDP-optimalpolicy}.
Therefore, we start by relaxing the per-slot constraint \eqref{eq_bw_st} into a time average constraint and subsequently model the \textit{relaxed problem} as a \textit{constrained MDP} (CMDP). The CMDP problem is then transformed into an unconstrained MDP problem through the Lagrangian approach \cite{altman1999constrained}.
{The MDP problem decouples along the sensors and, therefore, for a fixed value of the Lagrange multiplier, we can find a per-sensor optimal policy.} 
% We search for the optimal Lagrange multiplier to obtain
{The optimal value of the Lagrange multiplier is found via bisection. This provides}
an optimal policy for the relaxed problem, called \textit{optimal relaxed policy} hereinafter.
Finally, we propose an online \textit{truncation} procedure to ensure that the constraint \eqref{eq_bw_st} is satisfied at each slot. We remark that our optimality analysis in Section \ref{sec_optimality} shows that the proposed {relax-then-truncate approach} is \textit{asymptotically optimal} as the number of sensors goes to infinity.
% \brown{In addition, we analytically find an upper bound for the performance deviation of the proposed policy compared to the optimal performance obtained by $\pi^\star$ and show that the proposed algorithm is \textit{asymptotically optimal} as the number of sensors goes to infinity.}

\subsection{CMDP Formulation}\label{sec_cmdp_modeling}
% In this part, 
We relax the constraint \eqref{eq_bw_st} and formulate the relaxed problem as a CMDP. To this end, we define the
% \blue{expected}
average number of command actions under a policy $\pi_\R$ as
% Namely, under a policy $\pi_\R$, the time average number of command actions, which is denoted by $\bar{J}_{\pi_\R}$, is defined as
\begin{equation}\label{eq_average_transmision}
\bar{J}_{\pi_\R} \triangleq \lim_{T\rightarrow\infty}\frac{1}{KT}\sum_{t=1}^{T} \sum_{k=1}^{K}\mathbb{E}_{\pi_\R}[a_k(t)],
\end{equation} 
and express the relaxed problem as
\begin{equation}
    \begin{array}{ll}
    (\textbf{P2})~~\pi_{\mathrm{R}}^\star \in \argmin_{\pi_{\mathrm{R}}} & \bar{C}_{\pi_\R}\\
    % \mbox{minimize}   & (1/2)\|Ax-b\|_2^2 + \lambda \|x\|_1 \\
    \hspace{2cm} \mbox{subject to} & \bar{J}_{\pi_\R} \leq \Gamma
    \end{array}
    \label{st_opt2}
\end{equation}
where $\Gamma \triangleq \frac{M}{K}$ is the normalized transmission budget.
% , i.e., the proportion of sensors that can be commanded on the average. 

%, at each time slot.
% \brown{the maximum \st{allowable} average number of command actions in a normalized form}

% \blue{and \eqref{st_opt2} is the time average transmission constraint}
% \footnote{\brown{*Initial state?*}}. 
% \blue{*Initial state?*}

% normalized  maximum transmission resource 

We  model (\textbf{P2}) as a CMDP defined by the tuple ${\big(\mathcal{S}, \mathcal{A}_{\R},\Pr(\mathbf{s}(t + 1)|\mathbf{s}(t), \mathbf{a}(t)), c(\mathbf{s}(t), \mathbf{a}(t))\big)}$,
where the state space $\mathcal{S}$ and the relaxed action space $\mathcal{A}_{\R}$ were defined in Section~\ref{sec_state_action_cost_def}, and $\Pr(\mathbf{s}(t + 1)|\mathbf{s}(t), \mathbf{a}(t))$ and $c(\mathbf{s}(t), \mathbf{a}(t))$ were defined in Section~\ref{sec-MDP}. Note that the only difference between the CMDP tuple and the MDP tuple in Section~\ref{sec-MDP} is in the action space ($\mathcal{A}_{\R}$ vs $\mathcal{A}$). 
% \begin{subequations}\label{average_cost_relax_sp1}
% \begin{align}
% (\textbf{P2})~~\pi_{\mathrm{R}}^\star = \argmin_{\pi_{\mathrm{R}}\in \Pi_{\mathrm{R}}} \quad & \displaystyle\underset{T\rightarrow\infty}{\mathrm{lim}}\frac{1}{NKT}\mathbb{E}_{\pi_{\mathrm{R}}}\left[\sum_{t=1}^{T}\sum_{k=1}^{K} c_k(t)\right]\\
% \mathrm{s.t.} \quad &  \lim_{T\rightarrow\infty}\frac{1}{NKT}\mathbb{E}_{\pi_{\mathrm{R}}}\left[\sum_{t=1}^{T} \sum_{k=1}^{K}a_k(t)\right] \leq \frac{M}{NK},\label{st_opt2}
% \end{align}
% \end{subequations}
%%
% where $\{a_k(t)\}_{k\in\mathcal{K}}$, $t=1,2,\dots$ are the optimization variables.
% where a policy $\pi_{\mathrm{R}}$ is a mapping from the state space $\mathcal{S}$ to the \textit{relaxed} action space $\mathcal{A}_{\mathrm{R}} = \mathcal{A}_1 \times \dots \mathcal{A}_K$, i.e., $\pi_{\mathrm{R}}: \mathcal{S} \rightarrow \mathcal{A}_{\mathrm{R}}$.

It is worth noting that any policy $\pi$ that satisfies the per-slot transmission constraint \eqref{eq_bw_st} satisfies the time average transmission constraint \eqref{st_opt2}
in (\textbf{P2}). Thus, the average cost obtained by following policy $\pi_{\mathrm{R}}^\star$ is a \textit{lower bound} on the average cost obtained under policy $\pi^\star$, i.e.,
\begin{equation}\label{eq_lowerboundinequality}
 \bar{C}_{\pi^\star_\R} \leq \bar{C}_{\pi^\star}. 
\end{equation}
% \subsubsection{Lagrangian Function}

To solve the CMDP problem (\textbf{P2}), 
% \st{we start by rewriting it in its Lagrangian form.} 
we introduce a Lagrange multiplier $\mu$ and define the Lagrangian associated with problem (\textbf{P2}) as
% \red{** Following the Lagrangian approach to solve the formulated average cost CMDP problem (\textbf{P2}), we define the Lagrangian function
% for a policy $\pi$, given a Lagrange multiplier $\mu \geq 0$, as **}
% \yellow{To solve (\textbf{P2}), we rewrite it in its Lagrangian form. Namely, under a policy $\pi$, the Lagrange function is defined as}
\begin{equation}\label{lagrang_fcn}
    {\mathcal{L}}(\pi_{\mathrm{R}},\mu) \triangleq \lim_{T\rightarrow\infty} \frac{1}{NKT}\sum_{t=1}^{T}\sum_{k=1}^{K} \mathbb{E}_{\pi_{\mathrm{R}}}[c_k(t) + \mu a_k(t)] - \mu \frac{\Gamma}{N}.
\end{equation}
% \yellow{where $\mu \geq 0$ is the Lagrangian multiplier.}
For a given ${\mu \geq 0}$, we define
% \red{\st{the optimal Lagrangian}} --
the Lagrange dual function 
% -- as
${\mathcal{L}^\star({\mu}) = \min_{\pi_{\mathrm{R}}} {\mathcal{L}}({\pi_{\mathrm{R}},\mu})}$. A policy that achieves ${\mathcal{L}}^\star({\mu})$ is called \textit{$\mu$-optimal}, denoted by $\pi_{\R,\mu}^\star$, and it is a solution of the following (unconstrained) MDP problem
\begin{equation}\label{average_cost_relax_dual_sp2}
(\textbf{P3})~~\pi_{\R,\mu}^\star \in \argmin_{\pi_{\mathrm{R}}} {\mathcal{L}}({\pi_{\mathrm{R}},\mu}).
\end{equation}
% \begin{equation}\label{average_cost_relax_dual_sp2}
% \begin{aligned}
% \min \quad \displaystyle\underset{T\rightarrow\infty}{\mathrm{lim}}\frac{1}{KT}\mathbb{E}_{\pi}\Big[\sum_{t=1}^{T}\sum_{k=1}^{K} \Big[c_k(t) - \mu \Big(\frac{M}{K}-a_k(t)\Big)\Big]\mid s(0) = s\Big].
% \end{aligned}
% \end{equation}

% By checking the growth condition [17, Eq. 11.21], the
% optimal value of the CMDP problem (3), J∗(s[0]), and the
% optimal value of the MDP problem (4) which is indicated by
% L∗(λ, s[0]), ensures the following:

% \red{**Corollary 12.2 Altman conditions**}
% The state space dimension is finite, thus,
Since the dimension of the state space $\mathcal{S}$ is finite, the growth condition \cite[Eq.~11.21]{altman1999constrained} is satisfied. Moreover, the immediate cost function is bounded below, i.e.,  $c(\mathbf{s}, \mathbf{a}) \geq 0$, $\forall \mathbf{a},\mathbf{s}$. Having these conditions satisfied, 
% by \cite[Corollary~12.2]{altman1999constrained},
the optimal value of the CMDP problem (\textbf{P2}), $\bar{C}_{\pi^\star_{\R}}$, and the optimal value of the MDP problem (\textbf{P3}), $\mathcal{L}^\star({\mu})$, ensures the following relation \cite[Corollary~12.2]{altman1999constrained}
\begin{equation}\label{dultiy-relation}
    \bar{C}_{\pi^\star_{\R}} = \sup_{\mu \geq 0} 
    % \underbrace{\frac{1}{K}\sum_{k = 1}^{K} \bar{C}^\star_{R,\mu,k}}_{\bar{C}^\star_{R,\mu}} 
    {\mathcal{L}}^\star({\mu}).
    % - \mu \bar{J}_{\mathrm{max}}
\end{equation}
Therefore, an optimal policy for (\textbf{P2}) is found by a two-stage iterative algorithm:
%that iteratively executes the following two steps: 
1) for a given $\mu$, we find a $\mu$-optimal policy, and 2) we update $\mu$ in a direction that obtains  $\bar{C}_{\pi^\star_\R}$ according to \eqref{dultiy-relation}. 
These two steps are detailed in
Sections~\ref{sec_opt_fixed_mu} and \ref{sec_opt_mu}, respectively.
% \red{the following.}

%%%%%%%%%%%%%%%%%%%%%%%%%%%%%%5
\subsubsection{An Optimal Policy for a Fixed Lagrange Multiplier}\label{sec_opt_fixed_mu}
% \noindent\textbf{1) Find $\mu$-Optimal Policy}:
For a given $\mu$, the problem of finding an optimal policy $\pi_{\R,\mu}^\star$ in (\textbf{P3}) is \textit{separable} across sensors $k \in \mathcal{K}$. Thus, (\textbf{P3}) can be decoupled into $K$ per-sensor problems as follows. We express the Lagrangian in \eqref{lagrang_fcn} equivalently as ${\mathcal{L}}({\pi_{\mathrm{R}},\mu}) = \frac{1}{{N}K} \textstyle\sum_{k=1}^{K}{\mathcal{L}_k}({\pi_k,\mu}) {- \mu \frac{\Gamma}{N}}$,
where  $\mathcal{L}_{k}({\pi_k,\mu})$ is 
% \red{\st{the per-sensor Lagrangian,}}
defined as
\begin{equation}\label{per-sensor-lagrang-fcn}
    {\mathcal{L}}_k({\pi_k,\mu}) \triangleq \lim_{T\rightarrow\infty} \frac{1}{T}\sum_{t=1}^{T} \mathbb{E}_{\pi_k}[c_k(t) + \mu a_k(t)],~k = 1, \dots, K,
\end{equation}
where the per-sensor policy $\pi_k$ was defined in Section~\ref{sec_policy_def}. Thus, finding an optimal policy $\pi_{\R,\mu}^\star$ reduces to finding $K$ per-sensor optimal policies, denoted by $\pi_{\R,\mu,k}^\star$, $k = 1,\dots,K$, as
\begin{equation}\label{_persensor-average_cost_relax_dual_sp2}
(\textbf{P4})~~\pi_{\R,\mu,k}^\star \in \argmin_{\pi_k} {\mathcal{L}}_{k}({\pi_k,\mu}),~k = 1,\dots,K.
\end{equation}

Each  sub-problem (\textbf{P4}), for a particular $k$, can be modeled as an (unconstrained) MDP problem. Particularly, the MDP model associated with sensor $k$ is defined as the tuple $\left(\mathcal{S}_k, \mathcal{A}_k,\Pr(s_k(t + 1)|s_k(t), a_k(t)), c_k(s_k(t), a_k(t))+\mu a_k(t)\right)$, where {the per-sensor state space} $\mathcal{S}_k$ and {the per-sensor action space} $\mathcal{A}_k$ were defined in Section~\ref{sec_state_action_cost_def}, the per-sensor state transition probabilities $\Pr(s_k(t+1)|s_k(t),a_k(t))$ are calculated as in \eqref{eq_persensor_STP}, and the 
%\yellow{\st{one-step}} 
cost of taking action $a_k(t)$ in state $s_k(t)$ is 
%\yellow{\st{defined as}} 
$c_k(s_k(t), a_k(t))+\mu a_k(t)$, where $c_k(s_k(t), a_k(t))$ is {defined in Section~\ref{sec-MDP}.}
% calculated as in \eqref{eq_persensor_cost}. 
%The per-sensor state transition probability $\Pr(s_k(t+1)|s_k(t),a_k(t))$ were defined in maps a state-action pair at slot $t$ onto a distribution of states at  slot $t + 1$; the probability of transition from current state $s_k(t) = \{r_k,b_k,\Delta_k\}$ to next state $s_k(t+1) = \{r^\prime_k,b^\prime_k,\Delta^\prime_k\}$ under action ${a}_k(t) = a_k$ is calculated as
% \begin{equation}
%     \Pr \left(s_k(t+1)\mid s_k(t),a_k(t) \right) 
%     = \Pr(r^\prime_k) \Pr(b^\prime_k\mid b_k , a_k) \Pr(\Delta^\prime_k \mid b_k, \Delta_k,a_k),
% \end{equation}
% where $\Pr(r^\prime_k)$, $\Pr(b^\prime_k\mid b_k , a_k)$, and $\Pr(\Delta^\prime_k \mid b_k, \Delta_k,a_k)$ were calculated in \eqref{eq_per_sensor_STPs}.
\vspace{-2mm}\begin{Pro}\label{theorem_communicating_per_sensorMDP}
The per-sensor MDP formulated for (\textbf{P4}) is communicating, i.e., for every pair of states {$s, s^\prime \in \mathcal{S}_k$}, there exists a stationary policy under which $s^\prime$ is accessible from $s$.
\end{Pro}\vspace{-5mm}
\begin{proof}
The proof is presented in Appendix~\ref{sec_appendix_theorem_communicating_per_sensorMDP}.
\end{proof}\vspace{-2mm}

% \begin{remark}
% **Problem 3**
% The per-sensor optimization problem:

% \red{$\pi_{R,k}^*$}

% \begin{equation}\label{average_cost_relax_dual_persensor}
% \begin{aligned}
% \min \quad \displaystyle\underset{T\rightarrow\infty}{\mathrm{lim}}\frac{1}{T}\mathbb{E}_{\pi_k}\Big[\sum_{t=1}^{T} [c_k(t) - \mu (\frac{M}{K}-a_k(t))]\Big].
% \end{aligned}
% \end{equation}
% \end{remark}

% \red{ *** Now we can model the above optimization problem for each sensor as an MDP as the tuple $\{\mathcal{S}_k,...\}$ with the cost function .... ***}
%\begin{lemm}

% a per-sensor $\mu$-optimal policy $\pi_{R,\mu,k}^\star$ can be obtained by solving the following Bellman's equation 

By Proposition~\ref{theorem_communicating_per_sensorMDP} and Proposition~\ref{lemma_Bellman}, and rewriting the Bellman's equation in \eqref{eq_ballman} for the per-sensor MDP formulation, we have
\begin{equation}\label{eq_persensor_ballman}
  {\mathcal{L}}_{k}^\star({\mu}) + h_{\R,\mu,k}(s) = \textstyle\min_{a \in \mathcal{A}_k} \left[c_k(s,a)+ \mu a + \textstyle\sum_{s^\prime \in \mathcal{S}_k} \Pr(s^\prime|s,a) h_{\R,\mu,k}(s^\prime)\right],~s\in \mathcal{S}_k,
\end{equation}
where 
% \red{\st{$h_{\R,\mu,k}(s)$ is the per-sensor relative value function, ${\mathcal{L}_k^\star({\mu})}$ is the optimal per-sensor Lagrangian obtained by $\pi_{\R,\mu,k}^\star$ (i.e.,}}
${\mathcal{L}}_k^\star({\mu}) \triangleq \min_{\pi_k}{\mathcal{L}}_{k}({\pi_k,\mu})$. 
% We define the per-sensor action-value function $q_{\R,\mu,k}(s,a)$ as
% \begin{equation}
%     q_{\R,\mu,k}(s,a) = c_k(s,a)+ \mu a + \textstyle\sum_{s^\prime \in \mathcal{S}_k} \Pr(s^\prime|s,a) h_{\R,\mu,k}(s^\prime),~s\in \mathcal{S}_k,a\in\mathcal{A}_k.
% \end{equation}
In addition, an optimal policy in state $s \in \mathcal{S}_k$ is given by 
\begin{equation}
\pi_{\R,\mu,k}^\star(s) \in \textstyle\argmin_{a\in\mathcal{A}_k} \left[c_k(s,a)+ \mu a + \textstyle\sum_{s^\prime \in \mathcal{S}_k} \Pr(s^\prime|s,a) h_{\R,\mu,k}(s^\prime)\right],~ s\in \mathcal{S}_k.
\end{equation}
% \end{lemm}
By turning \eqref{eq_persensor_ballman} into an iterative procedure, $h_{\R,\mu,k}(s)$ and consequently $\pi^\star_{\R,\mu,k}(s)$, ${s}\in \mathcal{S}_k$, are obtained iteratively.
Particularly, at each iteration $i = 0,1,\dots$, we have
% let us denote the per-sensor state-value function in state $s$ at iteration $i = 1,2,\ldots$ by $v^{(i)}_{\R,\mu,k}(s)$, which is given by
\begin{equation}\label{eq-per-sensor-v-itr}
\begin{array}{ll}
&V^{(i+1)}_{\R,\mu,k}(s) =\textstyle\min_{a\in\mathcal{A}_k} [c_k(s,a) + \mu a+  \sum_{{s^\prime} \in \mathcal{S}_k} \Pr ({s}^\prime|{s},{a}) h^{(i)}_{\R,\mu,k}(s^\prime)],\\
&h^{(i+1)}_{\R,\mu,k}(s) = V^{(i+1)}_{\R,\mu,k}(s) - V^{(i+1)}_{\R,\mu,k}(s_{\mathrm{ref}}),
\end{array}
\end{equation}
where 
% \red{\st{$V^{(i)}_{\R,\mu,k}(s)$, and $h^{(i)}_{\R,\mu,k}(s)$ denote per-sensor value function and per-sensor relative value function at iteration $i$, respectively, and}}
${s_{\mathrm{ref}} \in \mathcal{S}_k}$ is an arbitrary reference state.
For any initialization $V^{(0)}_{\R,\mu,k}(s)$, the sequences $\{h^{(i)}_{\R,\mu,k}(s)\}_{{i=1,2,\ldots}}$ and $\{V^{(i)}_{\R,\mu,k}(s)\}_{{i=1,2,\ldots}}$ converge, i.e., ${\lim_{i\to\infty}h^{(i)}_{\R,\mu,k}(s) = h_{\R,\mu,k}(s)}$ ${\lim_{i\to\infty}V^{(i)}_{\R,\mu,k}(s) = V_{\R,\mu,k}(s)}$, $\forall s \in \mathcal{S}_k$. Thus, $h_{\R,\mu,k}(s) = V_{\R,\mu,k}(s) - V_{\R,\mu,k}(s_{\mathrm{ref}})$ satisfies \eqref{eq_persensor_ballman} and ${\mathcal{L}}_k^\star({\mu}) = V_{\R,\mu,k}(s_{\mathrm{ref}})$. The proposed RVIA is presented in Algorithm~\ref{alg_cmdp_RVIA}~(Lines~\ref{alg-lst-line-function-RVIA}--\ref{alg-last-line-function-RVIA}).
% , where $\theta$ is a sufficiently small positive real number for the RVIA termination criterion.

% Thus, \brown{similarly to the discussion expressed in Section~} RVIA can be used to find per-sensor optimal policies $\pi_{R,\mu,k}^\star$ by turning the Bellman's equation  \eqref{eq_persensor_ballman} into an iterative procedure. 
% Details are presented in Alg.~\ref{alg_cmdp_RVIA} (Lines~6-12), where $\theta$ is a sufficiently small positive real number for the RVIA termination criterion. 

Next, to give insight to optimal policies, we analyze the properties of a per-sensor optimal policy $\pi_{\R,\mu,k}^\star$ obtained by the proposed RVIA.
% , the inherent threshold-based structure of an optimal policy
We first prove that $V_{\R,\mu,k}(s)$ has monotonic properties. Then, exploiting this monotonicity, we prove that an optimal policy has a threshold-based structure with respect to the AoI.
\vspace{-2mm}\begin{lemm}\label{prop_struct_optimal_persensor_v}
The function $V_{\R,\mu,k}$ is non-decreasing with respect to the AoI{, i.e., for any two states $s = (r,b,\Delta) \in \mathcal{S}_k$ and $\sbar = (r,b,\Deltabar)\in \mathcal{S}_k$, where $\Deltabar \geq \Delta$, we have $V(\sbar) \geq V(s)$.}
% , and (ii) non-increasing with respect to the battery level.
\end{lemm}\vspace{-5mm}
\begin{proof}
The proof is presented in Appendix~\ref{sec-apndix-prop_struct_optimal_persensor_v}.
\end{proof}\vspace{-2mm}
\vspace{-2mm}\begin{theorem}\label{theorem_AoI_threshold-persensor}
% \normalfont
For any $\mu \geq 0$, a per-sensor optimal policy $\pi_{\R,\mu,k}^\star$ {obtained by RVIA} has a \mbox{threshold-based} structure with respect to the AoI{, i.e., if an optimal action in state $s = (r,b,{\Delta})$ is $\pi_{\R,\mu,k}^\star(s) = 1$, then for all states $\sbar = (r,b,\Deltabar)$, $\Deltabar \geq {\Delta}$, an optimal action is also $\pi_{\R,\mu,k}^\star(\sbar) = 1$.}
\end{theorem}\vspace{-5mm}
\begin{proof}
The proof is presented in Appendix~\ref{sec-apndix-theorem_AoI_threshold-persensor}.
\end{proof}\vspace{-2mm}

%%%%%%%%%%%%%%%%%%%%%%%%%%%%%%%
% Approximation/
% \red{Approximating the optimal dual variable}
% \noindent \textbf{2) approximating the optimal Lagrange Multiplier:}
\subsubsection{Determination of the Optimal Lagrange Multiplier}\label{sec_opt_mu}
Recall that the cost function associated with the per-sensor MDP formulation (established for (\textbf{P4})) is defined as $c_k(s_k(t),a_k(t)) + \mu a_k(t)$. Hence, by increasing $\mu$, the cost of taking action $a_k(t) = 1$ increases, and thus, the edge node tends to use the command action less. More precisely, $\bar{C}_{\pi_{\R,\mu}^\star}$ and ${\mathcal{L}}(\pi_{\R,\mu}^\star,\mu)$ are increasing in $\mu$, whereas $\bar{J}_{\pi_{\R,\mu}^\star}$ is decreasing in $\mu$
% \cite[Lemma~3.4]{sennott1993constrained}, 
\cite[Lemma~3.1]{beutler1985optimal}. Therefore, we are interested in
% \red{\st{the infimum (roughly speaking,}}
the smallest value of the Lagrange multiplier such that policy $\pi_{\R,\mu}^\star$ satisfies the time average transmission constraint \eqref{st_opt2}.
Formally,
% according to \cite{beutler1985optimal}
% \footnote{\brown{In \cite{beutler1985optimal}, the authors are interested in maximizing a reward an an objective, where as here we are interested in minimizing a cost function. Thus, we use $\inf$ in equation \eqref{eq-mu-star}}.}
% and \cite{sennott1993constrained},
we define the optimal Lagrange multiplier as \cite{beutler1985optimal}
\begin{equation}\label{eq-mu-star}
    \mu^* \triangleq \inf \left\{\mu \geq 0 \mid \bar{J}_{\pi_{\R,\mu}^\star} \leq \Gamma\right\},
\end{equation}
where $\bar{J}_{\pi_{\R,\mu}^\star}$ is the average number of command actions under policy $\pi_{\R,\mu}^\star$, which is calculated using \eqref{eq_average_transmision}.
% Let us define the per-sensor time average number of command actions under a per-sensor policy $\pi_k$ as
% \begin{equation}\label{eq_persensor_average_transmision}
% \bar{J}_{\pi_k} \triangleq 
% %\overset{\Delta}{=}
% \lim_{T\rightarrow\infty}\frac{1}{NT}\mathbb{E}_{\pi_k}\left[\sum_{t=1}^{T} a_k(t)\right].
% \end{equation} 
From \eqref{eq_average_transmision}
% , \eqref{eq_persensor_average_transmision}, 
and the fact that (\textbf{P3}) is decoupled into $K$ per-sensor problems (\textbf{P4}), $\bar{J}_{\pi_{\R,\mu}^\star}$ is calculated as $ \bar{J}_{\pi^\star_{\R,\mu}}  = \frac{1}{K}\sum_{k = 1}^{K}\bar{J}_{\pi_{\R,\mu,k}^\star}$, where $\bar{J}_{\pi_{\R,\mu,k}^\star}$ denotes the per-sensor time average number of command actions under the per-sensor policy $\pi_{\R,\mu,k}^\star$, which is defined as
\begin{equation}\label{eq_persensor_average_transmision}
\bar{J}_{\pi_{\R,\mu,k}^\star} \triangleq 
%\overset{\Delta}{=}
\lim_{T\rightarrow\infty}\frac{1}{T}\sum_{t=1}^{T} \mathbb{E}_{\pi_{\R,\mu,k}^\star}[a_k(t)].
\end{equation} 
Thus, \eqref{eq-mu-star} is rewritten as ${\mu^* = \inf \left\{\mu \geq 0: \textstyle\sum_{k =1}^{K}\bar{J}_{\pi_{\R,\mu,k}^\star} \leq K\Gamma \right\}}$.

\algrenewcommand\algorithmicrequire{\textbf{Input}}
\algrenewcommand\algorithmicensure{\textbf{Output}}

\begin{algorithm}[t!]
\caption{Policy design for the CMDP problem (\textbf{P2}) via RVIA and bisection search}\label{alg_cmdp_RVIA}
\begin{algorithmic}[1]  
\State \textbf{Initialize} Set $\mu \gets 0$, $\mu^{-} \gets 0$, $\mu^{+}$ as a large positive number, and determine a small $\epsilon > 0$
% \State $\mu \gets 0$
\State \Call{RVIA}{$\mu$} \Comment{\textit{run function RVIA for input $\mu = 0$}}
\If{$\bar{J}_{\pi^\star_{\R,\mu}} \leq \Gamma$}
\State $\pi^\star_{\R} = \pi^\star_{\R,\mu}$
\Else
\While{$|\mu^{+} - \mu^{-}| > \epsilon$}
% \State $\mu = \frac{\mu^{+} + \mu^{-}}{2}$
\State \Call{RVIA}{$\frac{\mu^{+} + \mu^{-}}{2}$} \Comment{\textit{run function RVIA for $\mu = \frac{\mu^{+} + \mu^{-}}{2}$}}
% \State Compute $\bar{J}_{\pi^\star_{\R,\mu}}$
% \If{$\bar{J}_{\pi^\star_{\R,\mu}} \geq \Gamma$}
% \State $\mu^{-} \gets \mu$
% \Else
% \State $\mu^{+} \gets \mu$
% \EndIf
\State \textbf{if} $\bar{J}_{\pi^\star_{\R,\mu}} \geq \Gamma$ \textbf{then} $\mu^{-} \gets \mu$ \textbf{else} $\mu^{+} \gets \mu$
\EndWhile
\State  $\mu^* \gets 1/2(\mu^{-}+\mu^{+})$, $\mu^{*-} \gets \mu^{-}$, and $\mu^{*+} \gets \mu^{+}$
% \If{$\bar{J}_{\pi^\star_{\R,\mu}} = \Gamma$}
% \State $\pi^\star_{\R} = \pi^\star_{\R,\mu^*}$
% \Else
% \State $\pi^\star_{\R}$ is a mixture of two deterministic policies $\pi_{\R,\mu^{*-}}^\star$ and $\pi_{\R,\mu^{*+}}^\star$.
% \EndIf
\State \textbf{if} $\bar{J}_{\pi^\star_{\R,\mu}} = \Gamma$ \textbf{then} $\pi^\star_{\R} = \pi^\star_{\R,\mu^*}$ \textbf{else} $\pi^\star_{\R} = \eta \pi_{\R,\mu^{*-}}^\star + (1-\eta) \pi_{\R,\mu^{*+}}^\star$
\EndIf
\vspace*{-.5\baselineskip}\Statex\hspace*{\dimexpr-\algorithmicindent-2pt\relax}\rule{\textwidth}{0.4pt}
% \Algphase{function RVIA($\mu$) \normalfont{\Comment{\textit{find optimal policies $\pi^\star_{\R,\mu,k}$ using RVIA for fixed $\mu$.}}}}
\Function{RVIA}{$\mu$}\label{alg-lst-line-function-RVIA}
\Comment{\textit{find optimal policies $\pi^\star_{\R,\mu,k}$ using RVIA for fixed $\mu$}} \vspace{1.5mm}
\State \textbf{Initialize} $V_{\R,\mu,k}(s) \leftarrow 0$, $h_{\R,\mu,k}(s) \leftarrow 0$, $\forall k, \forall s$, determine $s_{\mathrm{ref}} \in \mathcal{S}_k$ and a small $\theta > 0$
\For{$k = 1, \dots, K$}
\Repeat
\For{$s \in \mathcal{S}_k$}
\State $V_{\mathrm{tmp}}(s) \gets \underset{a\in\mathcal{A}_k}{\min} [c_k(s,a) + \mu a +\underset{s' \in \mathcal{S}_k}{\sum} \Pr (s^\prime \mid s,a)h_{\R,\mu,k}(s^\prime)]$
\EndFor
\State $\delta \gets \max_{s\in\mathcal{S}_k} (V_{\mathrm{tmp}}(s) - V_{\R,\mu,k}(s)) - \min_{s\in\mathcal{S}_k} (V_{\mathrm{tmp}}(s) - V_{\R,\mu,k}(s))$
\State $V_{\R,\mu,k}(s) \gets V_{\mathrm{tmp}}(s)$, for all $s\in\mathcal{S}_k$
\State $h_{\R,\mu,k}(s) \gets V_{\R,\mu,k}(s) - V_{\R,\mu,k}(s_{\mathrm{ref}})$, for all $s\in\mathcal{S}_k$
\Until{$\delta < \theta$}
\State $\pi^\star_{\R,\mu,k}(s) = \underset{a\in\mathcal{A}_k}{\argmin}[c_k(s,a) + \mu a + \sum_{s^\prime \in \mathcal{S}_k} \Pr \left(s^\prime \mid s,a\right) h_{\R,\mu,k}(s^\prime)]$, for all $s\in\mathcal{S}_k$
\EndFor 
% \State \textbf{Output}: optimal per-sensor policies $\pi^\star_{\R,\mu,k},~\forall k$.
\EndFunction \label{alg-last-line-function-RVIA}
\end{algorithmic}
\end{algorithm}

{We now characterize an optimal relaxed policy $\pi^\star_{\R}$ for (\textbf{P2})}. If the average number of command actions obtained by 
%\st{optimal per-sensor policies}
$\pi_{\R,\mu^*,k}^\star$ satisfies $\frac{1}{K}\sum_{k =1}^{K}\bar{J}_{\pi_{\R,\mu^*,k}^\star} = \Gamma$, then, $\pi_{\R,\mu^*,k}^\star$, $k\in \mathcal{K}$, form an optimal policy for (\textbf{P2}), i.e., $\pi_{\R}^\star = \pi_{\R,\mu^*}^\star$. 
Otherwise, $\pi_{\R}^\star$ is a mixture of two deterministic policies $\pi_{\R,\mu^{*-}}^\star$ and $\pi_{\R,\mu^{*+}}^\star$, which are defined by \cite[Theorem~4.4]{beutler1985optimal}
\begin{equation}\label{eq-mixingpolicies_def}
    \pi_{\R,\mu^{*-}}^\star  \triangleq \lim_{\mu \rightarrow \mu^{*-}} \pi_{\R,\mu}^\star ~\mathrm{and}~
    \pi_{\R,\mu^{*+}}^\star \triangleq \lim_{\mu \rightarrow \mu^{*+}} \pi_{\R,\mu}^\star,
\end{equation}
% \begin{equation}\label{eq-mixingpolicies_def}
% \begin{array}{ll}
%     &\pi_{\R,\mu^{*-}}^\star  \triangleq \lim_{\mu \rightarrow \mu^{*-}} \pi_{\R,\mu}^\star,\\
%     &\pi_{\R,\mu^{*+}}^\star \triangleq \lim_{\mu \rightarrow \mu^{*+}} \pi_{\R,\mu}^\star,
% \end{array}
% \end{equation}
{and is written symbolically as $\pi^\star_{\R} \triangleq \eta \pi_{\R,\mu^{*-}}^\star + (1-\eta) \pi_{\R,\mu^{*+}}^\star$, where $\eta$ is the mixing factor. This mixed policy is a  {stationary randomized} policy where the action at each state $\mathbf{s}$ is $\pi_{\R,\mu^{*-}}^\star(\mathbf{s})$ with probability $\eta$ and $\pi_{\R,\mu^{*+}}^\star(\mathbf{s})$ with probability $1 -\eta$, where $\eta$ is obtained\footnote{There is no closed-form expression for $\eta$ \cite{Shwartz1986Est}. Therefore, we can numerically search for such $\eta \in [0,1]$.} such that $\bar{J}_{\pi^\star_{\R}} = \Gamma$.}

% Even if the policies g and 3, and the value 7' were readily available,
% the computation of the optimal bias q* may prove to be a non-trivial
% task, for it requires solving for q in the implicitequation K(f,) = V on the
% interval [0,1], and makes it necessary to evaluate the expression K(f,) for
% each 0 5 q 5 1.

{To search for $\mu^*$ as defined in \eqref{eq-mu-star}, we apply the bisection method that exploits the monotonicity of $\bar{J}_{\pi_{\R,\mu}^\star}$ with respect to $\mu$.
% Let us denote the Lagrange multiplier in the $i$th iteration by $\mu^{(i)}$. 
Particularly, if $\frac{1}{K}\sum_{k =1}^{K}\bar{J}_{\pi_{\R,\mu,k}^\star} \leq \Gamma$ for $\mu = 0$, then the constraint \eqref{st_opt2} is inactive, and an optimal policy for (\textbf{P2}) is $\pi^\star_{\mathrm{R},0}$. Otherwise, we apply an iterative update procedure until $|\mu^{+} - \mu^{-}| < \epsilon$ and $\frac{1}{K}\sum_{k =1}^{K}\bar{J}_{\pi_{\R,\mu,k}^\star} \leq \Gamma$ are satisfied.
Details are expressed in Algorithm~\ref{alg_cmdp_RVIA}, where 
% we initialize $\mu^{-} = 0$ and $\mu^{+}$ as a large positive real number, and 
$\epsilon$ is a small constant for the bisection termination criterion.}

\begin{remark}\normalfont
It is worth noting that the complexity of finding an optimal relaxed policy $\pi^\star_{\R}$ increases linearly in the number of sensors, whereas the complexity of finding an optimal policy $\pi^\star$ grows exponentially in $K$. 
% For better clarification,
Consider a scenario with $K = 100$ sensors, $N = 7$ users, $\Delta^{\mathrm{max}} = 64$, and $B_{k} = 15$. The size of the state space $\mathcal{S}$ is $|\mathcal{S}| = \left(8\times16\times64\right)^{100} = 2^{1300}\approx 10^{400}$. However, the per-sensor state space size is $|\mathcal{S}_k| = 8\times16\times64 = 2^{13}\approx 10^{6}$.
\end{remark}

\subsection{{Truncation Procedure}}\label{sec_truncation}
Recall that there is no guarantee that the per-slot constraint \eqref{eq_bw_st} is satisfied under an optimal policy for the relaxed problem (\textbf{P2}) (i.e., $\pi^\star_{\R}$).
Here, %\st{based on $\pi^\star_{\R}$,}
we propose the following truncation procedure
%\st{policy  $\tilde{\pi}$}
that satisfies the constraint \eqref{eq_bw_st} at each slot.
More precisely, at slot $t$, let ${\mathcal{X}(t) = \{k\mid a_k(t) = 1, k \in \mathcal{K}\} \subseteq \mathcal{K}}$ denote the set of sensors that are commanded under $\pi^\star_{\R}$. The truncation step separates into two cases: 1) if $|\mathcal{X}(t)| \leq M$, the edge node simply commands all the sensors in $\mathcal{X}(t)$, and 2) otherwise, the edge node selects $M$ sensors from the set $\mathcal{X}(t)$ \textit{randomly} according to the discrete uniform
distribution and commands them to send status updates. 
{The online truncation  procedure is presented in Algorithm~\ref{alg_truncation}.}
% The optimality of the proposed truncation based algorithm is analyzed in the next section.

% \brown{We can represent $\tilde{\pi}$ as a randomized stationary policy.}

% \iffalse
\begin{algorithm}[t!]
\caption{Truncation procedure \label{alg_truncation}} % \st{Online} 
\begin{algorithmic}[1]
% \State \textbf{Input} $\pi^\star_{\R,\mu^\star,k}$
\Require Optimal relaxed policy $\pi^\star_{\R}$
\For{each slot $t = 1, 2, 3, \dots$}
\State Construct the set $\mathcal{X}(t)$ based on $\pi^\star_{\R}$
\State \textbf{if} $|\mathcal{X}(t)| \leq M$ \textbf{then} $a_k(t) = 1$, for all $k \in \mathcal{X}(t)$
\textbf{else} 
\State \hspace{5mm} Select $M$ sensors from $\mathcal{X}(t)$ \textit{randomly (uniform)} and command them
\State \textbf{end if}
% \If{$|\mathcal{X}(t)| \leq M$}
% \State $a_k(t) = 1$, for all $k \in \mathcal{X}(t)$
% \Else
% \State Select $M$ sensors from $\mathcal{X}(t)$ \textit{randomly (uniform)} and command them
% \EndIf
\EndFor
\end{algorithmic}
\end{algorithm}
% \fi

%%%%%%%%%%%%%%%%%%%%%%%%%%%
\section{Asymptotic Optimality of the Proposed Relax-then-Truncate Approach}\label{sec_optimality}
In this section, we analyze the optimality of {the proposed} relax-then-truncate policy  -- denoted by $\tilde{\pi}$ hereinafter -- developed in Section~\ref{sec-lowcomplxity_alg}. We first find an upper bound for the difference between the average cost obtained by the policy $\tilde{\pi}$ and the average cost obtained by an optimal policy $\pi^\star$.
% \yellow{, called the performance error -- performance deviation -- hereinafter.} 
Then, we present two lemmas that are used to show that the relax-then-truncate {approach} is \textit{asymptotically optimal} as the number of sensors goes to infinity. 
% Then, we exploit this upper bound to prove that the proposed algorithm is asymptotically optimal as the number of sensors goes to infinity.

\vspace{-2mm}\begin{theorem}\label{Theorem_upperbound_deviation}
% \normalfont
The difference between the average cost obtained by the relax-then-truncate policy $\tilde{\pi}$ and the average cost obtained by an optimal policy $\pi^\star$ is upper bounded as
\begin{equation}\label{eq_theorem_upperbound_deviation}
    {\bar{C}_{\tilde{\pi}} - \bar{C}_{\pi^\star} \leq \frac{\Delta^{\mathrm{max}}}{M} \lim_{T\rightarrow\infty}\frac{1}{T}\sum_{t=1}^{T} \underbrace{\mathbb{E}_{\pi_{\R}^\star}\left[|\mathcal{X}(t)| - \mathbb{E}_{\pi_{\R}^\star}[|\mathcal{X}(t)|]\right]}_{ \triangleq \mathrm{MAD}(|\mathcal{X}(t)|)}},
\end{equation}
where $\mathrm{MAD}(\cdot)$ denotes the Mean Absolute Deviation.
% , and, as defined earlier, $\Gamma = M/K$ is the normalized transmission budget and $|\mathcal{X}(t)|$ represents the number of sensors that are commanded under $\pi_{\R}^\star$ at slot $t$. 
% Since $\pi_{\R}^\star$ is stationary, i.e., an action taken at slot $t$ and state $\mathbf{s}(t) = \mathbf{s}$  depends only on $\mathbf{s}$, not the time index $t$, thus, we have $\lim_{T\rightarrow\infty}\frac{1}{T}\sum_{t=1}^{T} \mathrm{MAD}(|\mathcal{X}(t)|) = \mathrm{MAD}(|\mathcal{X}(t)|)$, $\forall t$. Therefore, \eqref{eq_theorem_upperbound_deviation} is rewritten as
% \begin{equation}\label{eq_theorem_upperbound_deviation-simpleform}
%     \bar{C}_{\tilde{\pi}} - \bar{C}_{\pi^\star} \leq \frac{\Delta^{\mathrm{max}}}{\Gamma K}\underbrace{\mathbb{E}_{\pi_{\R}^\star}\left[|\mathcal{X}(t)| - \mathbb{E}_{\pi_{\R}^\star}[|\mathcal{X}(t)|]\right]}_{ \triangleq \mathrm{MAD}(|\mathcal{X}(t)|)},
% \end{equation}
% where $\mathrm{MAD}(\cdot)$ denotes the Mean Absolute Deviation, and, as defined earlier, $\Gamma = M/K$ 
% % is the normalized transmission budget 
% and $|\mathcal{X}(t)|$ represents the number of sensors that are commanded under $\pi_{\R}^\star$ at slot $t$.
\end{theorem}\vspace{-5mm}
\begin{proof}
The proof is presented in Appendix~\ref{sec-apndix-Theorem_upperbound_deviation}.
\end{proof}\vspace{-3mm}

We next present two lemmas that will subsequently be used in Theorem~\ref{theorem-asymptotically-opt} to prove the asymptotic optimality of the relax-then-truncate approach.
\vspace{-2mm}\begin{lemm}\label{lemma_MAD_normal}
% \normalfont
For a random variable $X$ that follows a normal distribution with mean $\nu$ and variance $\sigma^2$, i.e., $X \sim \mathcal{N}(\nu,\sigma^2)$, the mean absolute  deviation is given as $\mathrm{MAD}(X) = \sqrt{\frac{2}{\pi}}\sigma$.
% \begin{align}
%     &\mathrm{MAD}(X) = \sqrt{\frac{2}{\pi}}\sigma
% \end{align}
\end{lemm}\vspace{-5mm}
\begin{proof}
The proof is presented in Appendix~\ref{sec-apndix-lemma_MAD_normal}.
\end{proof}\vspace{-2mm}
\vspace{-3mm}\begin{lemm}\label{lemma_MAD}
% \normalfont
% \blue{It is shown that $\mathrm{MAD}(|\mathcal{X}(t)|)$ is upper bounded by $\sqrt{\frac{K}{2\pi}}$ as the number of sensors goes to infinity, i.e., $\underbrcae{\lim}_{K\rightarrow \infty}\mathrm{MAD}(|\mathcal{X}(t)|) \leq \sqrt{\frac{K}{2\pi}}$, for all $t$, when $K\rightarrow \infty$.}
When $K\rightarrow \infty$, by following the policy $\pi_{\R}^\star$, {we have $\mathrm{MAD}\left(\frac{|\mathcal{X}(t)|}{\sqrt{K}}\right) \leq 1$}. 
% \st{the MAD of $|\mathcal{X}(t)|$ is upper bounded by $\sqrt{\frac{K}{2\pi}}$}.
% \begin{equation}
%     \mathrm{MAD}(|\mathcal{X}(t)|) \leq \sqrt{\frac{K}{2\pi}}, ~\forall{t}.\notag
% \end{equation}
\end{lemm}\vspace{-5mm}
\begin{proof}
The proof is presented in Appendix~\ref{sec-apndix-lemma_MAD}.
\end{proof}\vspace{-2mm}
\vspace{-3mm}\begin{theorem}\label{theorem-asymptotically-opt}
% \normalfont
% \red{For any $\Gamma = M/K > 0$},
{For a fixed 
%\yellow{normalized transmission budget} 
$\Gamma = M/K$}, {the relax-then-truncate policy} $\tilde{\pi}$ is asymptotically optimal with respect to the number of sensors, i.e., ${\lim_{K\rightarrow\infty} (\bar{C}_{\tilde{\pi}} - \bar{C}_{\pi^\star}) = 0}$.
\end{theorem}\vspace{-5mm}
% \textcolor{red}{If K tends to infinity in order to keep $\alpha$ constant it requires that M tends to infinity otherwise $\alpha$ tends to zero. If we allow M to grow that big then it means we have infinite resources, or?}
% \red{**Slides.tex contains the proof --- need some modifications**}
\begin{proof}
The proof is presented in Appendix~\ref{sec-apndix-theorem-asymptotically-opt}.
\end{proof}\vspace{-2mm}

% \brown{Concluding remarks -- Numerical results show that the optimality gap tends to vanish for some moderate numbers $K$.}

% \subsection{Unreliable Communication/Links Between the Edge Node and the Sensors}\label{Comm_link}

% ** Unreliable Feedback **

% We consider
% an \textit{error-free} binary/single-bit \textit{command} link  from the edge node to each sensor, and an \textit{error-prone} wireless link from each sensor to the edge node{, i.e., a transmission through the link is either \textit{successful} or \textit{failed}.}
% Let $h_k(t) = 1$  denote the  event that a status update from sensor $k$ is successfully received by the edge node at slot $t$. Otherwise, $h_k(t)=0$, which accounts {for both that} 
% 1) sensor $k$ sends a status update but the transmission is failed, or 2) the sensor does not send an update.
% Let $\xi_k$ be the conditional probability that, given that sensor $k$ transmits a status update, it is successfully received by the edge node, i.e.,  $\Pr\{ h_k(t) = 1 | d_k(t) = 1 \} = \xi_k$, 
% $\forall t$,
% {which}
% represents the \textit{transmit success probability} of the link from sensor $k$ to the edge node.

% \section{Optimal $M$}

%%%%%%%%%%%%%%%%%%%%%%%%%%%%%%%%
\section{Simulation Results}\label{sec_simulation}
In this section, we provide simulation results to demonstrate the performance of the low-complexity relax-then-truncate approach developed in Section~\ref{sec-lowcomplxity_alg}. In addition, simulation results are presented to demonstrate the structural properties of per-sensor optimal policies obtained by the RVIA in Algorithm~\ref{alg_cmdp_RVIA}.

\newcommand\fwL{0.325}
\begin{figure}[t]
\centering
\subfigure [$M =1$ and $K = 40$ ]{%
\includegraphics[width=\fwL \columnwidth]{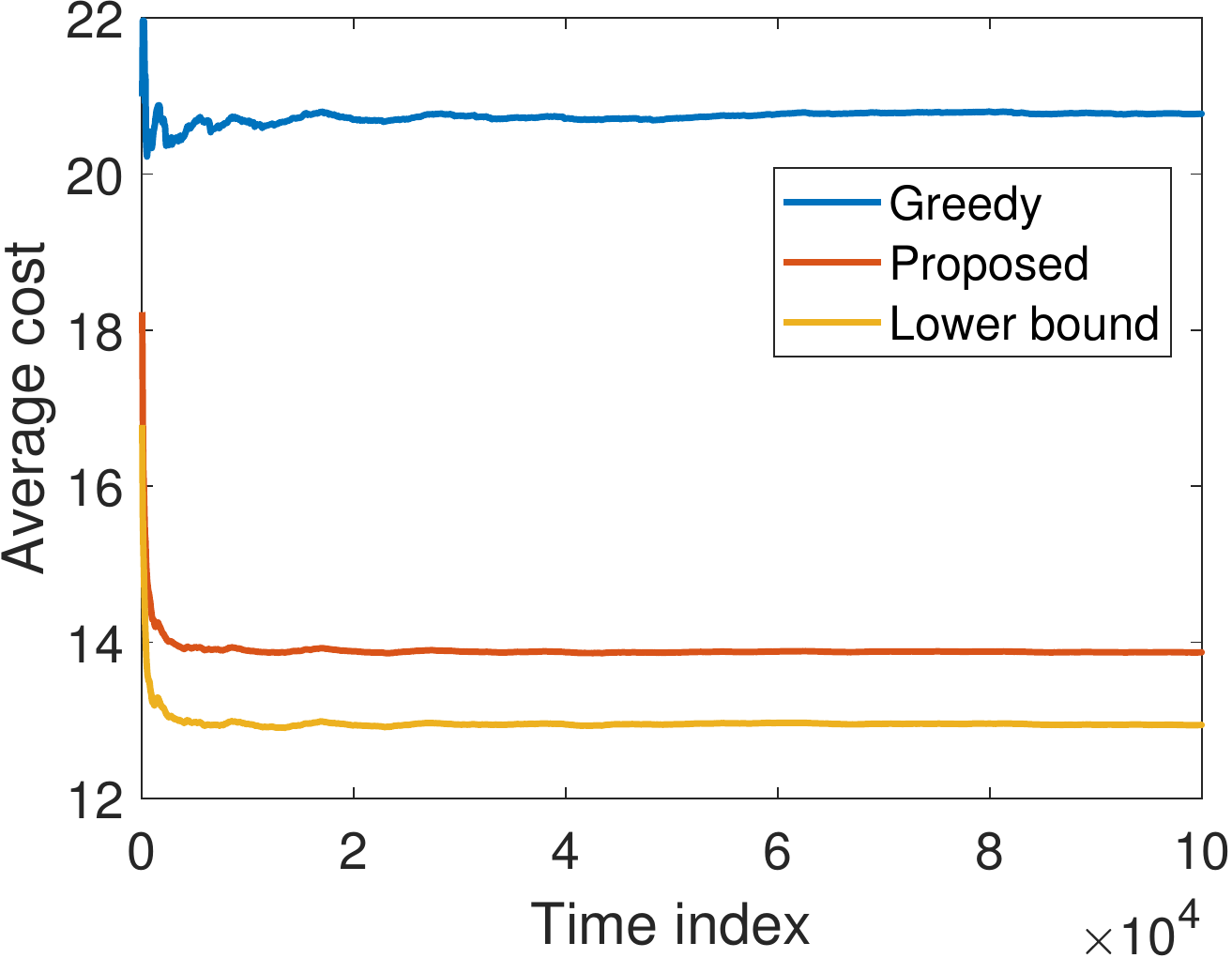}%
}\hspace{0.2mm}
\subfigure [$M =20$ and $K = 800$]{%
\includegraphics[width=\fwL \columnwidth]{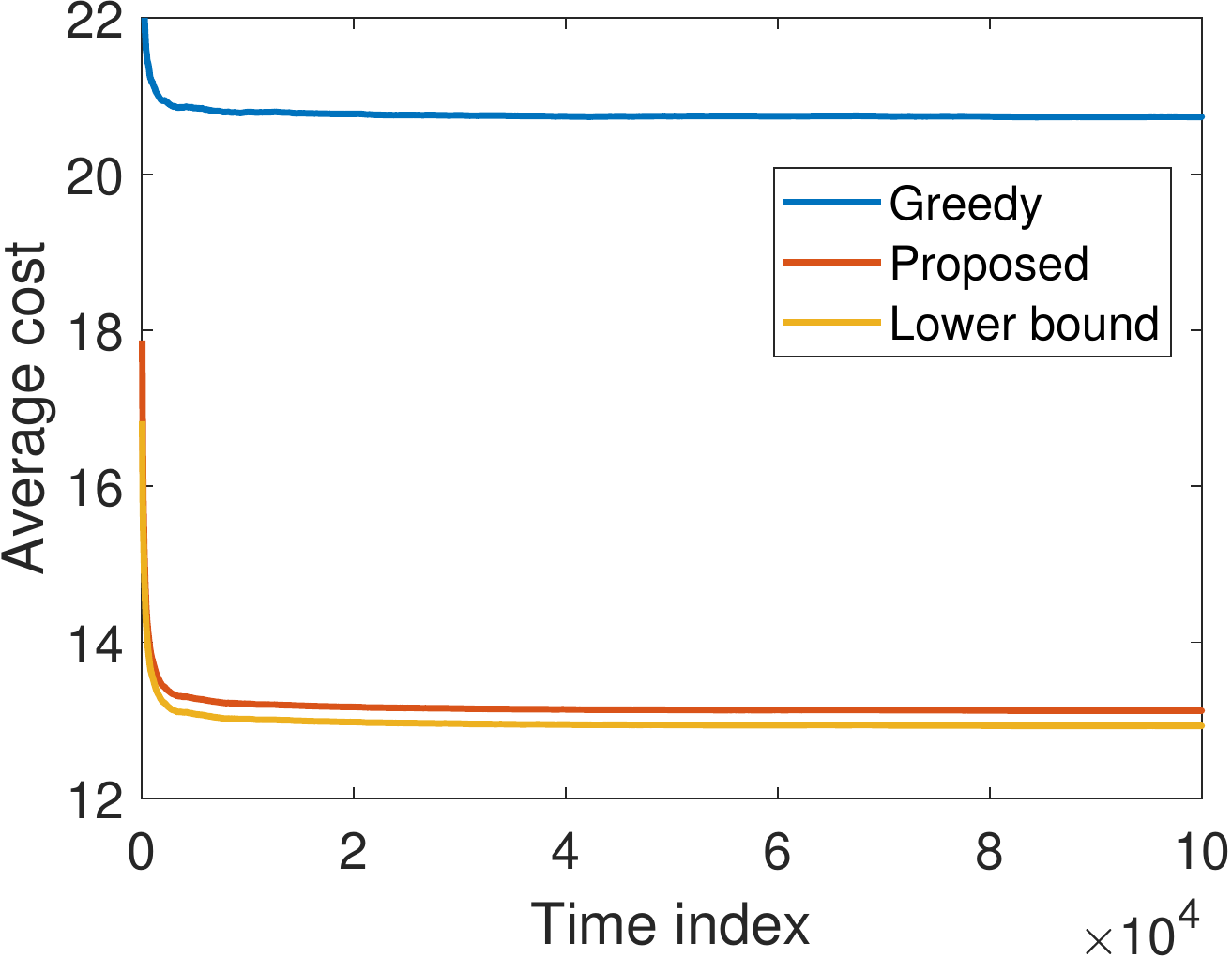}%
}\hspace{0.2mm}
\subfigure [$M = 200$ and $K = 8000$]{%
\includegraphics[width=\fwL \columnwidth]{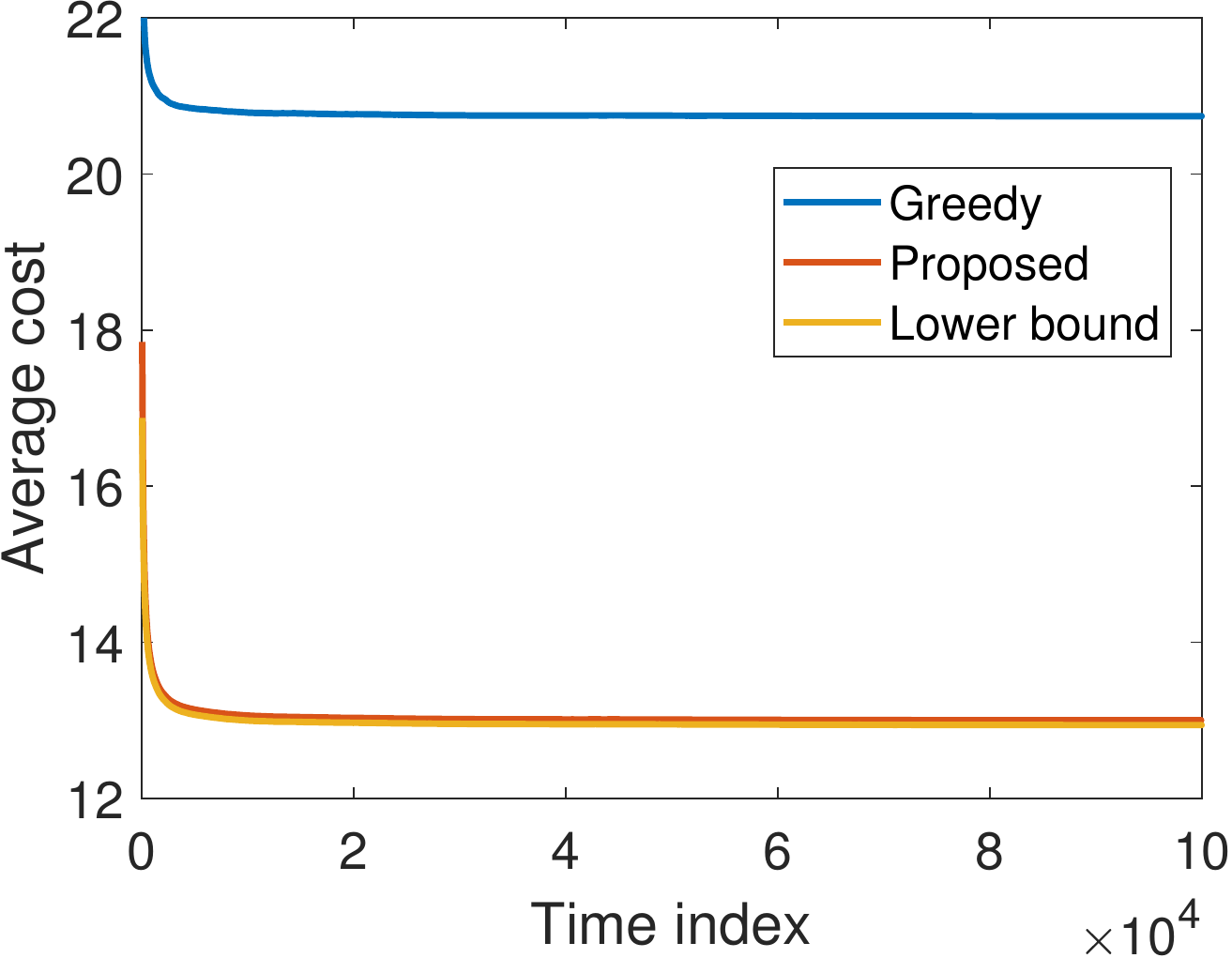}%
}
\vspace{-3mm}
\caption{Performance of the proposed relax-then-truncate algorithm in terms of average cost (i.e., average on-demand AoI over all the sensors and users) over time for different values of the number of sensors $K$ with a fixed normalized transmission budget ${\Gamma = 0.025}$. 
% (a) $K = 40$, (b) $K = 800$, (c) $K = 8000$.   ${\lambda_k \in [0.01,0.02,\dots, 0.1]}$.
}
\vspace{-7mm}
\label{fig_learning}
\end{figure}

\newcommand\fw{.43}
\begin{figure}[t]
\centering
\subfigure [$\Gamma = 0.025$]{%
\includegraphics[width=\fw \columnwidth]{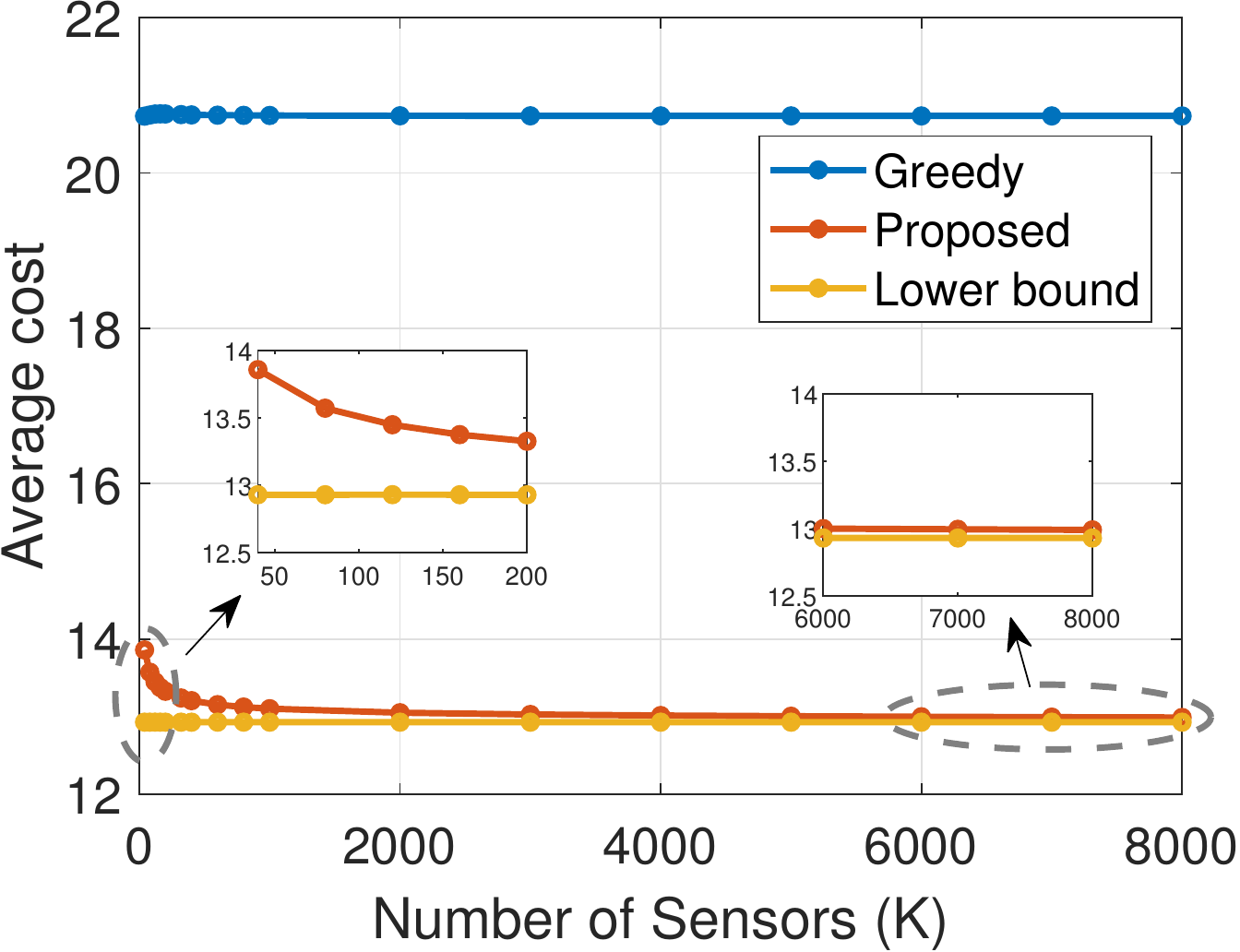}%
}\qquad
\subfigure [$\Gamma = 0.04$]{%
\includegraphics[width=\fw \columnwidth]{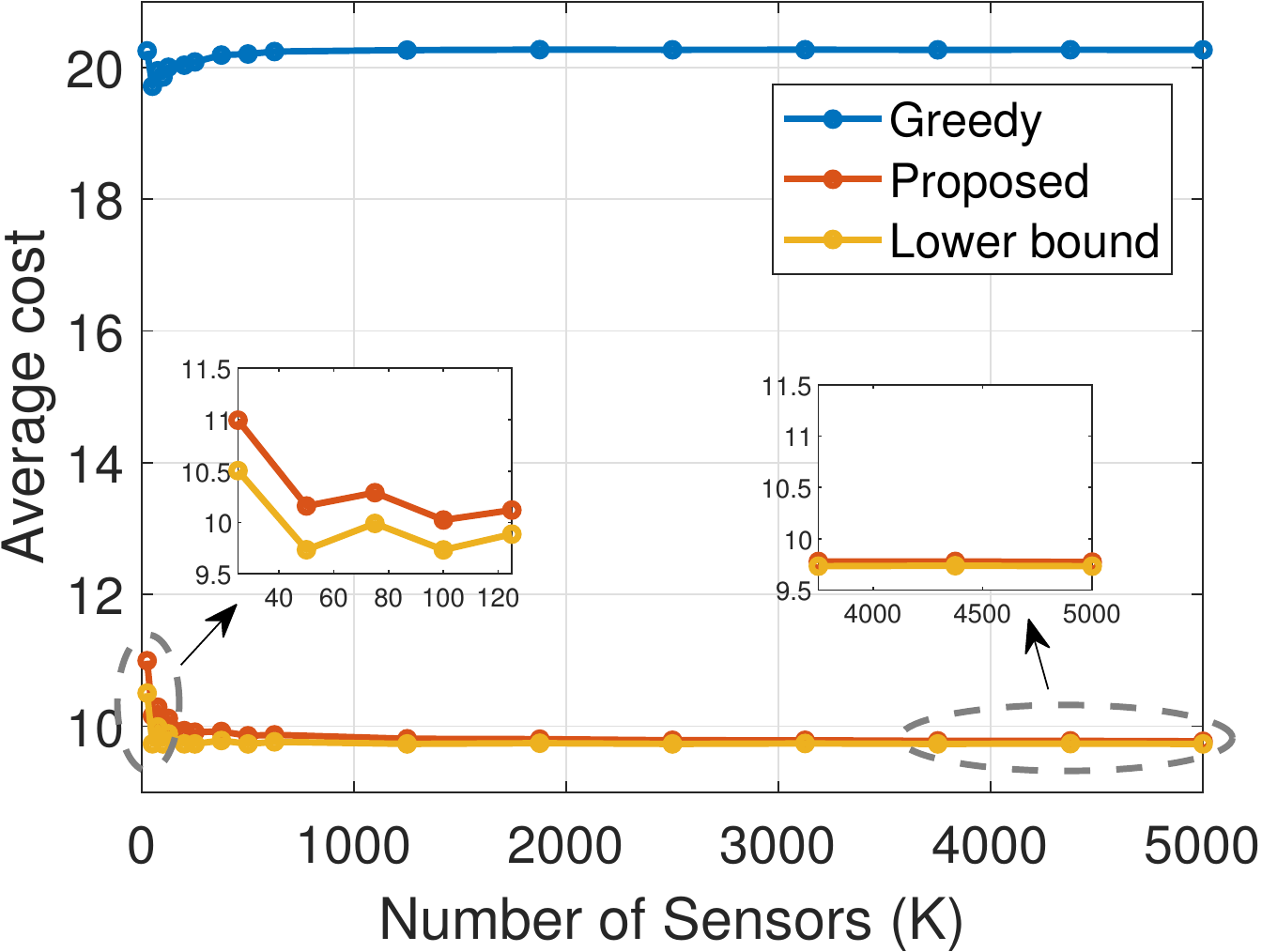}%
}\qquad
\subfigure [$\Gamma = 0.05$]{%
\includegraphics[width=\fw \columnwidth]{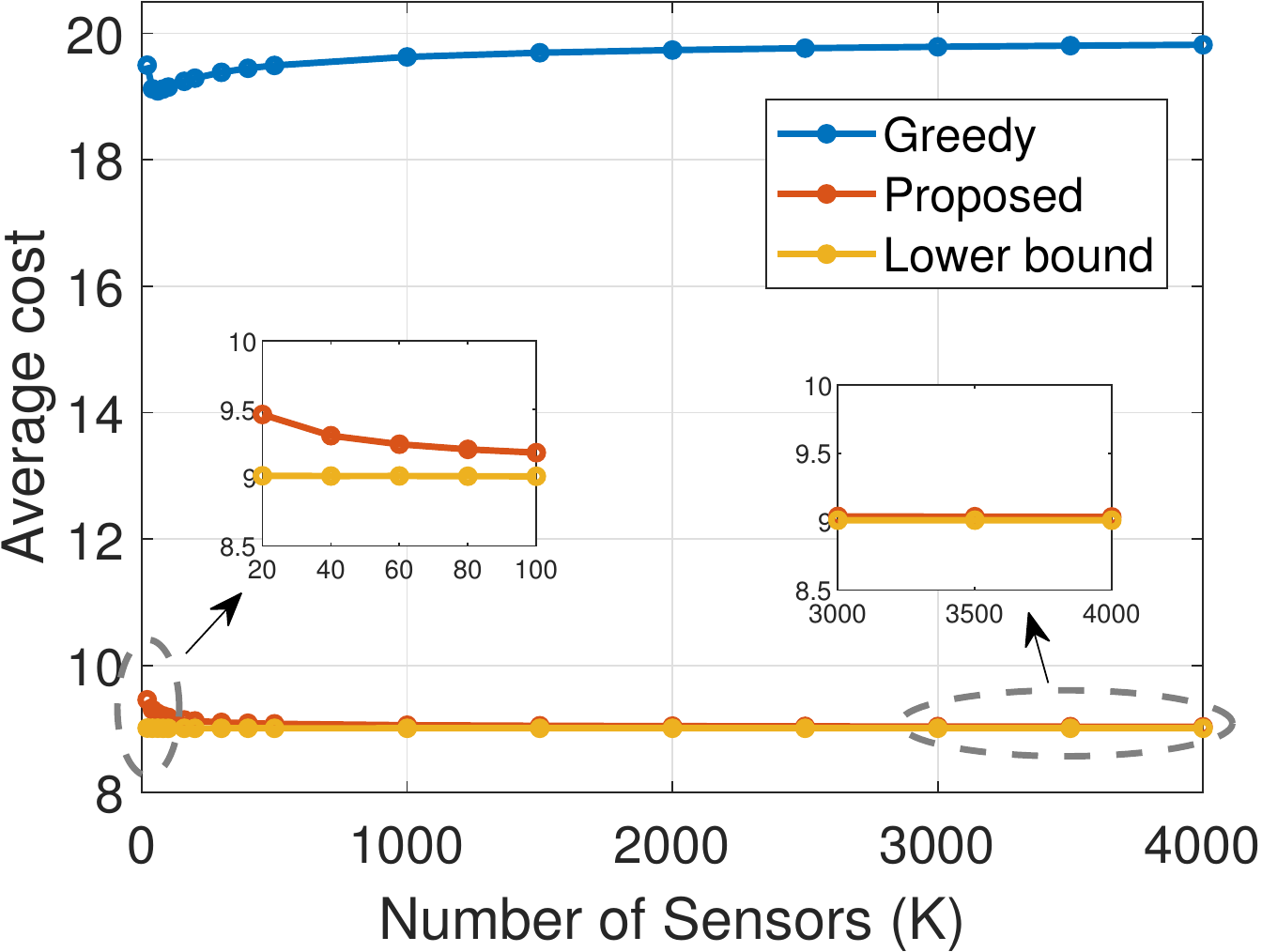}%
}\qquad
\subfigure [$\Gamma = 0.1$]{%
\includegraphics[width=\fw \columnwidth]{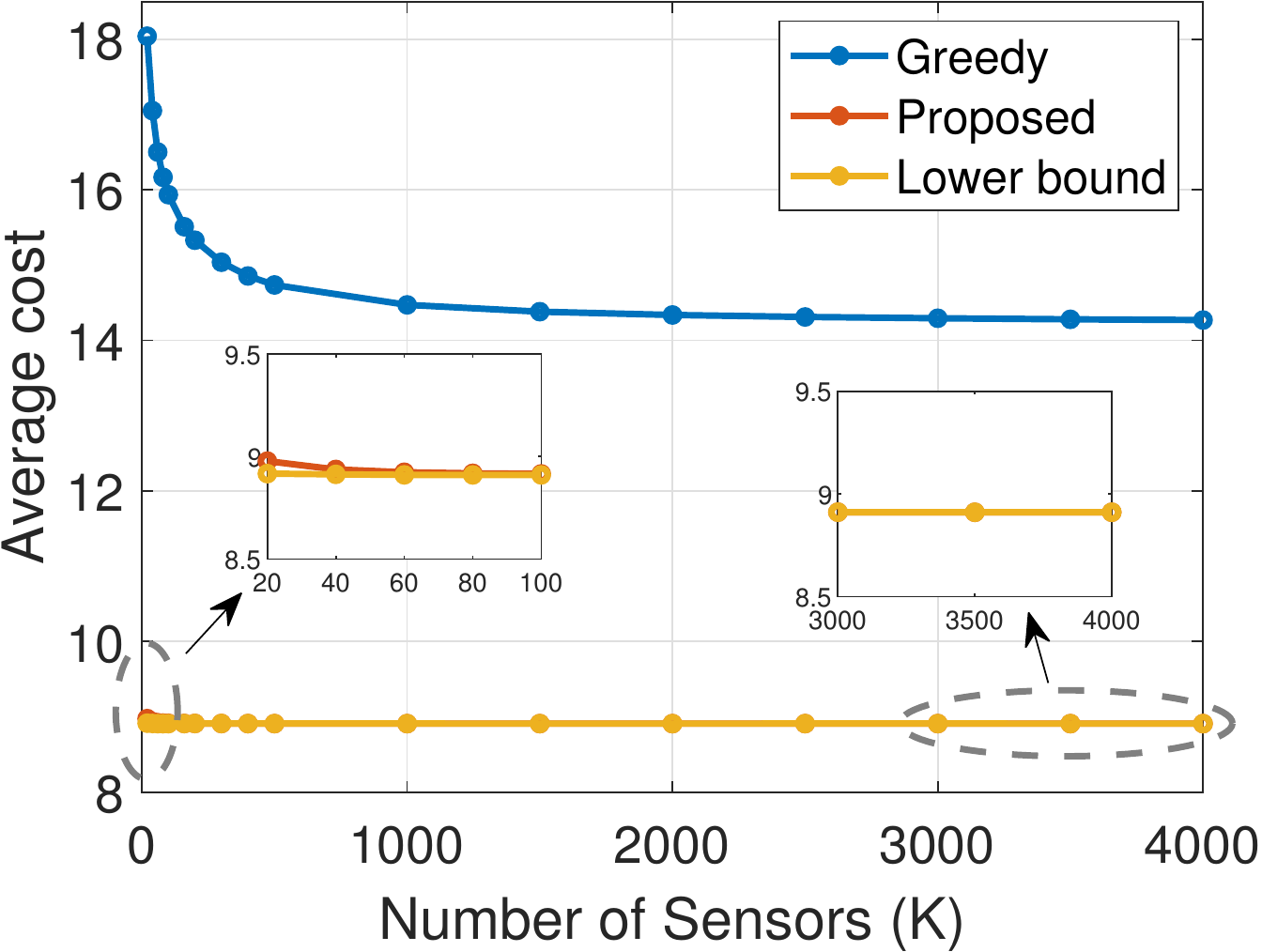}%
}
\vspace{-3mm}
\caption{Performance of the proposed {relax-then-truncate approach} in terms of average cost with respect to the number of sensors $K$ for different values of $\Gamma$. }\vspace{-7mm}
\label{fig_perf_alpha_fixed}
\end{figure}

%%%%%%%%%%%%% Performance 
\subsection{Performance of the Proposed Low-complexity Relax-then-Truncate Approach}
{We consider an IoT network with $N=3$ users, where in each slot, user $n$ requests a status of $f_k$ with probability $p_{k,n} = 0.6$. The battery capacity of each sensor is set to $B_k = 7$ units of
energy and the AoI upper-bound is set to $\Delta^{\mathrm{max}} = 64$.}
{Each sensor is assigned an energy harvesting rate $\lambda_k$ from the set $\{0.01,0.02,\dots, 0.1\}$ in the following sequential order:}

We compare the performance of the proposed relax-then-truncate policy with a greedy (myopic) policy and a lower bound. In the (request-aware) greedy policy, the edge node commands at most $M$ sensors with the largest AoI from the set $\mathcal{W}(t) \triangleq \{k \mid r_k(t) \geq 1, k\in\mathcal{K}\}$, i.e., the set of sensors whose measurements are requested by at least one user. In other words, this \textit{myopic} policy minimizes the expected one-step cost
% immediate cost function 
over all the sensors and users at each slot.
The lower bound is obtained by following an optimal relaxed policy $\pi^\star_{\R}$ (see \eqref{eq_lowerboundinequality}).
% , $\bar{C}_{\pi^\star_{\R}} \leq \bar{C}_{\pi^\star} $.}

% \yellow{; recall that the average cost obtained under policy $\pi^\star_{\R}$ is a lower bound for the average cost obtained by optimal policy $\pi^\star$.}

% \red{As stated in Section~\ref{sec-lowcomplxity_alg}, the average cost obtained under policy $\pi^\star_{\R}$ is a lower bound for the average cost obtained by optimal policy $\pi^\star$ (i.e., $\bar{C}_{\pi^\star_{\R}} \leq \bar{C}_{\pi^\star} $).}

% \brown{Fig.~\ref{fig_learning} depicts the} 

{$\text{Fig.\ \ref{fig_learning}}$} depicts the performance of the relax-then-truncate algorithm over time for different numbers of sensors $K$ with a fixed normalized transmission budget ${\Gamma = 0.025}$. As shown, the proposed algorithm reduces the average cost by approximately $50~\%$ compared to the greedy policy. Furthermore, the gap between the proposed policy and the lower bound is in general small and decreases as $K$ increases. The proposed policy approaches the lower bound for large $K$, which validates the asymptotic optimality of the proposed algorithm as proved in Theorem~\ref{theorem-asymptotically-opt}.

% \brown{The reason is that ...}

% \brown{
% Figure~\ref{fig_learning}:
% \begin{itemize}
%     \item learning and convergence behaviour over time
%     \item compare to the greedy policy
%     \item comparing these figures with each other, the gap reduces as $K$ increases
%     \item Asymptotic optimality behaviour of the proposed algorithm
% \end{itemize}
% }

$\text{Fig.\ \ref{fig_perf_alpha_fixed}}$ depicts the performance of the relax-then-truncate algorithm with respect to the number of sensors $K$ for different values of normalized transmission budget $\Gamma$. The results are obtained by averaging each algorithm over $50$ episodes where each episode takes $10^6$ slots. Due to asymptotic optimality of the proposed algorithm, for all values of $\Gamma$, the gap between the proposed policy and the lower bound is very small for large values of  $K$. Comparing \mbox{$\text{Figs.\ \ref{fig_perf_alpha_fixed}(a)--(d)}$} with each other, it can be seen that, as $\Gamma$ increases, the proposed policy converges to the optimal performance faster. This is because the proportion of the sensors that can be commanded by the edge node at each slot increases as $\Gamma$ increases, and hence, the proportion of the sensors that is truncated (i.e., the sensors that are not commanded under $\tilde{\pi}$ compared to $\pi^\star_{\R}$) decreases.
% , thus, the performance of the truncation based algorithm becomes closer to the lower bound. 

% \yellow{In addition, the average cost decreases as $\alpha$ increases, because the edge node has more resources, and thus, it  can command more sensors at each time slot and serves the users via fresh measurements more often.}

% \brown{
% Figure~\ref{fig_perf_alpha_fixed}:
% \begin{itemize}
%     \item average $10$ each one runs for $2e5$ time slot
%     \item the gap reduces as $K$ increases for different values of fixed $\frac{M}{K}$
%     \item Asymmetrical optimality behaviour of the proposed algorithm
%     \item As $\alpha$ increases, the proposed algorithm converges to the optimal performance faster. This is because of the  less truncation at each time slot.
%     \item comparing these figures, it is shown that the average cost reduces as $\alpha$ increase. 
% \end{itemize}
% }

\begin{figure}[t]
\centering
\subfigure [$K = 100$]{%
\includegraphics[width=\fw \columnwidth]{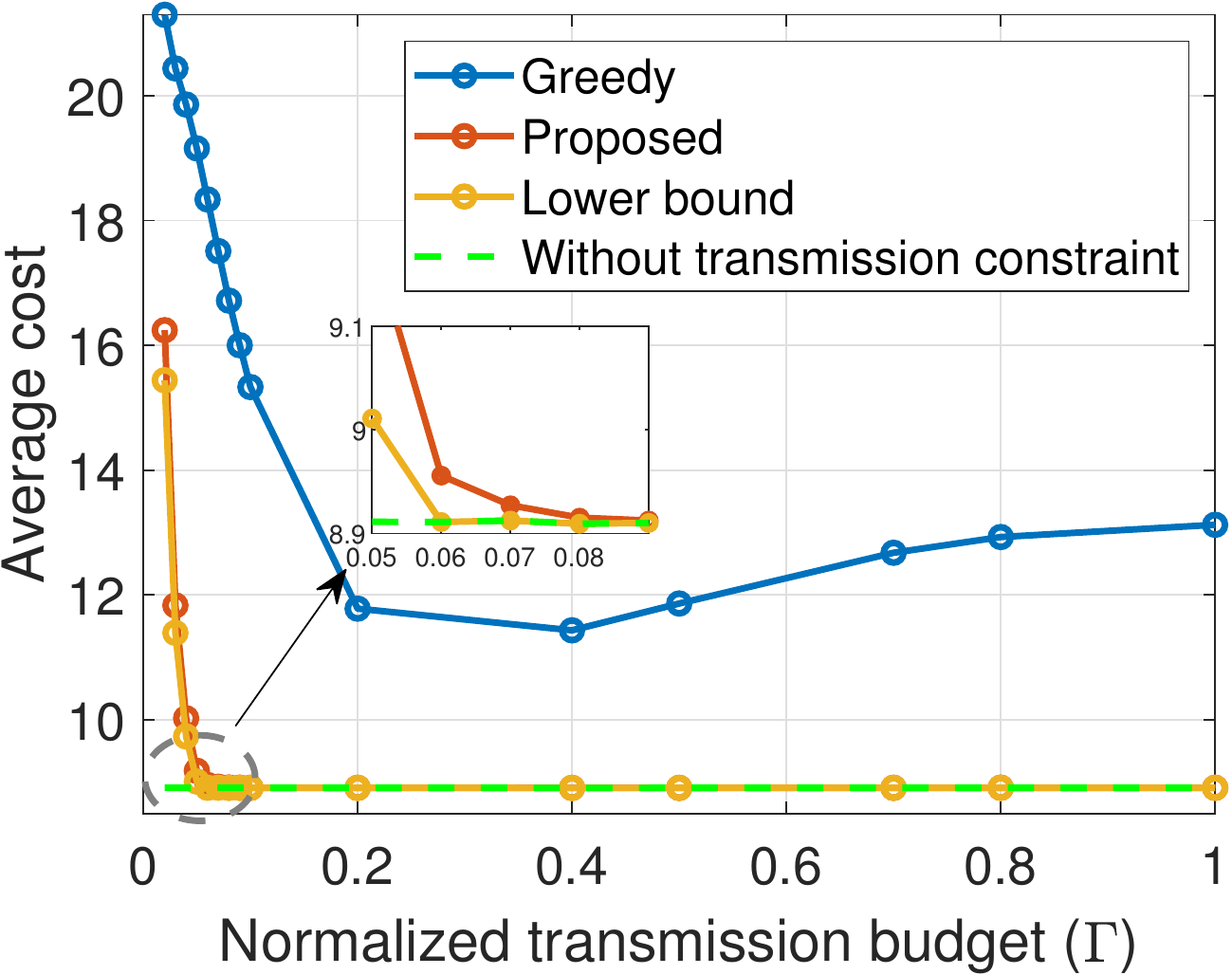}%
}\qquad
\subfigure [$K = 1000$]{%
\includegraphics[width=\fw \columnwidth]{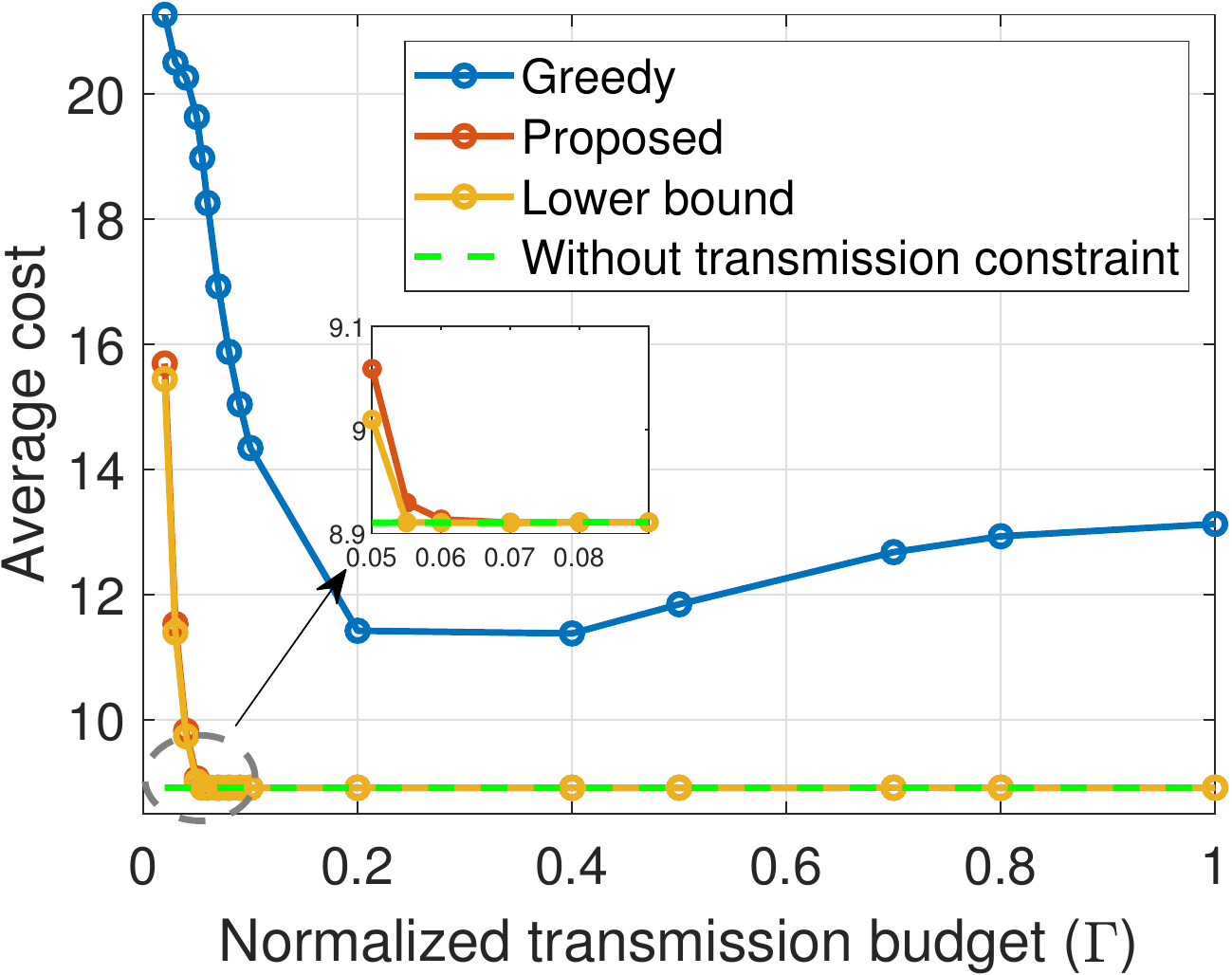}%
}
\vspace{-3mm}
\caption{Performance of the proposed algorithm in terms of average cost with respect to $\Gamma$.}
% \vspace{-7mm}
\label{fig_perf_K_fixed}
\centering
\subfigure [$K = 100$]{%
\includegraphics[width=\fw \columnwidth]{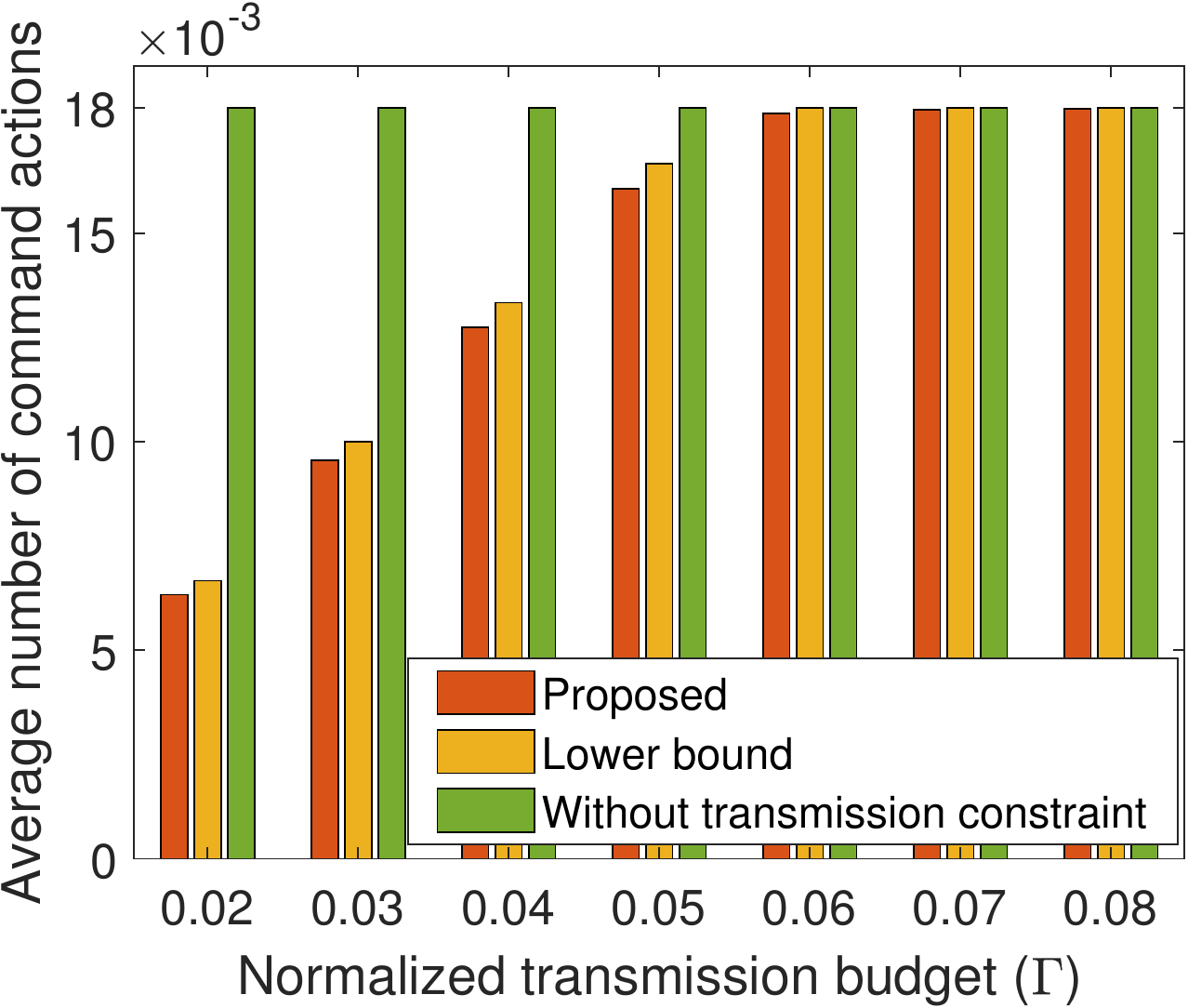}%
}\qquad
\subfigure [$K = 1000$]{%
\includegraphics[width=\fw \columnwidth]{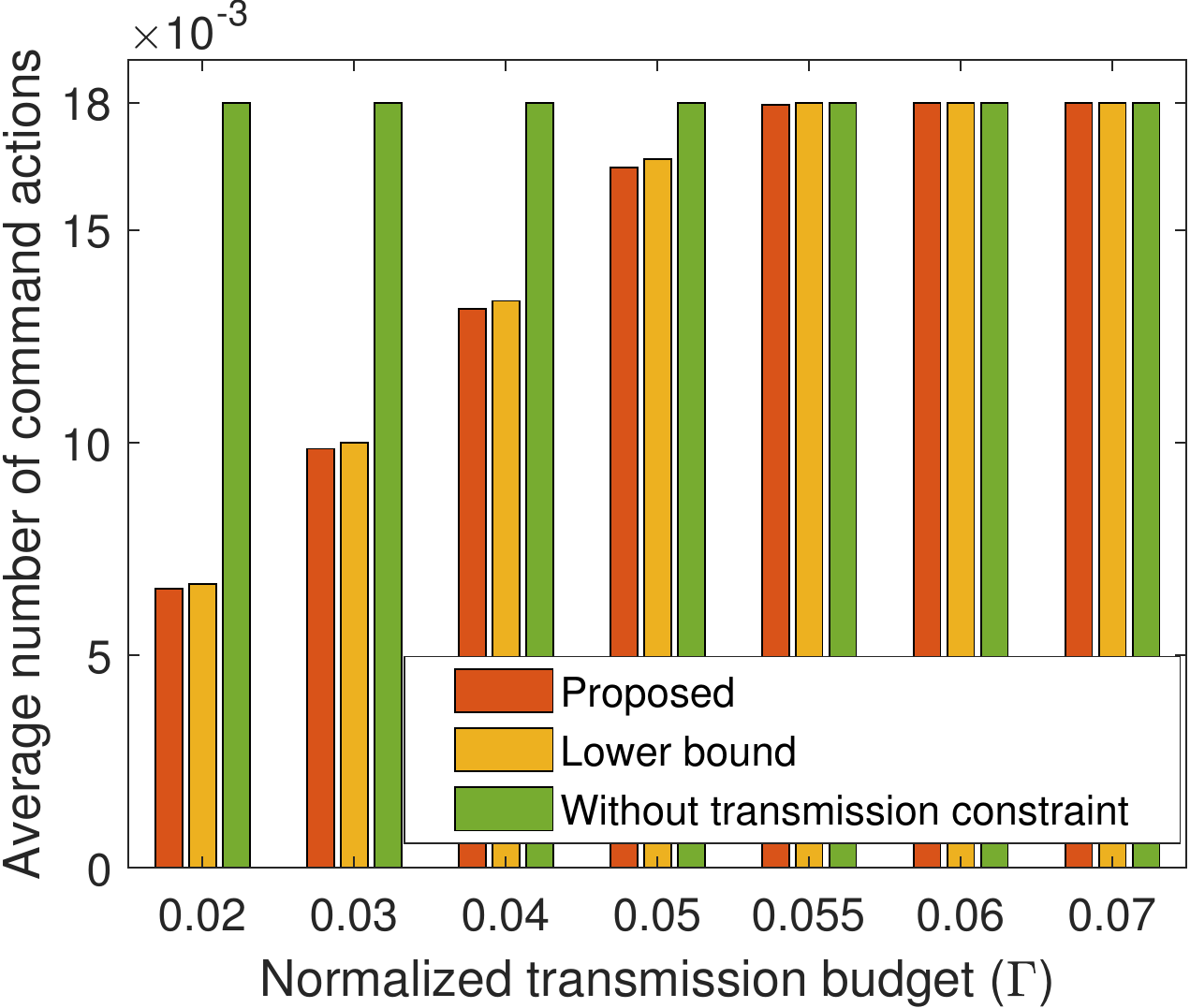}%
}
\vspace{-3mm}
\caption{Average number of command actions with respect to $\Gamma$. 
% \st{for fixed number of sensors $K$}.
}
\vspace{-7mm}
\label{fig_avg_command}
\end{figure}

$\text{Fig.\ \ref{fig_perf_K_fixed}}$ and $\text{Fig.\ \ref{fig_avg_command}}$ illustrate the average cost and the average number of command actions, respectively, with respect to the normalized transmission budget $\Gamma$. %\st{for a fixed number of sensors $K$}.
For the benchmarking, we also plot the performance of an optimal policy for the case without any transmission constraint (i.e., $M = K$) \cite{hatami2021spawc}.
As shown in $\text{Fig.\ \ref{fig_perf_K_fixed}}$, the average cost for the proposed algorithm decreases as $\Gamma$ increases.
This is because, for fixed $K$, the transmission budget $M$ increases by increasing $\Gamma$, and thus, the edge node can command more sensors at each slot, which results in serving the users via fresh measurements more often.
% , and thus, the average on-demand AoI (i.e., cost) decreases.}
% \red{By definition $\alpha = \frac{M}{K}$, thus, for fixed $K$, by increasing $\alpha$ the value of  $M$ increases as well.} 
% Therefore\yellow{, as $\alpha$ increases,} the edge node can command more sensors at each time slot and serve the users via fresh measurements more often, and thus, the average on-demand AoI (i.e., cost) decreases. 
Interestingly, from a certain point onward, increasing $\Gamma$ does not decrease the average cost. This is because the average number of command actions does not increase anymore, as shown in $\text{Fig.\ \ref{fig_avg_command}}$, i.e., the constraint \eqref{st_opt2} becomes inactive and the edge node has more transmission budget than it needs.
In these cases, the energy limitation of the sensors, which only rely on the energy harvesting, becomes dominant and does not allow the sensors to transmit more often. 
% \red{In this case,} the edge node has more transmission resource than it needs.

% \brown{
% Figure~\ref{fig_perf_K_fixed}:
% \begin{itemize}    
%     \item average $50$ each one runs for $1e6$ time slot
%     \item for the benchmarking, we also consider the case of no constraint 
%     \item The average cost decreases as $\alpha$ increase, because more bandwidth is equal to more command, then more fresh measurement
%     \item from a point onward, the average cost does not reduces any more, this is the point regarding the $\mu^\star = 0$. Because, ...
% \end{itemize}
% }

% \red{** Average resource usage can be plotted here! Average number of command action with respect to $\alpha$ for fixed number of sensors $K$ **}

% \newpage

% \footnote{All sensors are assumed to be identical in this part.}
%%%%%%%%%%%%%%%%%%%% Structure
\subsection{Structural Properties of Per-Sensor Deterministic Policies for the Relaxed Problem}\label{sec_sim_structure}
Here, we consider a setup with  $K = 400$ identical sensors with battery capacity $B_k = 15$ units of energy.
We analyze the structural properties of a per-sensor policy obtained by Algorithm~\ref{alg_cmdp_RVIA} for a particular sensor $k$, i.e., $\pi^\star_{\R,\mu^*,k}$, and investigate the effect of the transmission budget $M$, energy harvesting rate $\lambda_k$, and request probability $p_{k,n}$.
$\text{Fig.\ \ref{fig_structure_r}}$ illustrates the structure of $\pi^\star_{\R,\mu^*,k}$, where each point represents a potential per-sensor state as a three-tuple $s = (r,b,{\Delta})$. For each such state, a blue point indicates that the optimal action is to command the sensor (i.e., $\pi^\star_{\R,\mu^*,k}(s) = 1$), whereas a red point means not to command (i.e., $\pi^\star_{\R,\mu^*,k}(s) = 0$). The set of the blue points is referred to as the \textit{command region} hereinafter.

\newcommand\fwstruct{0.24}
\begin{figure}[t]
\centering
\subfigure [$r = 0$]{%
\includegraphics[width=\fwstruct \columnwidth]{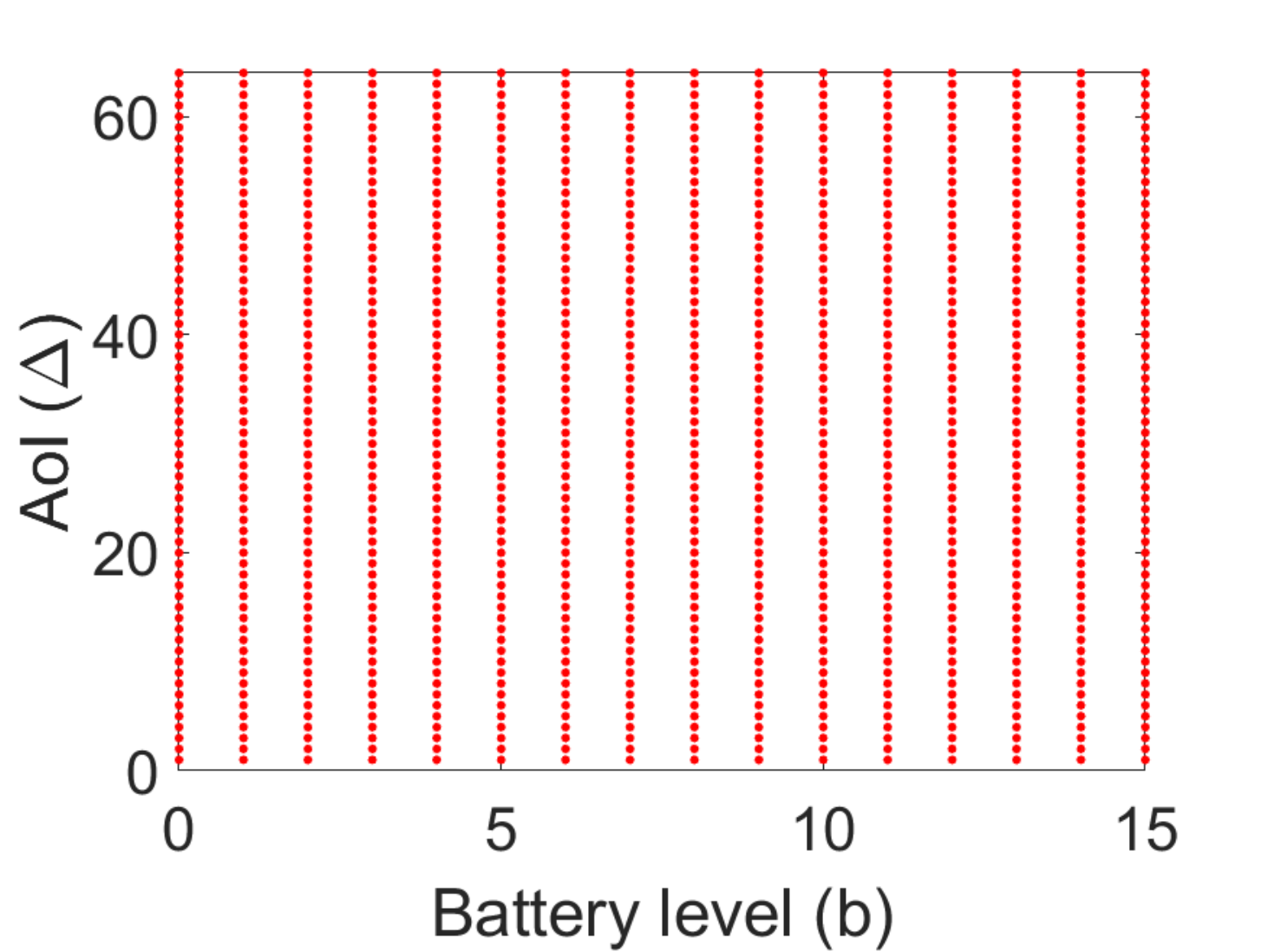}%
}
\subfigure [$r = 1$]{%
\includegraphics[width=\fwstruct \columnwidth]{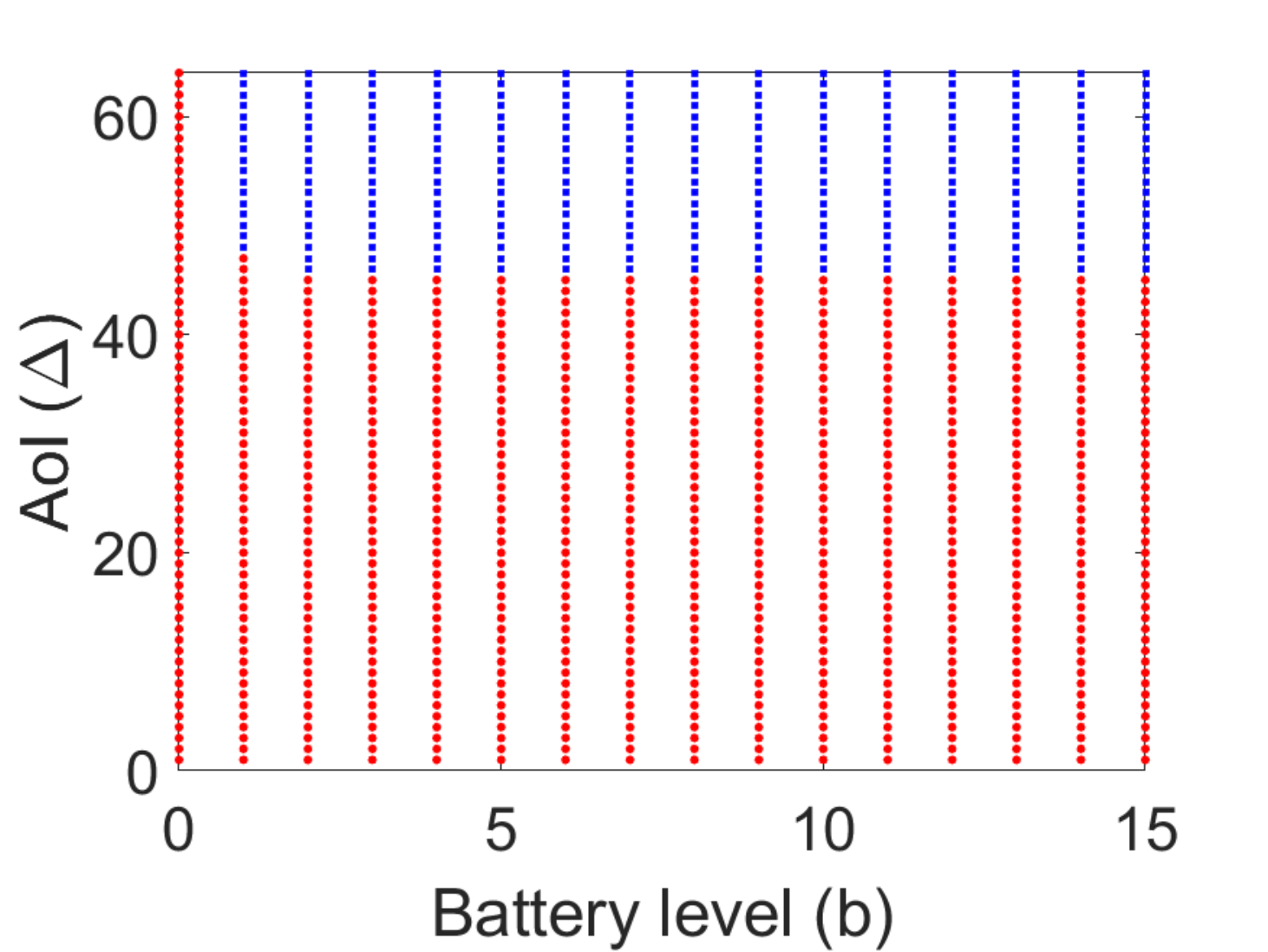}%
}
\subfigure [$r = 2$]{%
\includegraphics[width=\fwstruct \columnwidth]{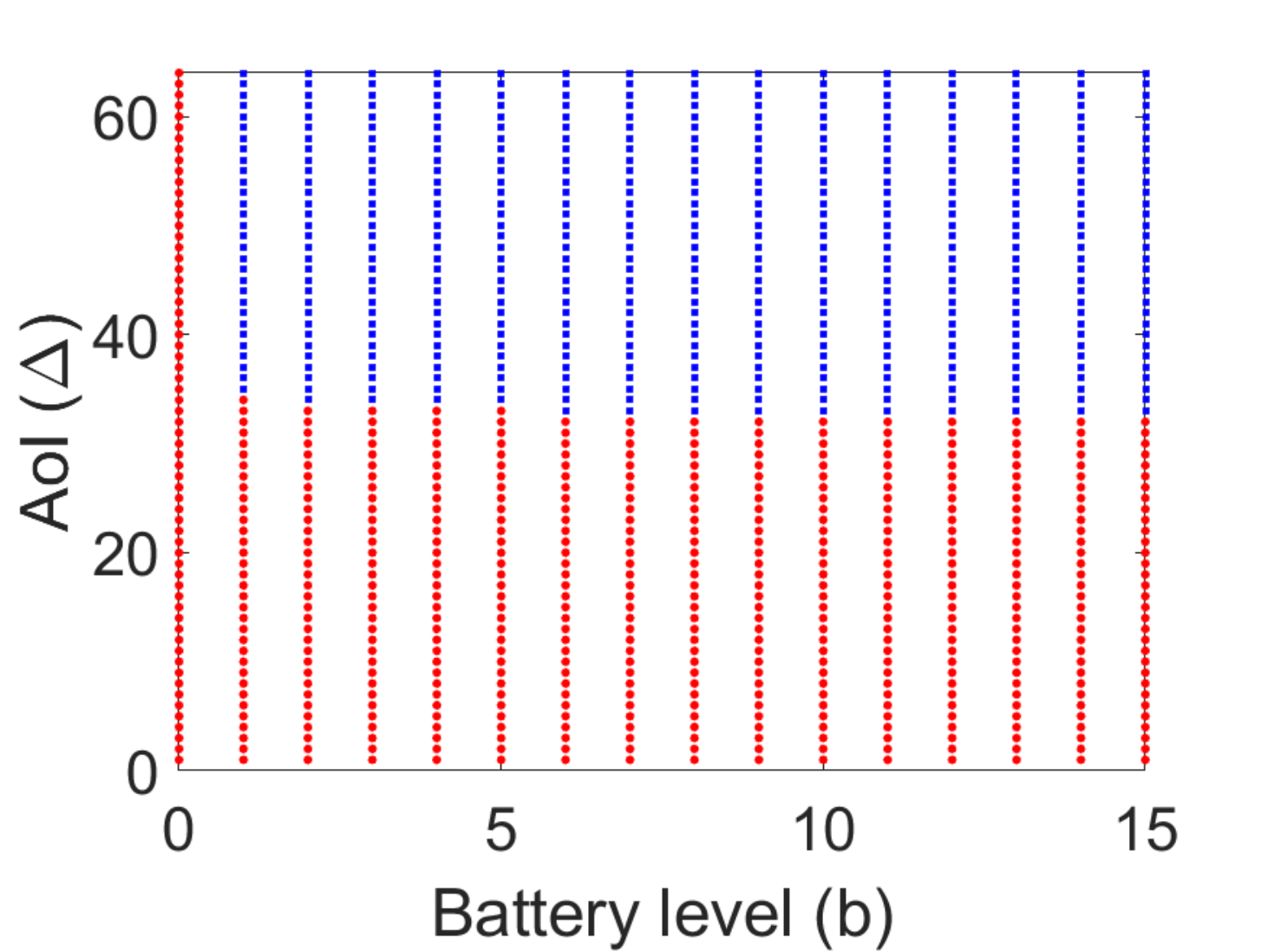}%
}
\subfigure [$r = 3$]{%
\includegraphics[width=\fwstruct \columnwidth]{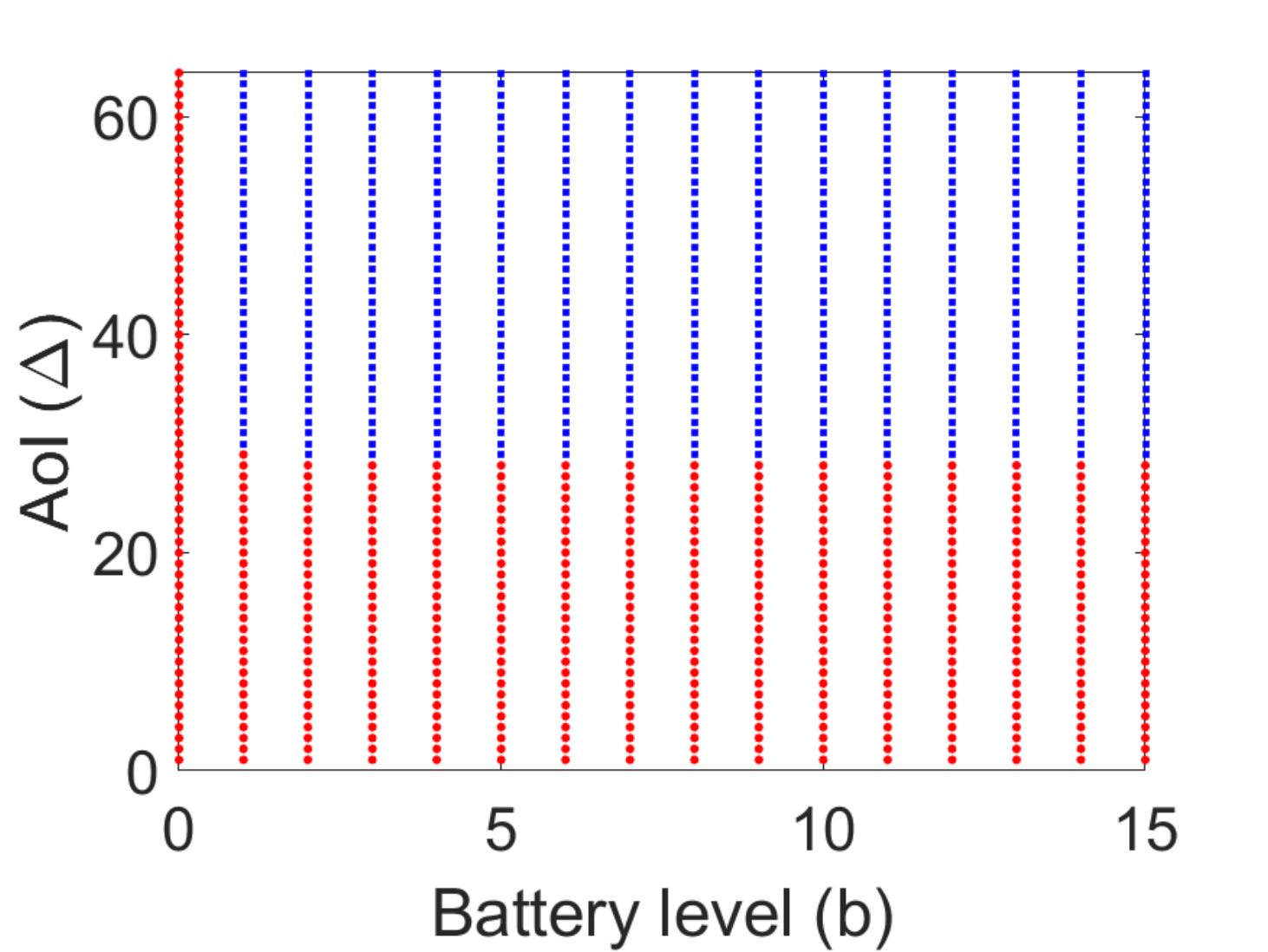}%
}
\vspace{-3mm}
\caption{Structure of an optimal policy for sensor~$k$ (i.e., $\pi^\star_{\R,\mu^*,k}$) for each state $s = \{r,b, {\Delta}\}$, where $M = 10$, $\lambda_k = 0.06$, and $p_{k,n} = 0.2$. Red: no command; blue: command.}
\vspace{-7mm}
\label{fig_structure_r}
\end{figure}

\begin{figure*}[t]
\centering
\subfigure [$M = 15$]{%
\includegraphics[width=\fwstruct \columnwidth]{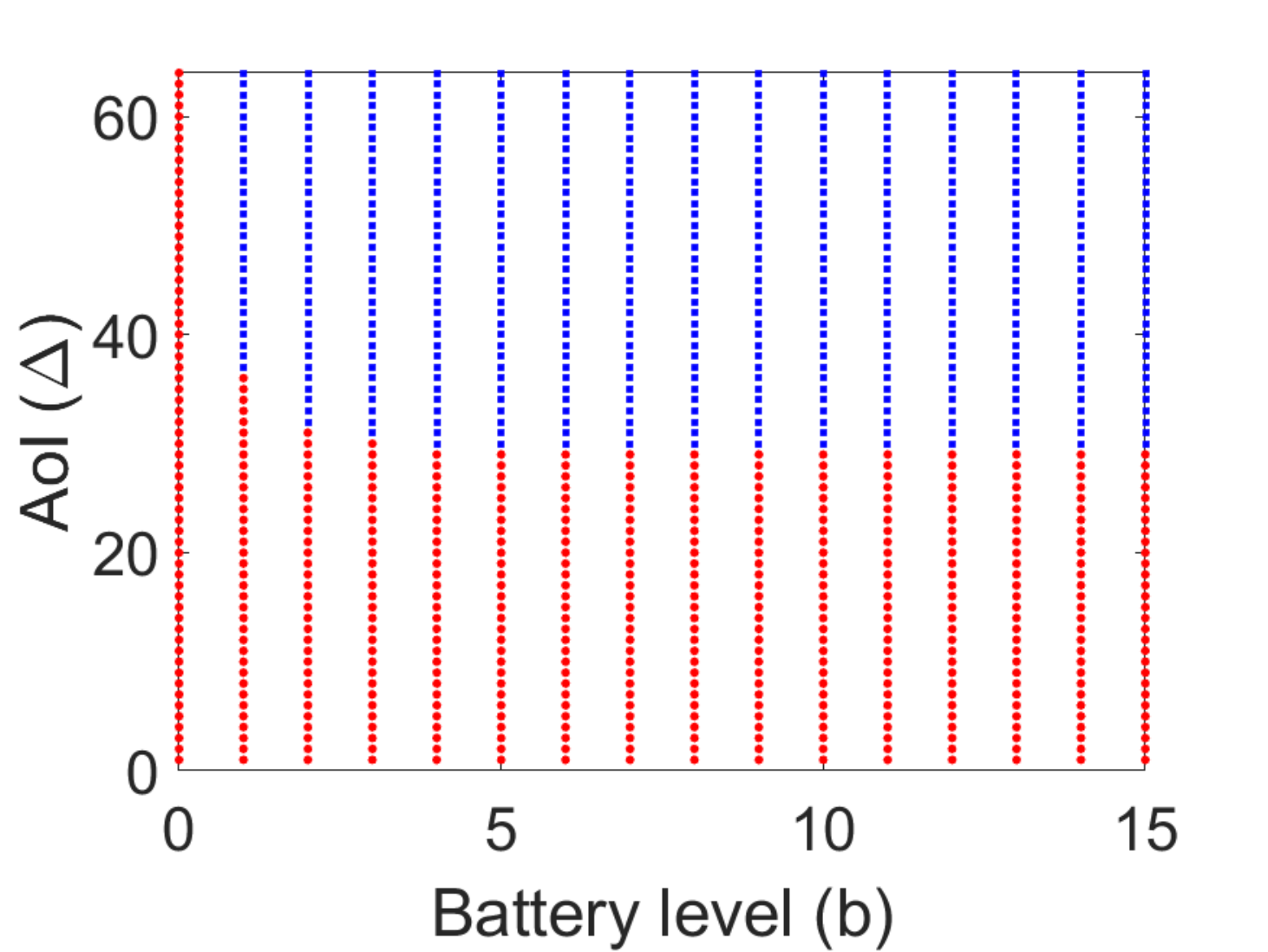}%
}
\subfigure [$M = 20$]{%
\includegraphics[width=\fwstruct \columnwidth]{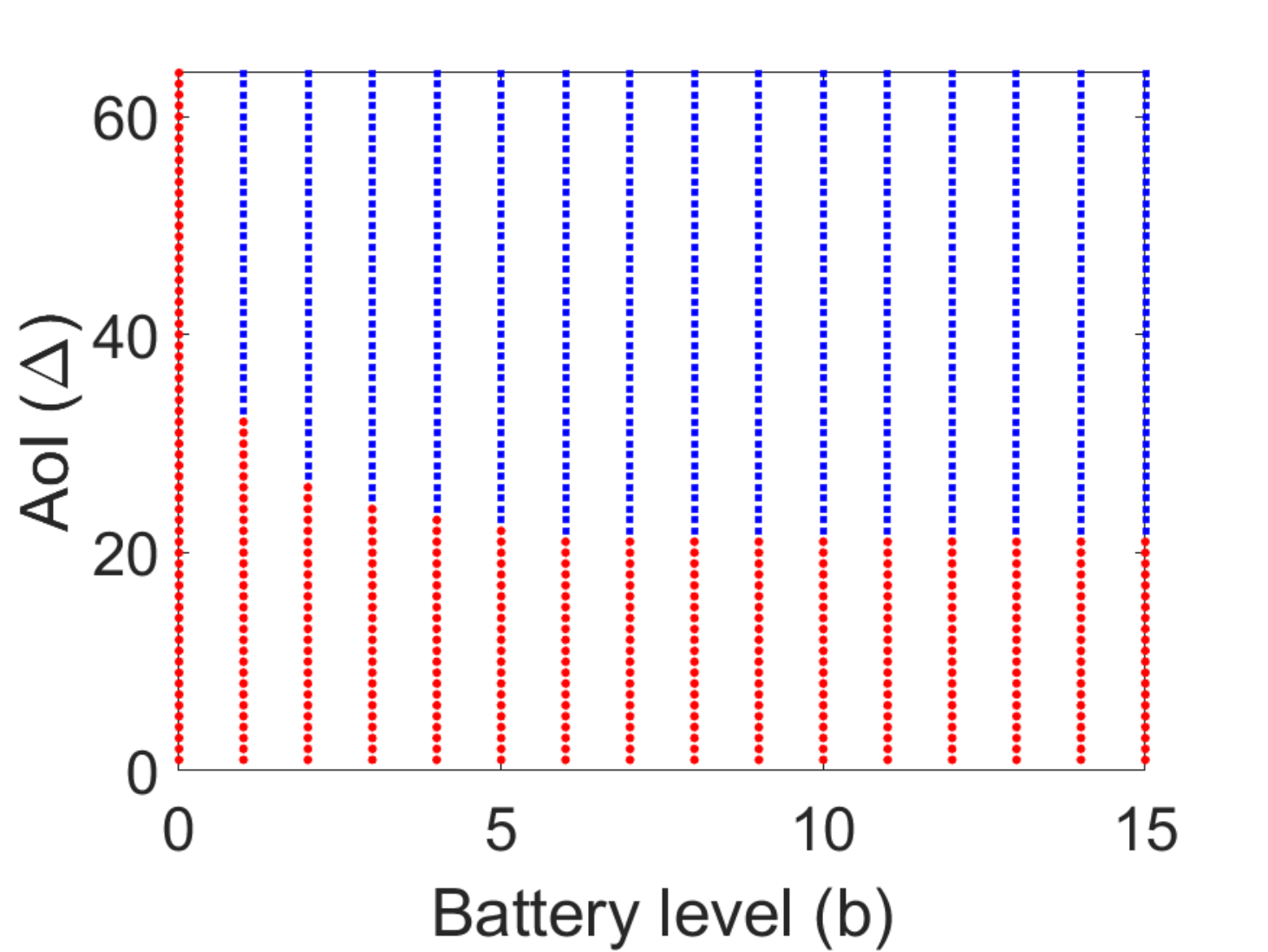}%
}
\subfigure [$M = 25$]{%
\includegraphics[width=\fwstruct \columnwidth]{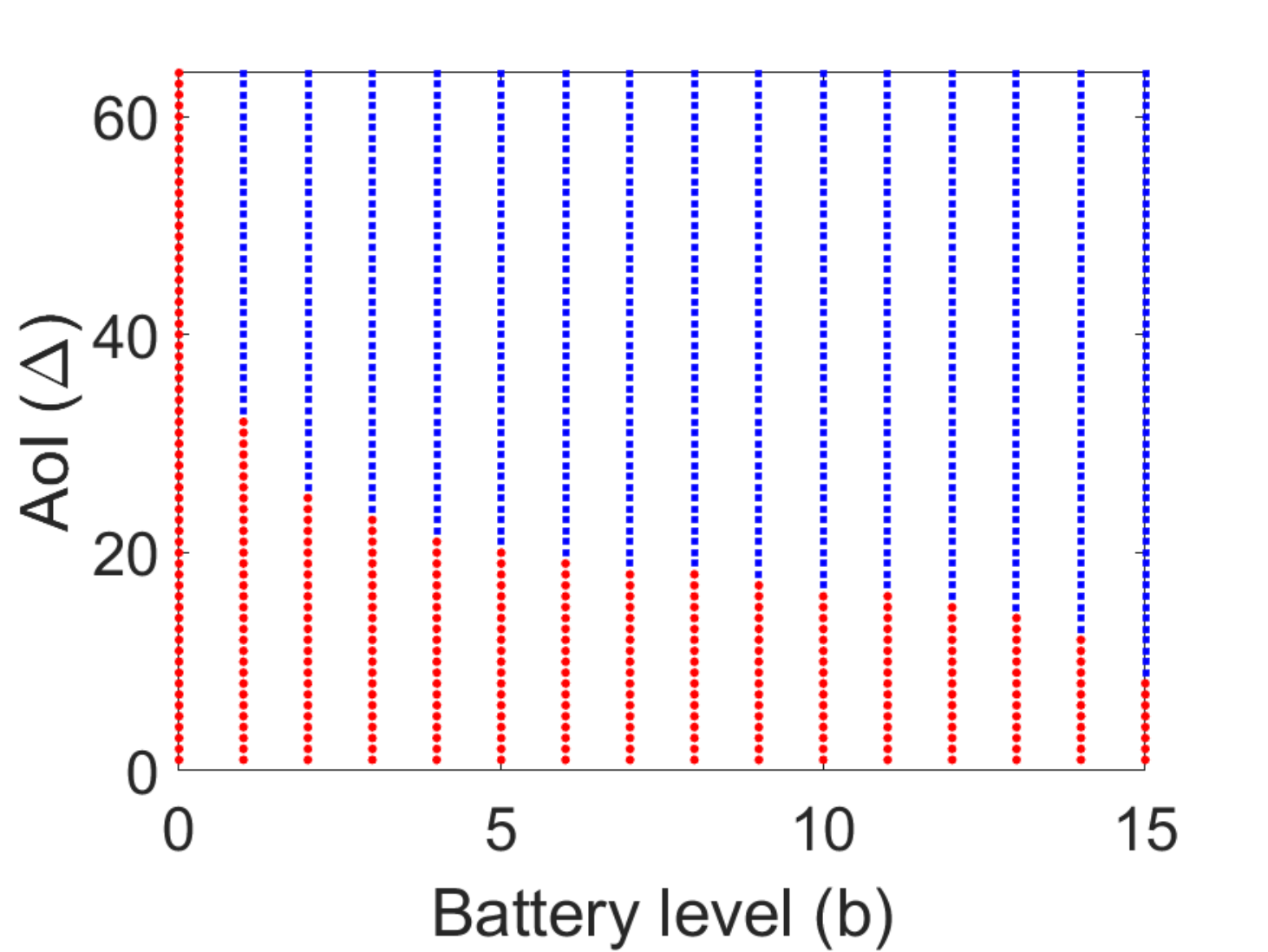}%
}
\subfigure [$M = 30$]{%
\includegraphics[width=\fwstruct \columnwidth]{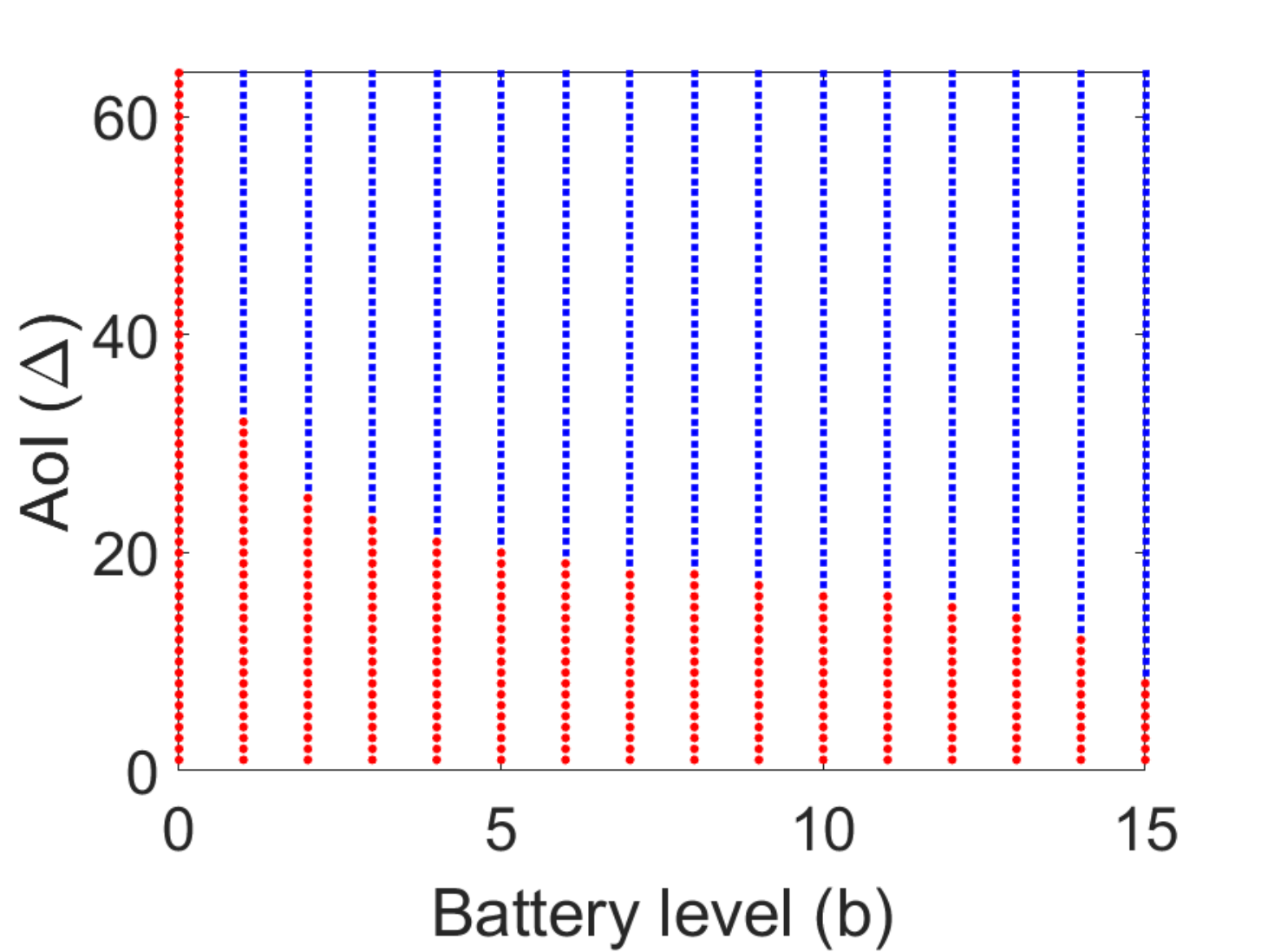}%
}
\vspace{-3mm}
\caption{Structure of an optimal policy for sensor~$k$ (i.e., $\pi^\star_{\R,\mu^*,k}$) in states $s = \{1,b, {\Delta}\}$ for different numbers of transmission budget $M$, where $\lambda_k = 0.06$ and $p_{k,n} = 0.2$.}
% \vspace{-7mm}
\label{fig_structure_M}
% \end{figure*}

% \begin{figure*}[t]
\centering
\subfigure [$\lambda_k = 0.04$]{%
\includegraphics[width=\fwstruct \columnwidth]{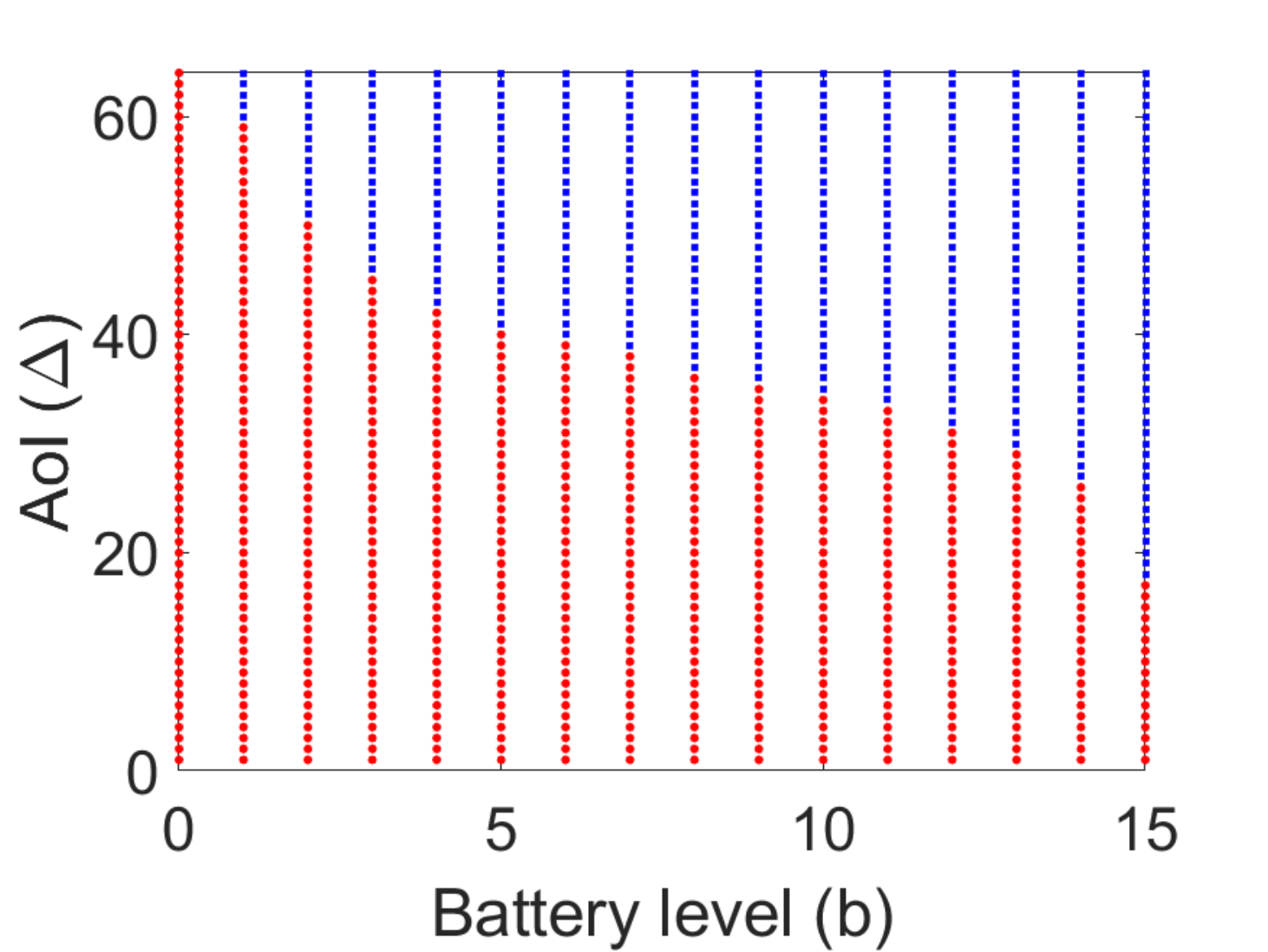}%
}
\subfigure [$\lambda_k = 0.06$]{%
\includegraphics[width=\fwstruct \columnwidth]{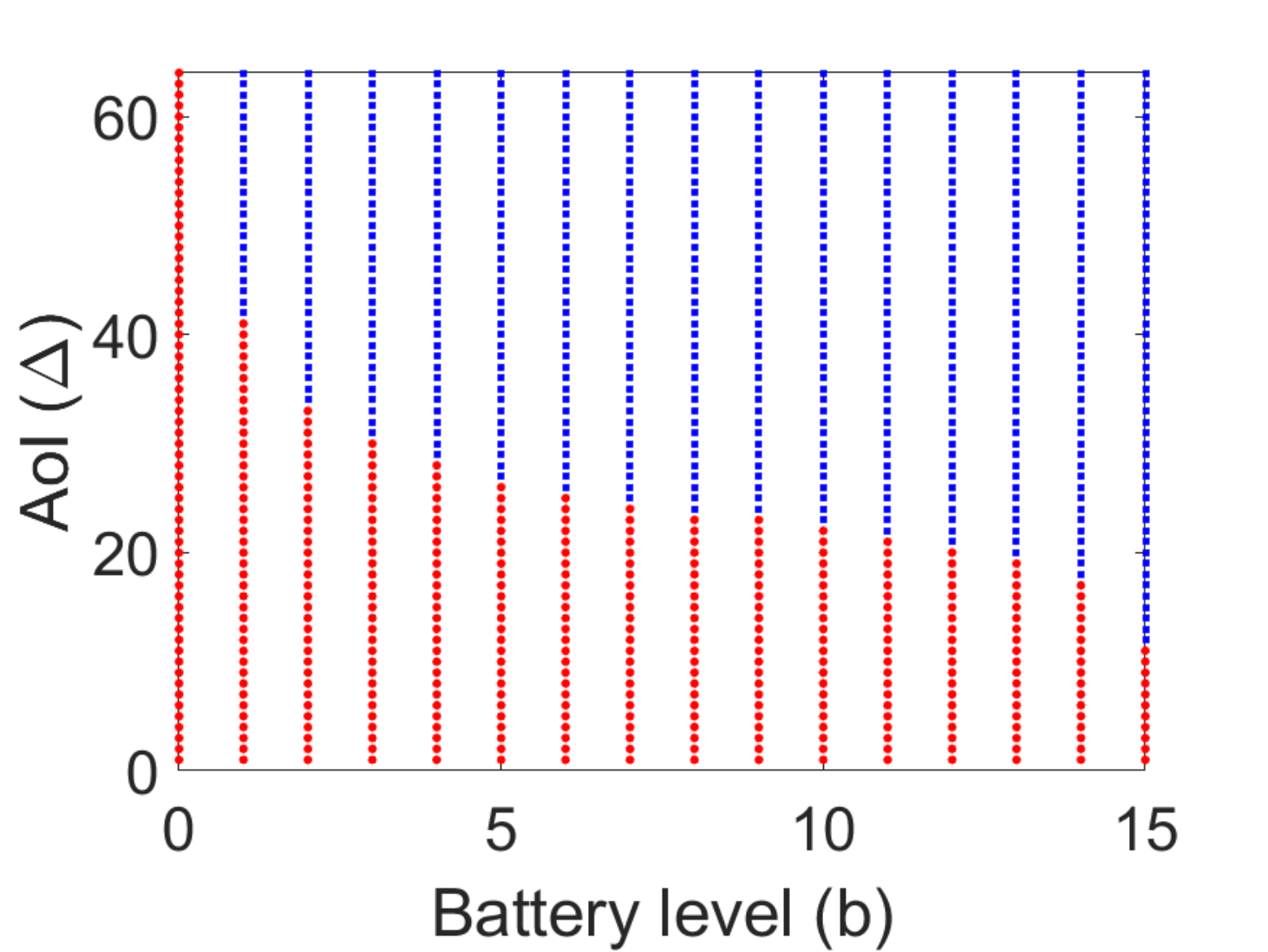}%
}
\subfigure [$\lambda_k = 0.08$]{%
\includegraphics[width=\fwstruct \columnwidth]{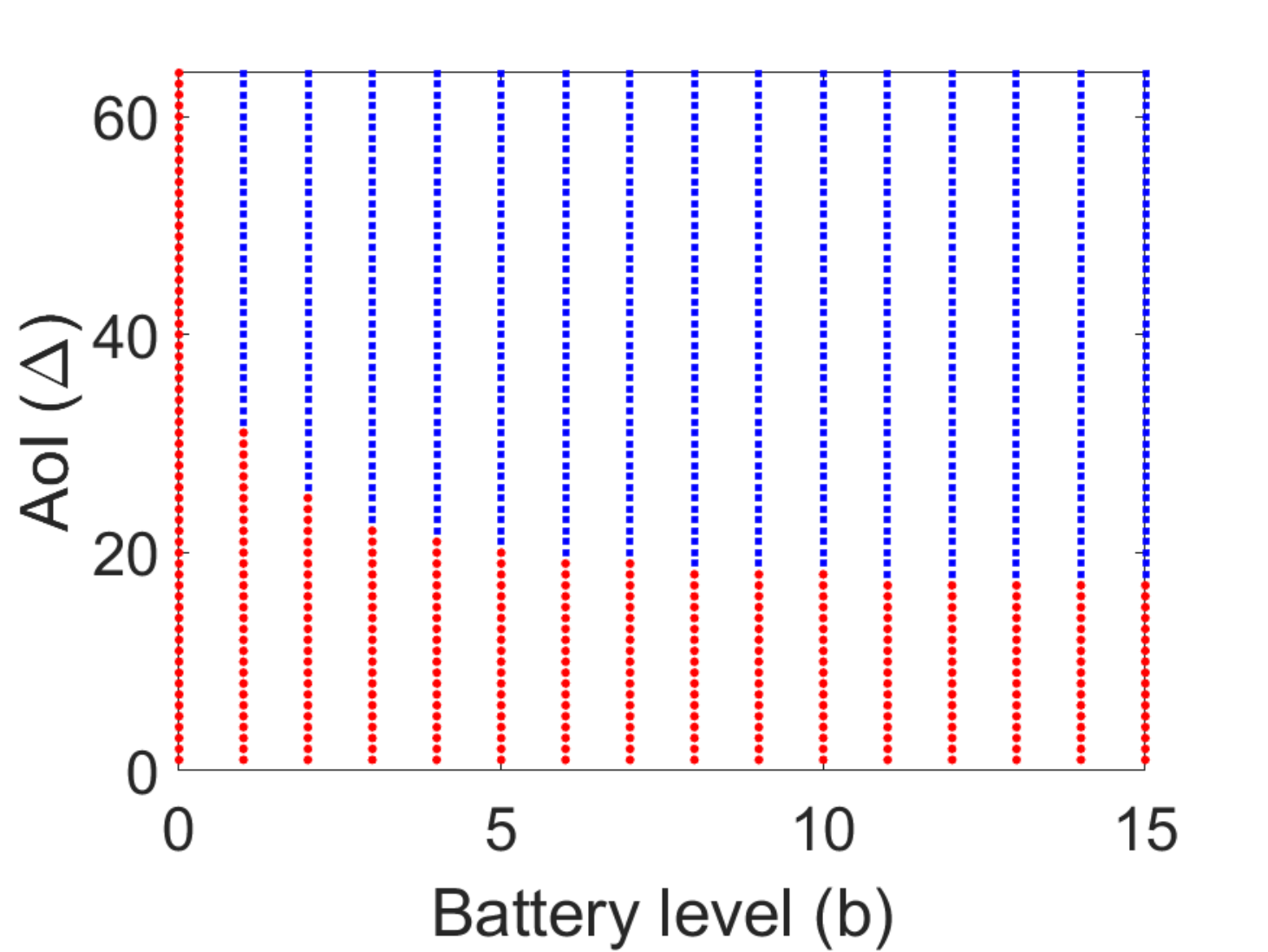}%
}
\subfigure [$\lambda_k = 0.10$]{%
\includegraphics[width=\fwstruct \columnwidth]{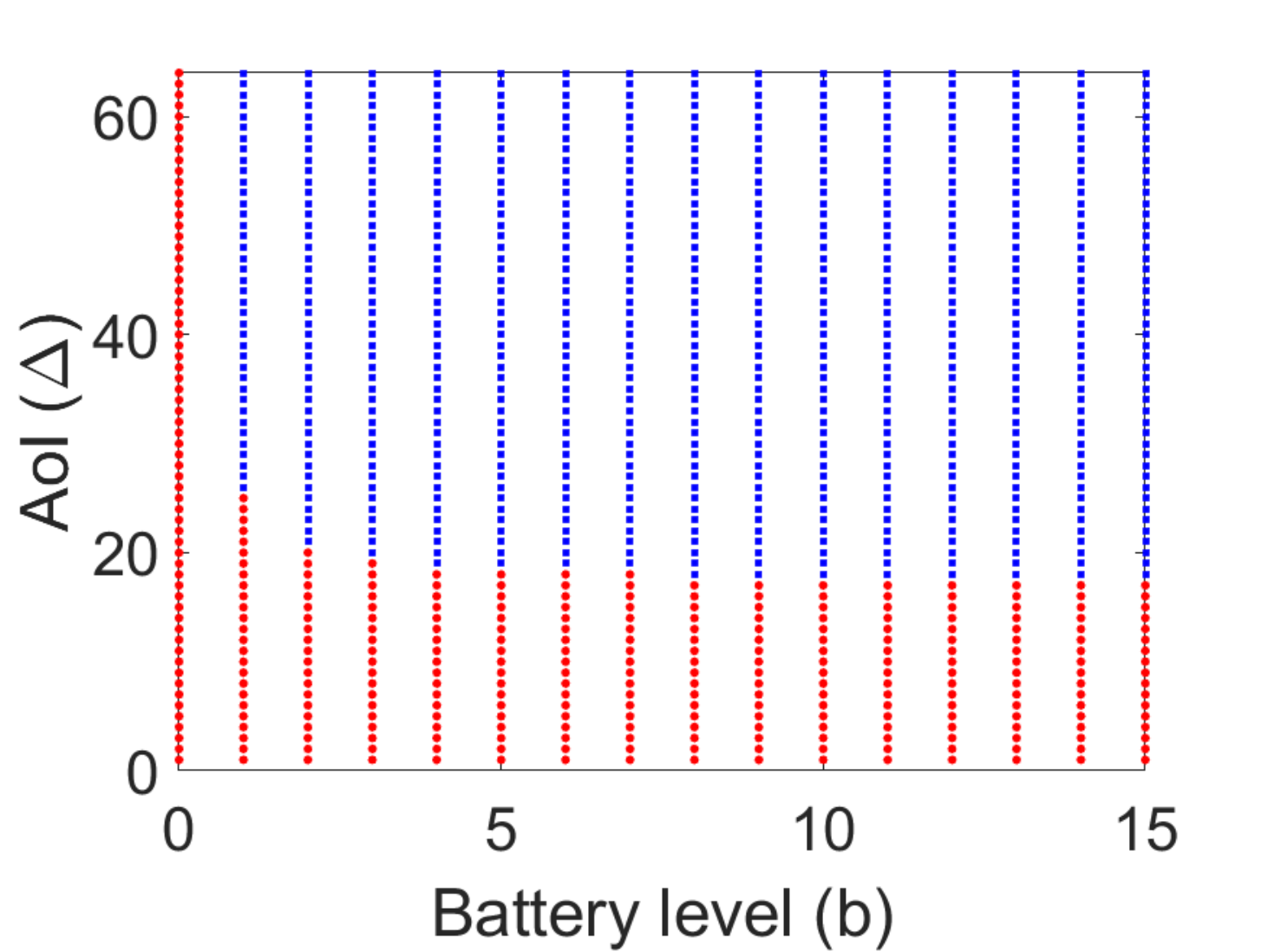}%
}
\vspace{-3mm}
\caption{Structure of an optimal policy for sensor~$k$ (i.e., $\pi^\star_{\R,\mu^*,k}$) in states $s = \{1,b, {\Delta}\}$ for different values of energy harvesting rate $\lambda_k$, where $M = 30$ and $p_{k,n} = 0.4$.}
% \vspace{-7mm}
\label{fig_structure_lambda}
% \end{figure*}

% \begin{figure*}[t]
\centering
\subfigure [$p_{k,n} = 0.1$]{%
\includegraphics[width=\fwstruct \columnwidth]{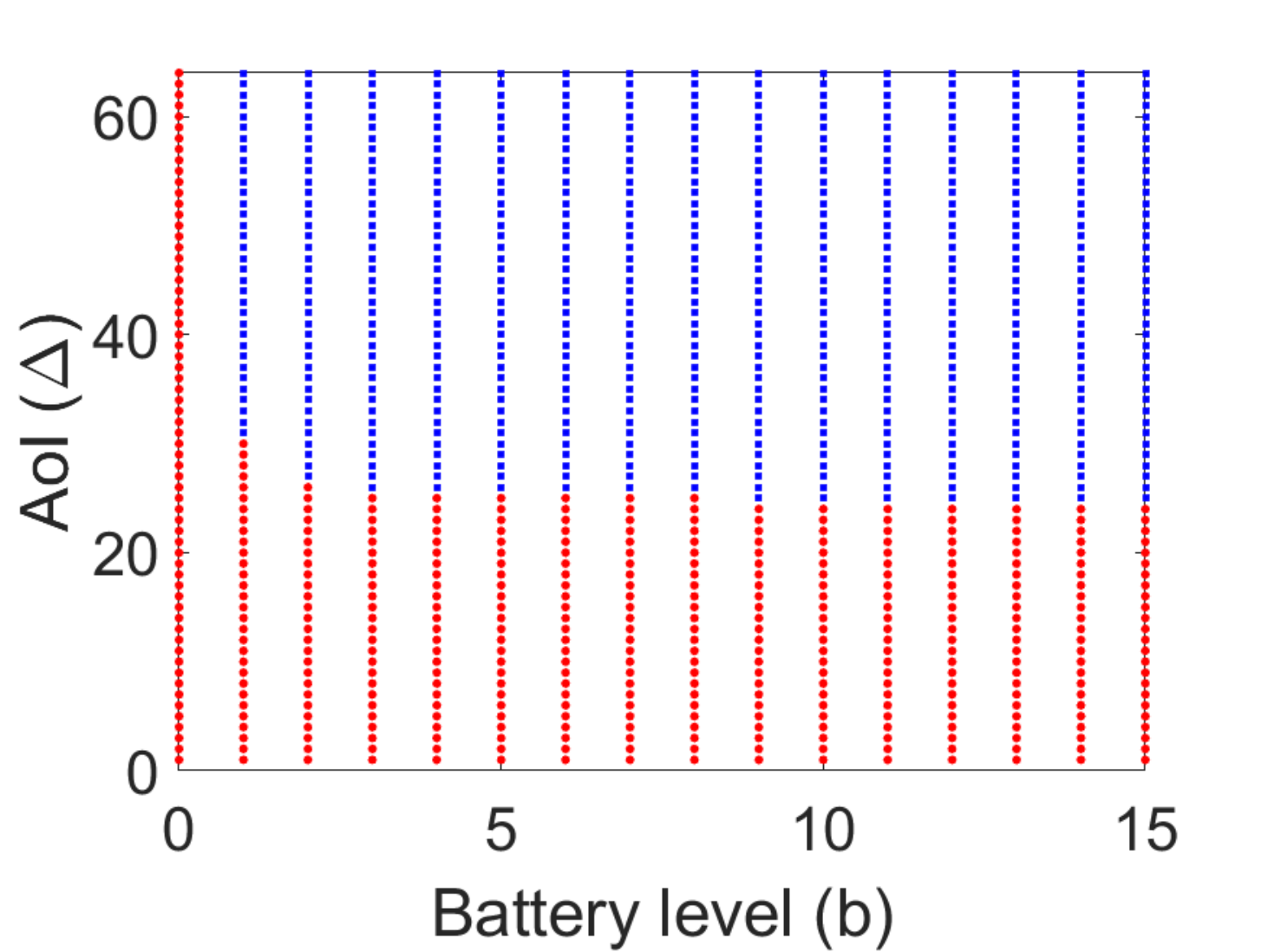}%
}
\subfigure [$p_{k,n} = 0.2$]{%
\includegraphics[width=\fwstruct \columnwidth]{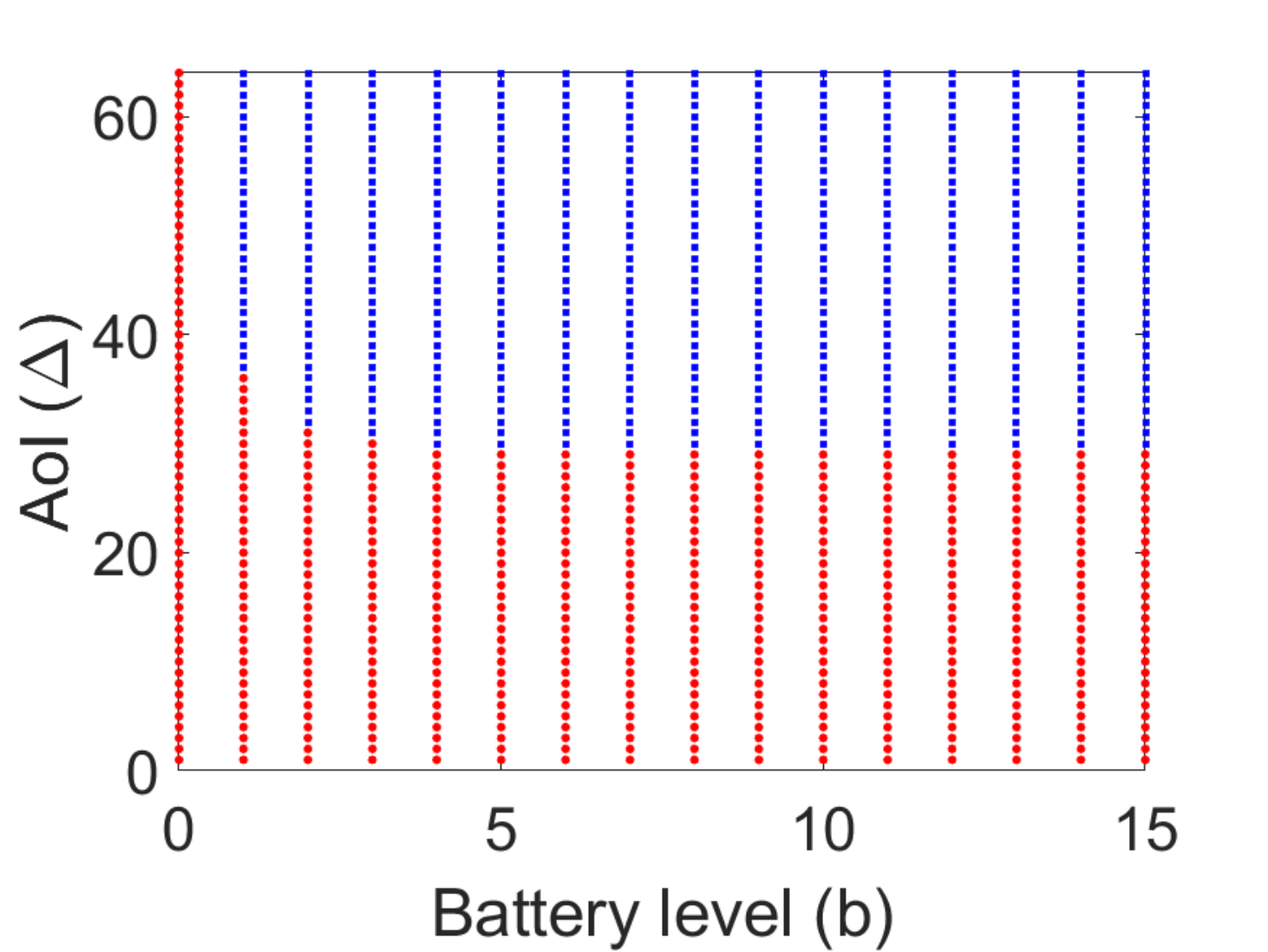}%
}
\subfigure [$p_{k,n} = 0.4$]{%
\includegraphics[width=\fwstruct \columnwidth]{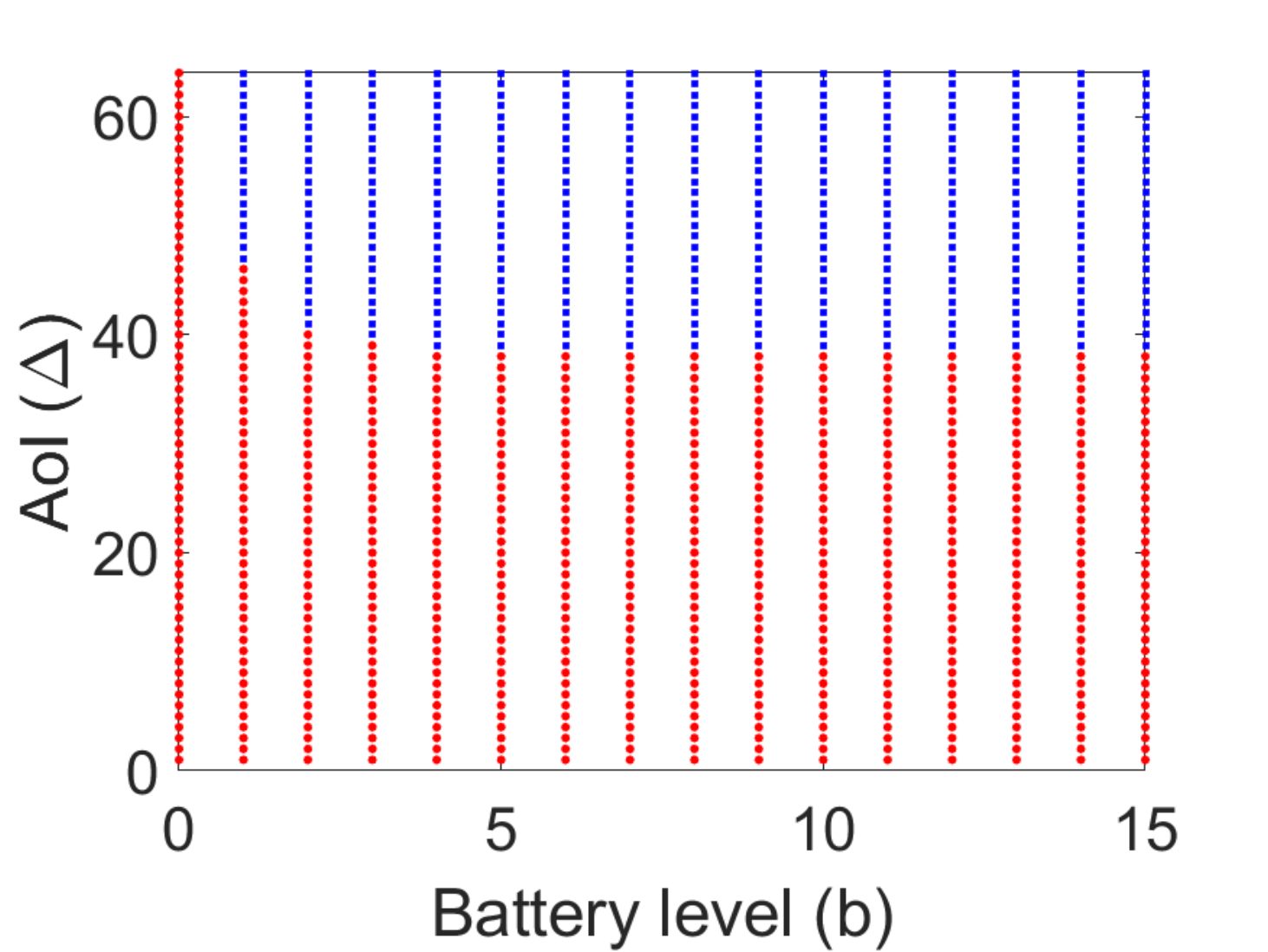}%
}
\subfigure [$p_{k,n} = 0.6$]{%
\includegraphics[width=\fwstruct \columnwidth]{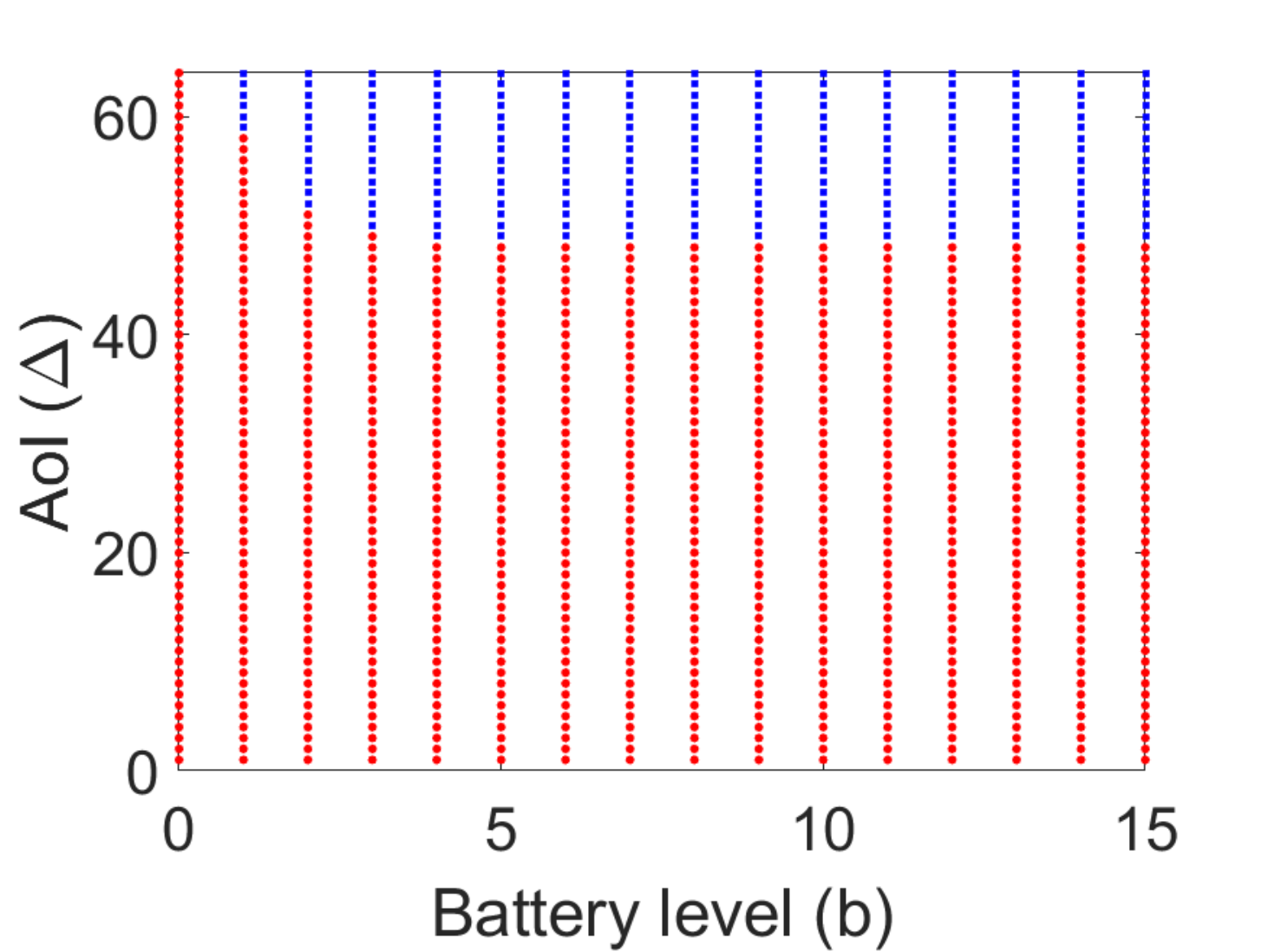}%
}
\vspace{-3mm}
\caption{Structure of an optimal policy for sensor~$k$ (i.e., $\pi^\star_{\R,\mu^*,k}$) in states $s = \{1,b, {\Delta}\}$ for different values  of requesting probability $p_{k,n}$, where $M = 15$ and $\lambda_k = 0.06$.}
\vspace{-7mm}
\label{fig_structure_pr}
\end{figure*}

% {Figure~\ref{fig_structure_r} shows that} 
From $\text{Fig.\ \ref{fig_structure_r}}$, we observe that the per-sensor policy $\pi^\star_{\R,\mu^*,k}$ has \textit{threshold-based} structure with respect to the number of requests $r$, battery level $b$, and  AoI $\Delta$. Consider a state $s = (1, 5, 50)$ in which $\pi^\star_{\R,\mu^*,k}(s) = 1$; then, by the threshold-based structure,  $\pi^\star_{\R,\mu^*,k}(\underline{s}) = 1$ for all states $\underline{s} = (r,b , \Delta )$, $r \geq 1$, $b \geq 5$, $\Delta \geq 50$. {Furthermore, $\text{Fig.\ \ref{fig_structure_r}}$ manifests the impact of considering the on-demand AoI (instead of conventional AoI) as the objective cost. Namely, since the cost function \eqref{persensor_cost} is (linearly) increasing with $r_k(t)$, the edge node has more incentive to command a sensor that is associated with a large number of requests. On the other hand, if there are no requests for $f_k$ (i.e., $r_k = 0$), the optimal action is not to command the sensor, regardless of the battery level and AoI, i.e., $\pi^\star_{\R,\mu^*,k}(0,b,\Delta) = 0$. This leads to energy saving for sensor $k$, which can be used later to serve the users with fresh measurements.} 

%In particular, since the cost function \eqref{persensor_cost} is (linearly) increasing with $r_k(t)$, the edge node has more incentive to command a sensor that is associated with a large number of requests. \yellow{Furthermore, $\text{Fig.\ \ref{fig_structure_r}(a)}$ shows that when there are no requests for $f_k$ (i.e., $r_k = 0$), the optimal action is not to command the sensor, regardless of the battery level and AoI, i.e., $\pi^\star_{\R,\mu^*,k}(0,b,\Delta) = 0$; in this case, the immediate cost \eqref{persensor_cost} becomes zero (i.e., $c_k(t)=0$) and the action $a_k(t) = 0$ leads to energy saving for sensor $k$, which can be used later to serve the users with fresh measurements.}

$\text{Fig.\ \ref{fig_structure_M}}$, $\text{Fig.\ \ref{fig_structure_lambda}}$, and $\text{Fig.\ \ref{fig_structure_pr}}$ depict the action under
% \yellow{policy}
$\pi^\star_{\R,\mu^*,k}$ in each state $s = \{1,b,\Delta \}$ for different values of the transmission budget $M$, energy harvesting rate $\lambda_k$, and request probability $p_{k,n}$, respectively. By comparing $\text{Figs.\ \ref{fig_structure_M}(a)--(d)}$, it is inferred that the command region enlarges as $M$ increases, because the edge node can command more sensors at each slot. Further, from a certain point onward ($M\!\geq\!25$), the command region does not increase anymore, because the energy limitation becomes dominant and  the edge node can not command more often.
By comparing $\text{Figs.\ \ref{fig_structure_lambda}(a)--(d)}$, it is inferred that the command region enlarges by increasing 
% the EH rate 
$\lambda_k$. This is expected because when a sensor harvests energy more often, it can send updates more often. As shown in $\text{Figs.\ \ref{fig_structure_pr}(a)--(d)}$, when the sensors are requested more often (i.e., $p_{k,n}$ increases), 
the command region shrinks; the edge node commands the sensor less to save its energy for the future requests.

\section{Conclusion}\label{sec_conclusions}
We investigated on-demand AoI minimization problem in a resource-constrained IoT network, where multiple users make on-demand requests to a cache-enabled edge node to send status updates about various random processes, each monitored by an EH sensor. We first modeled the problem as an MDP and proposed an iterative algorithm that obtains an optimal policy. Since the complexity of finding an optimal policy increases exponentially in the the number of sensors, we developed a low-complexity relax-then-truncate algorithm and then analytically showed that it is asymptotically optimal as the number of sensors goes to infinity. Numerical results illustrated that the relax-then-truncate algorithm significantly reduces the average cost (i.e., average on-demand AoI over all sensors and users) compared to a request-aware greedy policy and performs close to the optimal solution for moderate numbers of sensors.

\section{Acknowledgments}
The work has been financially supported in part by Infotech Oulu, the Academy of Finland (grant 323698), and Academy of Finland 6Genesis Flagship (grant 318927). M. Hatami would like to acknowledge the support of Nokia Foundation. The work of M. Leinonen has also been financially supported in part by the Academy of Finland (grant 340171 and 319485). The work of N. Pappas and Z. Chen have been supported in part by the Swedish Research Council (VR), ELLIIT, and CENIIT. Z. Chen would like to acknowledge the support of Knut and Alice Wallenberg (KAW) Foundation.

%%%%%%%%%%%%%%%%%%%%%%%%%%%%%%%%%%%%%%%%%%%%
%\section*{Appendix}
\begin{appendix}
\subsection{Proof of Proposition \ref{theorem_communicating_mdp}}\label{sec_appendix_theorem_communicating_mdp}
\begin{proof}
For any state ${\mathbf{s} = \left(s_1,\dots,s_K\right)}$, where $s_k = (r_k,b_k,\Delta_k)$, $k = 1,\dots,K$, we define the request vector $\mathbf{r} = (r_1,\dots,r_K)$, the battery vector $\mathbf{b} = (b_1,\dots,b_K)$, and the age vector ${\mathbf{\Delta} = (\Delta_1,\dots,\Delta_K)}$. Recall that at most $M$ sensors can send a fresh status update at each slot.
Thus, any state whose age vector has more than $M$ identical entries with values strictly less than $\Delta^{\mathrm{max}}$ is a transient state.
{We consider two non-transient states $\mathbf{s}, \mathbf{s}^\prime \in \mathcal{S}_{\mathrm{c}}$ and show that  $\mathbf{s}^\prime \triangleq (\mathbf{r}^\prime,\mathbf{b}^\prime,\mathbf{\Delta}^\prime)$ is accessible from $\mathbf{s} \triangleq (\mathbf{r},\mathbf{b},\mathbf{\Delta})$ under a stationary randomized policy $\pi$ in which, at each state $\mathbf{s}$, the edge node randomly selects an action $\mathbf{a}\in\mathcal{A}$ according to the discrete uniform distribution, i.e., $\pi(\mathbf{a}|\mathbf{s}) = \frac{1}{|\mathcal{A}|}$.}
{Let $\delta$ denote the largest element of the age vector $\mathbf{\Delta}^\prime$
% (associated with the state $\mathbf{s}^\prime$)
(i.e., $\max_{k}\Delta_k^\prime = \delta$). Let $\mathbf{e}_i$ denote a unit vector of length $K$  having a single $1$ at the $i$th entry and all other entries $0$. Let $\mathbf{e}_0$ denote a zero vector (i.e., all entries are $0$) of length $K$. We define a vector $\mathbf{a}_{i} = (a_{i,1},\ldots,a_{i,K})$ with elements $a_{i,k} = \mathds{1}_{\{\Delta_k^\prime = i,~ \Delta_k^\prime <\Delta^\mathrm{max}\}}$.
% , where $\mathds{1}_{\{\cdot\}}$ is the indicator function.
First, since the requests processes are independent from other variables in the system (e.g., actions), a state with a request vector $\mathbf{r}^\prime$ is accessible from any other state.
% in any length of slots. 
% Second, we show that $(\mathbf{b}^\prime,\mathbf{\Delta}^\prime)$ is accessible from $(\mathbf{b},\mathbf{\Delta})$. 
Second, realizing the actions $\mathbf{e}_1$ for $(b_1 - b^\prime_1)^+$  slots, $\mathbf{e}_2$ for $(b_2 - b^\prime_2)^+$ slots, $\dots$, $\mathbf{e}_K$ for $(b_K - b^\prime_K)^+$ slots, and $\mathbf{e}_0$ for $\tau = \max_{k} |b^\prime_k - b_k| - \sum_{k} (b_k - b^\prime_k)^+$ slots, the system reaches a state whose battery vector is $\mathbf{b}^\prime$ with a positive probability (w.p.p.). 
% \brown{Intuitively, we command those sensors ... to discharge the batteries and let the others stay idle to harvest some energy}.
{Note that, regardless of the actions happening next, the system reaches a state whose battery vector is still $\mathbf{b}^\prime$ w.p.p.}
% \red{Note that, realizing any action $\mathbf{a} = (a_1,\ldots,a_K)$, the system reaches a state whose battery vector is still $\mathbf{b}^\prime$ with probability $\prod_{k} \lambda_k^{a_k} (1-\lambda_k)^{(1-a_k)}$; if $a_k = 0$, $b_k$ does not change with probability $1-\lambda_k$ and if $a_k = 1$, $b_k$ does not change with probability $\lambda_k$.}
Third, realizing the consecutive actions $\mathbf{a}_{\delta}$, $\mathbf{a}_{\delta-1}$, $\ldots$, $\mathbf{a}_{1}$ leads the system reach a state whose age vector is $\mathbf{\Delta}^\prime$ w.p.p.
In summary, the system reaches a state with request vector $\mathbf{r}^\prime$, age vector $\mathbf{\Delta}^\prime$, and battery vector $\mathbf{b}^\prime$ w.p.p.. Thus, $\mathbf{s}^\prime$ is accessible from $\mathbf{s}$.}
\end{proof}\vspace{-4mm}
% \newpage

\subsection{Proof of Proposition \ref{theorem_communicating_per_sensorMDP}}\label{sec_appendix_theorem_communicating_per_sensorMDP}
\begin{proof}
We consider two arbitrary states $s, s^\prime \in \mathcal{S}_k $ and show that  $s^\prime = (r^\prime,b^\prime,\Delta^\prime)$ is accessible from $s = (r,b,\Delta)$ under a (per-sensor) stationary randomized policy $\pi_k$ in which, at each state $s$, the edge node randomly selects an action $a\in\mathcal{A}_k = \{0,1\}$ according to the discrete uniform distribution, i.e., $\pi_k(0|s) = \pi_k(1|s) = 1/2$. 
For the case where $b^\prime \geq b$, realizing the action $a = 0$ for $\tau = b^\prime - b+1$ consecutive slots leads to state ${(r^\prime,b^\prime+1,\min \{ \Delta+\tau,\Delta^{\mathrm{max}}\})}$ w.p.p.; then the action $a=1$ leads to state $(r^\prime,b^\prime,1)$ w.p.p., and subsequently action $a = 0$ for $\Delta^\prime - 1$ consecutive slots leads to state $s^\prime = (r^\prime,b^\prime,\Delta^\prime)$ w.p.p. Similarly, for the case 
where $b^\prime < b$, the action $a = 1$ for $\tau = b - b^\prime$ consecutive slots leads to state $(r^\prime,b^\prime,1)$ w.p.p., and subsequently $a = 0$ for $\Delta^\prime - 1$ consecutive slots leads to state $s^\prime = (r^\prime,b^\prime,\Delta^\prime)$ w.p.p.
\vspace{-5mm}
\end{proof}

% \newpage

\subsection{Proof of Lemma \ref{prop_struct_optimal_persensor_v}}\label{sec-apndix-prop_struct_optimal_persensor_v}
\begin{proof}
Here, we drop the unnecessary subscripts for the sake of notational convenience, e.g., $V_{\R,\mu,k}$ is simply shown by $V$.
To prove that $V$ is non-decreasing with respect to the AoI, we {consider} two states $s = (r,b,\Delta )$ and $\sbar = (r,b,\Deltabar)$, where $\Deltabar \geq \Delta$, and show that $V(\sbar) \geq V(s)$. Since the sequence $\{V^{(i)}(s)\}_{{i=1,2,\ldots}}$ converges to 
$V(s)$ for any initialization, it suffices to prove that $V^{(i)}(\sbar) \geq V^{(i)}(s)$, $\forall{i}$. We prove this by mathematical induction. 
{The  initial values is chosen arbitrarily, e.g., $V^{(0)}(s) = 0$ and $V^{(0)}(\sbar) = 0$, thus, the relation $V^{(i)}(\sbar) \geq V^{(i)}(s)$ holds for $i = 0$.} Assume that
$V^{(i)}(\sbar) \geq V^{(i)}(s)$ for some $i$; we need to prove that $V^{(i+1)}(\sbar) \geq V^{(i+1)}(s)$ as well. 
{Let us define $Q^{(i+1)}(s,a) \triangleq c_k(s,a)+ \mu a + \textstyle\sum_{s^\prime \in \mathcal{S}_k} \Pr(s^\prime|s,a) h^{(i)}(s^\prime),~s\in \mathcal{S}_k,a\in\mathcal{A}_k$. Thus, $V^{(i+1)}(s) = \min_{a\in\mathcal{A}_k} Q^{(i+1)}(s,a)$ (see \eqref{eq-per-sensor-v-itr}).}
Let us denote an optimal action taken in state $s$ at iteration $i = 1,2,\dots$ by $\pi^{(i)}(s)$, which is  given by $\pi^{(i)}(s) = \argmin_{a\in\mathcal{A}_k}Q^{(i)}(s,a)$. We have 
\begin{equation}
\begin{array}{ll}
& V^{(i+1)}(s) - V^{(i+1)}(\sbar) = \min_{a \in \mathcal{A}_k} Q^{(i+1)}(s,a) -  \min_{a \in \mathcal{A}_k} Q^{(i+1)}(\sbar,a) \notag\\ & =   Q^{(i+1)}(s,\pi^{(i+1)}(s)) -  Q^{(i+1)}(\sbar,\pi^{(i+1)}(\sbar)) \overset{(a)}{\leq} Q^{(i+1)}(s,\pi^{(i+1)}(\sbar)) -  Q^{(i+1)}(\sbar,\pi^{(i+1)}(\sbar)),\notag
\end{array}
\end{equation}
where $(a)$ follows from the fact that taking action $\pi^{(i+1)}(\sbar)$ in state $s$ is not necessarily optimal. We show that  ${Q^{(i+1)} (s,\pi^{(i+1)}(\sbar))-  Q^{(i+1)}(\sbar,\pi^{(i+1)}(\sbar)) \leq 0}$ for all possible actions ${\pi^{(i+1)} (\sbar)\in \{0,1\}}$. We present the proof for the case where $b < B_k$ and $\pi^{(i+1)}(\sbar) = 0$; the proof follows similarly for the other three cases, i.e., $b = B_k$ and $\pi^{(i+1)}(\sbar) = 0$, $b \geq 1$ and $\pi^{(i+1)}(\sbar) = 1$, and $b =0$ and $\pi^{(i+1)} (\sbar)= 1$. We have
% \begin{equation}\label{eq_proof_monotonicity_v}
% \begin{array}{ll}
%     &Q^{(i+1)}(s,0) -  Q^{(i+1)}(\sbar,0)  =  c_k(s,0) + \sum_{s^\prime \in \mathcal{S}_k} \Pr(s^\prime|s,0) V^{(i)}(s') - c_k(\sbar,0) - \notag \\ 
%     &\sum_{\sbar^\prime \in \mathcal{S}_k} \Pr(\sbar^\prime|\sbar,0) V^{(i)}(\sbar^\prime) \overset{(a)}{=} r\underbrace{\left(\min\{\Delta+1,\Delta^{\mathrm{max}}\}  - \min\{\Deltabar+1,\Delta^{\mathrm{max}}\} \right)}_{(b)\leq 0} + \notag
%     \end{array}
% \end{equation}
% \begin{equation}
% \begin{array}{ll}
%     & \sum_{n = 0}^N \sum_{l = 0}^{1} \Pr(r' = n) (l\lambda_k+(1-l)(1-\lambda_k))\times\notag \\ & \hspace{20mm}\underbrace{\left( V^{(i)}(n,b+l,\min\{\Delta+1,\Delta^{\mathrm{max}}\}) - V^{(i)}(n,b+l,\min\{\Deltabar+1,\Delta^{\mathrm{max}}\}) \right)}_{(c)\leq 0} \leq 0,\notag
%     \end{array}
% \end{equation}    
\begin{equation}\label{eq_proof_monotonicity_v}
\begin{array}{ll}
    &Q^{(i+1)}(s,0) -  Q^{(i+1)}(\sbar,0)  =  c_k(s,0) + \sum_{s^\prime \in \mathcal{S}_k} \Pr(s^\prime|s,0) V^{(i)}(s') - c_k(\sbar,0) - \notag \\ &\sum_{\sbar^\prime \in \mathcal{S}_k} \Pr(\sbar^\prime|\sbar,0) V^{(i)}(\sbar^\prime) \overset{(a)}{=} r\underbrace{\left(\min\{\Delta+1,\Delta^{\mathrm{max}}\}  - \min\{\Deltabar+1,\Delta^{\mathrm{max}}\} \right)}_{(b)\leq 0} + \notag \\ & \sum_{n = 0}^N \sum_{l = 0}^{1} \Pr(r' = n) (l\lambda_k+(1-l)(1-\lambda_k))\times\notag \\ & \hspace{20mm}\underbrace{\left( V^{(i)}(n,b+l,\min\{\Delta+1,\Delta^{\mathrm{max}}\}) - V^{(i)}(n,b+l,\min\{\Deltabar+1,\Delta^{\mathrm{max}}\}) \right)}_{(c)\leq 0} \leq 0,\notag
    \end{array}
\end{equation}    
% \end{align}
% \end{subequations}
where in step $(a)$ we used \eqref{eq_persensor_STP}--\eqref{eq_persensor_STP_AoI}, step $(b)$ follows from the assumption $\Delta \leq \Deltabar$, and step $(c)$ follows from the induction assumption.\vspace{-4mm}
\end{proof}

\subsection{Proof of Theorem \ref{theorem_AoI_threshold-persensor}}\label{sec-apndix-theorem_AoI_threshold-persensor}
\begin{proof}
% We remove the unnecessary subscripts from $\pi_{\R,\mu,k}^\star$, $q_{\R,\mu,k}^\star$, and $v_{\R,\mu,k}^\star$, and denote them by $\pi^\star$, $q^\star$, and $v^\star$, respectively, hereinafter.
Here, we drop the unnecessary subscripts for the sake of notational convenience, e.g., $V_{\R,\mu,k}$ is simply shown by $V$. {Let us define $Q(s,a) \triangleq c_k(s,a) + \mu a + \sum_{s' \in \mathcal{S}_k} \Pr(s'|s,a)  h(s')$. Thus, $V(s) = \min_{a\in\mathcal{A}_k}Q(s,a)$.}
Proving that $\pi^\star$ has a threshold-based structure with respect to the AoI is equivalent to showing the following: if the optimal action in state $s = (r,b,{\Delta})$ is $\pi^\star(s) = 1$, i.e., $Q(s,1)  - Q(s,0)\leq 0$, then for all states $\sbar = (r,b,\Deltabar)$ with $\Deltabar \geq {\Delta}$ the optimal action is also $\pi^\star(\sbar) = 1$, i.e., $Q(\sbar,1)  - Q(\sbar,0)\leq 0$. This is equivalent to showing that $Q(\sbar,1) - Q(\sbar,0) \leq Q(s,1) - Q(s,0)$. 
% $Q(s,1) - Q(\sbar,1) - Q(s,0) + Q(\sbar,0) \geq 0$. 
We present the proof for the case where $1 \leq b < B_k$; {for the other two cases, i.e., $b = 0$ and $b = B_k$, the proof follows similarly. We have}
\begin{equation}
    \begin{array}{lllll}
    & Q(s,1) - Q(\sbar,1) - Q(s,0) + Q(\sbar,0)\notag  = c_k(s,1) + \sum_{s' \in \mathcal{S}_k} \Pr(s'|s,1)  V(s') \notag \\& - c_k(\sbar,1) -  \sum_{\sbar^\prime \in \mathcal{S}_k} \Pr(\sbar^\prime|\sbar,1) V(\sbar^\prime) - c_k(s,0) - \sum_{s' \in \mathcal{S}_k} \Pr (s'|s,0) V(s')\notag \\&+c_k(\sbar,0) +  \sum_{\sbar^\prime \in \mathcal{S}_k} \Pr (\sbar^\prime|\sbar,0)  V(\sbar^\prime)\notag
     = r \underbrace{\left(\min\{\Deltabar+1,\Delta^{\mathrm{max}}\}  - \min\{\Delta+1,\Delta^{\mathrm{max}}\} \right)}_{(a)\geq 0} + \notag \\&\sum_{n=0}^{N} \sum_{l=0}^{1} \Pr(r^\prime = n) (l\lambda_k+(1-l)(1-\lambda_k)) \times \notag \\ & \hspace{30mm}\underbrace{\left( V(n,b+l,\min\{\Deltabar+1,\Delta^{\mathrm{max}}\}) - V(n,b+l,\min\{\Delta+1,\Delta^{\mathrm{max}}\})\right)}_{(b)\geq 0}
     \geq 0,\notag
    \end{array}
\end{equation}
where step $(a)$ follows from the assumption $\Delta \leq \Deltabar$ and step $(b)$ follows from {Lemma}~\ref{prop_struct_optimal_persensor_v}.
\end{proof}

% \newpage

\subsection{Proof of Theorem \ref{Theorem_upperbound_deviation}}\label{sec-apndix-Theorem_upperbound_deviation}
\begin{proof}
% Recall that the average cost obtained by an optimal relaxed policy $\pi^\star_\R$ is a lower bound for the average cost obtained by an optimal policy $\pi^\star$ (see \eqref{eq_lowerboundinequality}). Moreover, the proposed relax-then-truncate policy $\tilde{\pi}$ is a sub-optimal solution for  (\textbf{P1}). Therefore, we have 
% \begin{equation}\label{eq_policy_order}
%     \bar{C}(\pi^\star_\R) \leq  \bar{C}(\pi^\star) \leq \bar{C}(\tilde{\pi}).
% \end{equation}
Let $\mathcal{T}(t) \subset \mathcal{X}(t)$ denote the set of \textit{truncated} sensors at slot $t$, i.e., the sensors that are not commanded under the relax-then-truncate policy $\tilde{\pi}$, given that they are commanded under policy $\pi_{\mathrm{R}}^\star$. 
By the truncation procedure, if ${|\mathcal{X}(t)|> M}$, $M$ sensors are chosen randomly (uniform) from the set $\mathcal{X}(t)$ and commanded (i.e., ${|\mathcal{X}(t)| - M}$ sensors are not commanded). Thus, the probability that sensor $k$ belongs to $\mathcal{T}(t)$ is $\mathds{1}_{\{|\mathcal{X}(t)|>M|\}}\left(\frac{|\mathcal{X}(t)|-M}{|\mathcal{X}(t)|}\right)$. At each slot, the additional per-sensor cost under $\tilde{\pi}$ compared to $\pi^\star_{\R}$ is no more than $N \Delta^{\mathrm{max}}$ (see \eqref{persensor_cost}). Therefore, the expected additional cost over all sensors under $\tilde{\pi}$ compared to $\pi^\star_{\R}$ is upper bounded by
\begin{equation}\label{eq_extra_penalized}
     {\sum_{k = 1}^{K}} \underbrace{\mathds{1}_{\{\mathcal{X}(t) > M\}} \frac{|\mathcal{X}(t)| - M}{|\mathcal{X}(t)|}}_{\Pr(k \in \mathcal{T}(t))} N\Delta^{\mathrm{max}} = NK\Delta^{\mathrm{max}} {\frac{(|\mathcal{X}(t)| - M)^{+}}{|\mathcal{X}(t)|}},
\end{equation}
where $(\cdot)^+ \triangleq \max \{0,\cdot\}$.

% Based on the above discussions, 
We introduce the following (penalized) {strategy} $\hat{\pi}_{\mathrm{R}}$: at each slot, command the sensors based on $\pi^\star_{\R}$ but add a penalty $NK\Delta^{\mathrm{max}} \frac{(|\mathcal{X}(t)| - M)^{+}}{|\mathcal{X}(t)|}$ to the cost over all sensors (see \eqref{eq_extra_penalized}).
It is clear that the average cost obtained under $\hat{\pi}_{\mathrm{R}}$ is not less than that obtained by $\tilde{\pi}$, i.e., $\bar{C}_{\tilde{\pi}} \leq \bar{C}_{\hat{\pi}_{\mathrm{R}}}$. Also, recall from \eqref{eq_lowerboundinequality} that the average cost obtained under policy $\pi^\star_\R$ is a lower bound for the average cost obtained by an optimal policy $\pi^\star$, i.e., $\bar{C}_{\pi^\star_\R} \leq  \bar{C}_{\pi^\star}$. Moreover, policy $\tilde{\pi}$ is a sub-optimal solution for  (\textbf{P1}), i.e., $\bar{C}_{\pi^\star} \leq \bar{C}_{\tilde{\pi}}$.  Therefore, we have 
\begin{equation}\label{eq_policy_order_final}
\bar{C}_{\pi^\star_\R} \leq  \bar{C}_{\pi^\star} \leq \bar{C}_{\tilde{\pi}} \leq \bar{C}_{\hat{\pi}_\R}.
\end{equation}
% Let us define $\Delta_{\mathrm{max}} \doteq \max_{k \in \mathcal{K}} \Delta_{k,\mathrm{max}}$, and $(\cdot)^{+} \doteq \max\{{0,\cdot}\}$.

Using \eqref{eq_policy_order_final}, the difference between the average cost obtained by the proposed relax-then-truncate policy $\tilde{\pi}$ and the average cost obtained by an optimal policy $\pi^\star$ is upper bounded as
\begin{equation}
\begin{array}{ll}
    \bar{C}_{\tilde{\pi}} - \bar{C}_{\pi^\star} &\overset{(a)}{\leq} \bar{C}_{\hat{\pi}_\R} - \bar{C}_{\pi^\star_\R}
    \\& =\lim_{T\rightarrow\infty} \frac{1}{NKT} \sum_{t=1}^{T}\mathbb{E}_{\pi^\star_\R}\left[NK\Delta^{\mathrm{max}} \frac{(|\mathcal{X}(t)| - M)^{+}}{|\mathcal{X}(t)|} \right]
    \\& \overset{(b)}{\leq} \frac{\Delta^{\mathrm{max}}}{M} \lim_{T\rightarrow\infty}\frac{1}{T} \sum_{t=1}^{T} \mathbb{E}_{\pi^\star_\R}\left[(|\mathcal{X}(t)| - M)^+\right]
    \\& \overset{(c)}{\leq} \frac{\Delta^{\mathrm{max}}}{M}  \lim_{T\rightarrow\infty}\frac{1}{T}\sum_{t=1}^{T} \mathbb{E}_{\pi^\star_\R}\left[(|\mathcal{X}(t)| - \mathbb{E}_{\pi^\star_\R}[|\mathcal{X}(t)|])^+\right]
    \\& \overset{(d)}{\leq} \frac{\Delta^{\mathrm{max}}}{M}  \lim_{T\rightarrow\infty}\frac{1}{T}\sum_{t=1}^{T} \mathbb{E}_{\pi^\star_\R}\Big[\big||\mathcal{X}(t)| - \mathbb{E}_{\pi^\star_\R}[|\mathcal{X}(t)|]\big|\Big]
    \\ & = \frac{\Delta^{\mathrm{max}}}{M}  \lim_{T\rightarrow\infty}\frac{1}{T}\sum_{t=1}^{T} \mathrm{MAD}(|\mathcal{X}(t)|)
\end{array}
\end{equation}
where $(a)$ follows from \eqref{eq_policy_order_final}, $(b)$ follows from $\frac{(|\mathcal{X}(t)| - M)^+}{|\mathcal{X}(t)|} \leq \frac{(|\mathcal{X}(t)| - M)^+}{M}$, $(c)$ follows from $\mathbb{E}_{\pi^\star_{\R}}[|\mathcal{X}(t)|] \leq M$, {for sufficiently large $t$}, and $(d)$ follows from $[\cdot]^{+} \leq |\cdot|$.
\vspace{-4mm}
\end{proof}

% \newpage

% Blank
% \newpage

% Blank

% \newpage

\subsection{Proof of Lemma \ref{lemma_MAD_normal}}\label{sec-apndix-lemma_MAD_normal}
\vspace{-6mm}
% \begin{proof}
% We have
\begin{equation}
    \begin{array}{lll}
    &\mathrm{MAD}(X) = \mathbb{E}[|X - \nu|] = \int_{-\infty}^{\infty} |x-\nu| \frac{1}{\sigma\sqrt{2\pi}}e^{-\frac{1}{2}\left(\frac{x-\nu}{\sigma}\right)^2} \mathrm{d}x 
    = \int_{-\infty}^{\nu} (\nu-x) \frac{1}{\sigma\sqrt{2\pi}}e^{-\frac{1}{2}\left(\frac{x-\nu}{\sigma}\right)^2} \mathrm{d}x \\& + \int_{\nu}^{\infty} (x-\nu) \frac{1}{\sigma\sqrt{2\pi}}e^{-\frac{1}{2}\left(\frac{x-\nu}{\sigma}\right)^2} \mathrm{d}x
    = \sqrt{\frac{2}{\pi}}\sigma \underbrace{\int_{0}^{\infty} y e^{-\frac{1}{2}y^2} \mathrm{d}y}_{=1}
    % \int_{0}^{\infty} y e^{-\frac{1}{2}y^2} \mathrm{d}y
    = \sqrt{\frac{2}{\pi}}\sigma.
\end{array}\notag\vspace{-7mm}
\end{equation}
% \end{proof}
% \newpage
\subsection{Proof of Lemma \ref{lemma_MAD}}\label{sec-apndix-lemma_MAD}
\begin{proof}
The cardinality of set $\mathcal{X}(t)$ (i.e., the set of sensors that are commanded under $\pi_{\R}^\star$) can be written as ${|\mathcal{X}(t)| = \sum_{k = 1}^{K} a_k(t)}$, where $a_k(t)\in \{0,1\}$, $k\in \mathcal{K}$, are $K$ independent binary random variables.
% \blue{the sum of independent random variables as ${|\mathcal{X}(t)| = \sum_{k = 1}^{K} a_k(t)}$.}
Let $\omega_k(t)$ be the probability that sensor $k$ is commanded at slot $t$ under policy $\pi_{\R}^\star${, i.e., $\omega_k(t) \triangleq \Pr(a_k(t) = 1)$.} 
% {This mean $a_k(t)\in \{0,1\}$ is a Bernoulli random variable with mean $\omega_k(t)$.}
We define a random variable $Z(t) \triangleq \frac{|\mathcal{X}(t)| - \sum_{k}\omega_k(t)}{\sqrt{\sum_{k}\omega_k(t)(1-\omega_k(t))}}$. We have 
\begin{equation}\label{eq_MAD_ineq_Z}
\begin{array}{ll}
    \mathrm{MAD}\left(Z(t)\right) &= \mathrm{MAD}\left(\frac{|\mathcal{X}(t)| - \sum_{k}\omega_k(t)}{\sqrt{\sum_{k}\omega_k(t)(1-\omega_k(t))}}\right) \overset{(a)}{=} \mathrm{MAD}\left(\frac{|\mathcal{X}(t)|}{\sqrt{\sum_{k}\omega_k(t)(1-\omega_k(t))}}\right) \\ & \overset{(b)}{\geq} \mathrm{MAD}\left(\frac{|\mathcal{X}(t)|}{\sqrt{K/4}}\right) {\geq} \mathrm{MAD}\left(\frac{|\mathcal{X}(t)|}{\sqrt{K}}\right),
    \end{array}
\end{equation}
where $(a)$ follows because the MAD does not change by adding a constant to all values of the variable (similar to variance) and $(b)$ follows from $\textstyle\sum_{k=1}^{K}\omega_k(t)(1-\omega_k(t)) \leq \frac{K}{4}$.
% ; the equality holds when ${\omega_k(t) = \frac{1}{2}}$.}
% \brown{where $(b)$ follows from MAD (similar to variance) is invariant with respect to changes in a location parameter. That is, if a constant is added to all values of the variable, the MAD is unchanged.}
% Let $\omega_k(t)$ be the probability that sensor $k$ is commanded at slot $t$ under policy $\pi_{\R}^\star$, i.e., $\Pr(a_k(t) = 1) \triangleq \omega_k(t)$. This mean $a_k(t)\in \{0,1\}$ is a Bernoulli random variable with mean $\omega_k(t)$.
% Note that the random variables $a_k(t)$, $k\in \mathcal{K}$, are independent across $k$.
% The random variable $|\mathcal{X}(t)|$ is the sum of independent random variables, and thus, 

% \begin{equation}
%     \lim_{K \to \infty} \frac{\sum_{k}\mathbb{E}[|a_k(t) - \omega_k(t)|^4]}{{[\sum_{k}\omega_k(t)(1-\omega_k(t)]}^2} = \lim_{K \to \infty} \frac{\sum_{k}[(1-\omega_k(t))^5+(\omega_k(t))^5]}{{[\sum_{k}\omega_k(t)(1-\omega_k(t)]}^2} = \lim_{K \to \infty} \frac{\mathcal{O}(K)}{\mathcal{O}(K^2)} = 0
% \end{equation}

By the Lyapunov central limit theorem \cite[Theorem~27.3]{billingsley1995probability},
% \yellow{the random variable} 
$Z(t)$ converges in distribution to a standard normal distribution, i.e., $ Z(t) \sim \mathcal{N}(0,1)$, as $K$ goes to infinity. Thus, we have
\begin{equation}
    \lim_{K\to \infty} \mathrm{MAD}\left(\frac{|\mathcal{X}(t)|}{\sqrt{K}}\right) \overset{(a)}{\leq} \lim_{K\to \infty} \mathrm{MAD}(Z(t)) \overset{(b)}{=} \sqrt{\frac{2}{\pi}} \leq 1,
\end{equation}
where $(a)$ follows from \eqref{eq_MAD_ineq_Z} and $(b)$ follows from Lemma~\ref{lemma_MAD_normal}.
% \begin{align}
%     \mathrm{MAD}(|\mathcal{X}(t)|) = \textstyle\sqrt{\frac{2}{\pi}}\sqrt{\textstyle\sum_{k=1}^{K}\omega_k(t)(1-\omega_k(t))} \overset{(a)}{\leq} \sqrt{\frac{K}{2\pi}},
%     % =  \mathcal{O}\left(\sqrt{K}\right)
% \end{align}
% where $(a)$ follows from $\textstyle\sum_{k=1}^{K}\omega_k(t)(1-\omega_k(t)) \leq \frac{K}{4}$; the equality holds when ${\omega_k(t) = \frac{1}{2}}$.
\end{proof}

% \newpage

% Blank

% \newpage

\subsection{Proof of Theorem \ref{theorem-asymptotically-opt}}\label{sec-apndix-theorem-asymptotically-opt}
\begin{proof}
We have
% \begin{subequations}
\begin{equation}
    \lim_{K\rightarrow\infty} \bar{C}_{\tilde{\pi}} - \bar{C}_{\pi^\star} \overset{(a)}{\leq} \lim_{K\rightarrow\infty}\left[\frac{\Delta^{\mathrm{max}}}{\Gamma \sqrt{K}} \lim_{T\rightarrow\infty}\frac{1}{T}\sum_{t=1}^{T} \mathrm{MAD}\left(\frac{|\mathcal{X}(t)|}{\sqrt{K}}\right)\right] \overset{(b)}{\leq} \lim_{K\to\infty}
    % \frac{\Delta^{\mathrm{max}}}{\Gamma K} \sqrt{\frac{K}{2\pi}} =
    \frac{\Delta^{\mathrm{max}}}{\Gamma\sqrt{K}} = 0, \notag
\end{equation}
% \end{subequations}
where $(a)$ follows from Theorem~\ref{Theorem_upperbound_deviation} and $M = \Gamma K$, and $(b)$ follows from Lemma~\ref{lemma_MAD}.
\end{proof}

\end{appendix}

\bibliographystyle{IEEEtran}
% \mybibliography
%	\bibliography{conf_short,IEEEabrv.bib,jour_short,Bibliography}
\begin{spacing}{1.33} %% COMPRESSION TWEAK
\bibliography{Bib/conf_short,Bib/IEEEabrv,Bib/Bibliography}
\end{spacing}
\end{document}